\newcommand{\pcmq}{\mbox{cm$^{-2}$}}
\newcommand{\psec}{\mbox{s$^{-1}$}}
\newcommand{\psr}{\mbox{sr$^{-1}$}}
\newcommand{\pmev}{\mbox{MeV$^{-1}$}}

\newcommand{\funit}{\mbox{ph~\pcmq~\psec}}
\newcommand{\fmev}{\mbox{ph~\pcmq~\psec~\pmev}}

\newcommand{\fmevster}{\mbox{ph~\pcmq~\psec~\pmev~\psr}}
\newcommand{\eunit}{\mbox{erg~\pcmq~\psec}}

\def\degrees{\ensuremath{^\circ}}
\def\al26{\mbox{$^{26}$\hspace{-0.2em}Al}}

\documentclass{aa}  

\usepackage{graphicx}
\usepackage{txfonts}
\usepackage[draft]{hyperref}
\usepackage{amsmath}
\usepackage{natbib}
\usepackage[switch]{lineno}
\usepackage{xspace}
\usepackage{xcolor} 
\usepackage{wasysym}
\DeclareMathOperator{\sgn}{sgn}
\begin{document} 

\bibpunct{(}{)}{;}{a}{}{,} 

\title{COMPTEL data analysis using GammaLib and ctools}

\author{
J. Kn\"odlseder\inst{1} \and
W. Collmar\inst{2} \and
M. Jarry\inst{1} \and
M. McConnell\inst{3,4}
}
\institute{
Institut de Recherche en Astrophysique et Plan\'etologie, Universit\'e de Toulouse, CNRS, CNES, 
9 avenue Colonel Roche, 31028 Toulouse, Cedex 4, France
\and
Max-Planck-Institut f\"ur extraterrestrische Physik, Postfach 1603, 85740 Garching, Germany
\and
University of New Hampshire, Space Science Center, Durham, NH 03824, U.S.A.
\and
Southwest Research Institute, Dept. of Earth, Oceans, and Space, Durham, NH. 03824, U.S.A.
}

\date{Received 20 April, 2022; Accepted 3 June, 2022}

\abstract{
More than 20 years after the end of NASA's Compton Gamma-Ray Observatory mission, the data 
collected by its Imaging Compton Telescope (COMPTEL) still provide the most comprehensive and 
deepest view of our Universe in MeV gamma rays.
While most of the COMPTEL data are archived at NASA's High Energy Astrophysics Science Archive 
Research Center (HEASARC), the absence of any publicly available software for their analysis 
means the data cannot benefit from the scientific advances made in the field of gamma-ray astronomy at 
higher energies.
To make this unique treasure again accessible for science, we developed open source software
that enables a comprehensive and modern analysis of the archived COMPTEL telescope data.
Our software is based on a dedicated plug-in to the GammaLib library, a community-developed toolbox
for the analysis of astronomical gamma-ray data.
We implemented high-level scripts for building science analysis workflows in ctools, a 
community-developed gamma-ray astronomy science analysis software framework.
We describe the implementation of our software and provide the underlying algorithms.
Using data from the HEASARC archive, we demonstrate that our software reproduces derived data 
products that were obtained in the past using the proprietary COMPTEL software.
We furthermore demonstrate that our software reproduces COMPTEL science results published
in the literature.
This brings the COMPTEL telescope data back into life, allowing them to benefit from recent 
advances in gamma-ray astronomy, and gives the community a means to unveil its still hidden 
treasures.
}

\keywords{
methods: data analysis --
gamma rays: general --
stars: neutron --
binaries: general --
nucleosynthesis
}

\maketitle

\section{Introduction}
\label{sec:intro}

The Imaging Compton Telescope (COMPTEL) was one of the four telescopes aboard NASA's Compton
Gamma-Ray Observatory (CGRO) satellite, which was operated in low Earth orbit from April 1991 to 
May 2000 \citep{schoenfelder1993}.
The telescope scrutinised the gamma-ray sky in the 0.75 -- 30 MeV energy range, and its observations
still present the most sensitive survey of the MeV sky ever performed.
Since the end of the CGRO mission, our view of the gamma-ray Universe has evolved considerably,
including observations of very high-energy gamma rays from a plethora of objects and the discovery
of new source classes, such as novae, gamma-ray binaries, star-forming regions, globular clusters, and
starburst galaxies.
Today, the latest catalogue based on data from the Fermi Large Area Telescope lists more than 5000
gamma-ray sources \citep{abdollahi2020}, and exploring the COMPTEL archive in light of this
knowledge has the potential to provide new insights into the MeV sky \citep{collmar2014}.

While most of the COMPTEL data accumulated during the nine-year mission are stored at NASA's High 
Energy Astrophysics Science Archive Research Center (HEASARC)\footnote{
  \url{https://heasarc.gsfc.nasa.gov/docs/cgro/archive/}},
no publicly available software had existed to analyse these data.
During CGRO operations the COMPTEL data were analysed using the COMPTEL Processing and 
Analysis Software System (COMPASS) \citep{devries1994}, but this system was only available at the 
COMPTEL collaboration institutes and was decommissioned after the end of the mission.
Nevertheless, a few science results were still obtained from COMPTEL data after the end of the mission
thanks to efforts at some COMPTEL collaboration institutes to keep Linux ports of the COMPASS
data-analysis system alive \citep{strong2019,coleman2020}.

In order to preserve the capability to analyse the unique COMPTEL archive and to make COMPTEL data
analysis accessible to the astronomical community at large, we have implemented a dedicated COMPTEL
plug-in for the GammaLib library, a community-developed toolbox for the analysis of astronomical
gamma-ray data \citep{gammalib2011}.
The plug-in enables a reanalysis of COMPTEL data from the HEASARC archive, including the computation
of the COMPTEL instrument response functions that until now had not been publicly available.
The plug-in benefits from generic data analysis capabilities provided by GammaLib,
including the joint maximum-likelihood fitting of multiple energy bands and the combination of COMPTEL
data with data from other gamma-ray telescopes, enabling novel analysis approaches that go beyond
the capabilities of the COMPASS system.

We furthermore extended the ctools gamma-ray astronomy science analysis software \citep{ctools2016}
with the addition of several dedicated Python scripts. These scripts provide basic building blocks that each perform 
well-defined COMPTEL data analysis tasks.
These building blocks can be combined to create flexible COMPTEL data analysis workflows for
the exploration of the MeV sky.

In this paper we present the GammaLib plug-in and the ctools COMPTEL science analysis scripts,
including the algorithms that were implemented, so that any science analysis that makes use of 
the software has a solid reference.
We demonstrate that COMPTEL data analysis products derived using GammaLib and ctools,
such as data cubes and response functions, are identical to equivalent products produced with 
the COMPASS software.
We show that the background model that is implemented in GammaLib and ctools provides a reliable
description of the COMPTEL instrumental background, and we discuss potential biases and
limitations.
Finally, we apply the software to a number of science cases and demonstrate that our software
reproduces results published in the literature.
All the algorithms and results presented in this paper were produced with GammaLib and
ctools version 2.0.0.
The analysis scripts and data presented in this work are available for download from
Zenodo.\footnote{\url{https://doi.org/10.5281/zenodo.6462842}}

\section{The COMPTEL telescope}
\label{sec:comptel}

Before describing the software implementation, we recall some COMPTEL fundamentals that are
important for understanding the remainder of the paper.
COMPTEL was an imaging spectrometer that was sensitive to gamma rays in the energy range
0.75 -- 30 MeV with an energy-dependent energy and angular resolution of $5-8\%$ (full width half 
maximum) and $1.7\degrees - 4.4\degrees$ (full width half maximum), respectively.
The telescope had a large field of view of about one steradian and was sensitive to detected
gamma-ray sources at a $1-30$~MeV flux level of $10^{-9}$ \eunit\ within an observing time of $10^6$ s
\citep{schoenfelder1993}.

COMPTEL was composed of two modular detector layers D1 and D2, separated by $158$ cm, where
an incident gamma-ray photon was first Compton scattered in one of the 7 modules of D1 and then
eventually interacted in one of the 14 modules of D2.
The Compton scatter direction $(\chi,\psi)$ was obtained from the interaction locations in D1 and D2,
the Compton scattering angle $\bar{\varphi}$ was computed from the measured energy deposits $E_1$
and $E_2$ in D1 and D2, respectively, using
\begin{equation}
\bar{\varphi} = \arccos \left( 1 - m_e c^2 \left( \frac{1}{E_2} - \frac{1}{E} \right) \right)
,\end{equation}
where $E = E_1 + E_2$ is the total energy deposit in the detectors, and $m_e c^2=0.511$ MeV is the
rest energy of the electron.

The detector layers were surrounded by an active anti-coincidence shield, composed of two veto domes 
for each layer, that allowed for the reduction of instrumental background\footnote{
  Throughout this paper we denote by instrumental background all events that are not related to
  celestial gamma-ray emission.}
due to charged particles.
A further strong discriminator of instrumental background was the time of flight (ToF) measurement, which 
is the time difference between the interactions in D1 and D2 (cf.~Sect.~\ref{sec:cube}).

COMPTEL data were often analysed in four standard energy bands, covering $0.75-1$, $1-3$, $3-10$
and $10-30$~MeV, and if not otherwise stated, the same energy bands will be adopted in the
present paper.
Event selection parameters used in this paper are discussed in Sect.~\ref{sec:cube}.

\section{Software implementation}
\label{sec:implementation}

\subsection{GammaLib plug-in}

The COMPTEL support was implemented in GammaLib as an instrument plug-in that provides
instrument-specific implementations of abstract virtual C++ base classes defining the data format
and the instrument response functions.
The plug-in also comprises wrapper functions providing access to all COMPTEL-specific C++ classes 
through the {\tt gammalib} Python module.
All COMPTEL-specific classes begin with the letters {\tt GCOM}.

\subsection{COMPTEL data}

COMPTEL data are available in the HEASARC archive and are grouped by so-called 
viewing periods with typical durations of two weeks during which the CGRO satellite had a stable 
pointing.
In total, the HEASARC archive contains 255 exploitable viewing periods, which is about $71\%$ 
of the total number of 359 viewing periods that were executed by COMPTEL.\footnote{
  Some viewing periods in the HEASARC archive are unreadable, and data from the end
  of the mission were not archived. Details about issues encountered in the HEASARC COMPTEL 
  data archive are provided in Appendix \ref{app:heasarc}.}
All data are provided in FITS format.

The input data that are relevant to GammaLib for a given viewing period are event files, good time 
intervals, and orbit aspect data.
The orbit aspect data comprise satellite orbit and telescope pointing information that are needed for coordinate
transformations.
COMPTEL data types are identified by a three-letter code, which is 
{\tt EVP} for event files,
{\tt TIM} for good time 
intervals,
and {\tt OAD} for orbit aspect data.
Each viewing period comprises a single {\tt EVP} and {\tt TIM} file, {\tt OAD} files are provided per
day.
{\tt EVP} data are handled by the GammaLib class {\tt GCOMEventList},
{\tt TIM} data by {\tt GCOMTim},
and {\tt OAD} data by {\tt GCOMOad} and {\tt GCOMOads}, where {\tt GCOMOad} implements a
single {\tt OAD} record (or superpacket), and {\tt GCOMOads} implements a collection of {\tt OAD}
records.
These data structures are combined in an unbinned COMPTEL observation, implemented by the
GammaLib class {\tt GCOMObservation}.
To construct an unbinned COMPTEL observation, the input data for a viewing period can be
specified either by their file names or via a so-called observation definition XML file.

The HEASARC archive mixes different versions of {\tt EVP} files that have different levels of
processing for the ToF values.
GammaLib will automatically correct for these differences, assuring that ToF values accessed
through GammaLib are always ToF$_{\rm III}$ values (see Appendix \ref{app:tof}).

\subsection{Event cubes}
\label{sec:cube}

COMPTEL data analysis is performed on binned event data using three-dimensional event cubes spanned
by the Compton scatter direction $(\chi,\psi)$ and the Compton scattering angle $\bar{\varphi}$.
The binning is performed for events within a given interval of total energy, and each energy interval
is treated as an individual COMPTEL observation.
Combining multiple energy intervals for spectral analysis can be achieved by collecting the relevant
COMPTEL observations in an observation container.
The same holds for the combination of multiple viewing periods for a joint maximum likelihood analysis.

The COMPTEL data space is implemented by the {\tt GCOMDri} class, and the 
{\tt GCOMDri::compute\_dre} method bins the events found in an {\tt EVP} file into an event cube.
In this process, event selection parameters are specified through the {\tt GCOMSelection} class,
comprising selection intervals for D1 energy deposit, D2 energy deposit, ToF channel
value, pulse shape discriminator (PSD) channel value, rejection flag and veto flag.
Standard event selection parameters that were used throughout this paper are summarised in
Table \ref{tab:select}.

\begin{table}[!h]
\caption{Standard event selection parameters that are suitable for most analysis situations
and that were used throughout this paper.
\label{tab:select}}
\centering
\begin{tabular}{l c c}
\hline\hline
Parameter & Minimum & Maximum \\
\hline
D1 energy deposit & 70 keV & 20 MeV \\
D2 energy deposit & 650 keV & 30 MeV \\
ToF value & 115 & 130 \\
PSD value & 0 & 110 \\
Rejection flag & 1 & 1000 \\
Veto flag & 0 & 0 \\
\hline
\end{tabular}
\tablefoot{ToF channel values are given in units of 0.25 ns.}
\end{table}

\begin{figure}[!t]
\centering
\includegraphics[width=8.8cm]{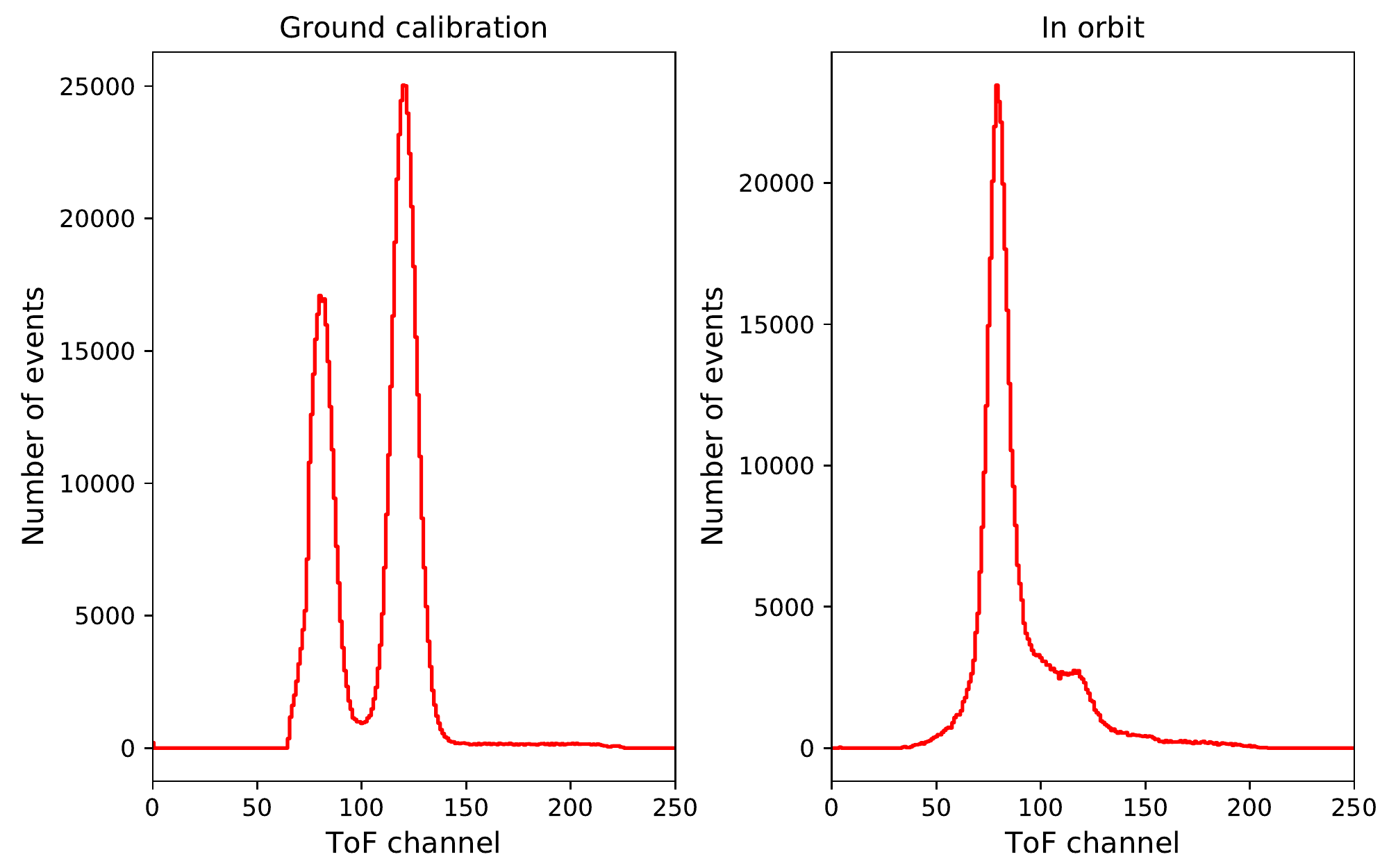}
\caption{
Time of flight spectrum of COMPTEL events registered during ground calibration (left)
and in orbit (right).
One ToF channel corresponds to 0.25 ns.
Downward moving photons interacting first in D1 and then in D2 lead to a peak around 
channel 120, while upward moving photons interacting first in D2 and then in D1 produce a peak 
around channel 80.
In the calibration data the upward moving photons can be clearly distinguished from the downward 
moving photons, while in the flight data the downward moving photons are blended with the tail of 
the dominating upward moving photons.
The figure is adapted from \citet{knoedlseder1994}.
\label{fig:tof}
}
\end{figure}

The ToF value measures the time between the interaction of the photons in the D1 and D2
modules, and provides a strong discriminator against instrumental background.
Figure \ref{fig:tof} illustrates that the ToF distribution of the events shows two distinct features: a
forward peak associated with celestial photons that first interact in D1 followed by an interaction in
D2, and a background peak due to events arising from a first interaction in D2 followed by an
interaction in D1.
While both peaks were clearly distinguishable in calibration data taken on ground (left panel
of Fig.~\ref{fig:tof}), they are heavily blended in orbit due to the dominance of upward moving
background events (right panel of Fig.~\ref{fig:tof}).
Selecting only events from the forward peak considerably improves the signal-to-noise ratio,
and Fig.~\ref{fig:tof} illustrates that the minimum ToF value has a strong impact on the background 
discrimination, with larger values removing more background at the expense of rejecting also 
an increasing number of celestial photons.

To correct for the photon rejection by the ToF selection an energy-dependent correction factor is
applied to the instrument response.
This ToF correction is performed internally by GammaLib (see Appendix \ref{app:tofcor} for details).
Consequently, flux values returned by GammaLib are always corrected for ToF selection, in contrast to 
the COMPASS system where the correction had to be applied by the user posterior to the analysis.

An additional, but less important, event selection parameter is the PSD value, which discriminates
between neutron and photon interactions in D1, with gamma rays found around channel 80 and
neutron-induced events above channel 100 \citep{schoenfelder1993}.
By default, the PSD selection interval is sufficiently large so that no flux correction factor needs to
be applied, yet more stringent selection intervals that lead to improved performance \citep{collmar1997}
eventually require the application of a PSD selection correction factor that is not implemented
in GammaLib.

Furthermore, the Earth atmosphere being a strong gamma-ray source, an important background
reduction is obtained by excluding events that may originate from the Earth atmosphere.
This is achieved by requiring that event circles have a minimum angular distance $\zeta_{\rm min}$ 
from the Earth horizon.
In the process of event binning, this requirement is translated into a constraint on the so-called 
Earth horizon angle (EHA), which is defined as the angular distance between the scatter direction 
of an event and the Earth horizon.
Specifically, only the events are retained that satisfy
${\rm EHA} \ge {\rm EHA}_{\rm min}(\bar{\varphi})$ with
\begin{equation}
{\rm EHA}_{\rm min}(\bar{\varphi}) = \bar{\varphi}_{\rm min} + \zeta_{\rm min}
\label{eq:eha}
,\end{equation}
where $\bar{\varphi}_{\rm min}$ is the lower boundary of the event cube layer that comprises the
Compton scattering angle $\bar{\varphi}$.
By default, $\zeta_{\rm min}=5\degrees$.

Event selection and binning in GammaLib is done per superpacket, which is a bunch of eight spacecraft
telemetry packets of 16.384 s duration that define the temporal granularity on which spacecraft orbital data
information is available.
Orbital data per superpacket are provided by the orbit aspect data ({\tt OAD}) files, and events will 
be used only if their times are contained in the validity interval of any of the available superpackets.
Furthermore, superpackets are considered valid only if their start and stop times are fully enclosed
in one of the good time 
intervals, specified by the {\tt TIM} file.\footnote{
  We note that COMPTEL times are specified by two values:
  the truncated Julian days (TJD), which is defined as the number of modified Julian days (MJD)
  minus 40\,000, and the COMPTEL tics, which are $1/8$ milliseconds long.
  As an example, TJD = 8393 and zero tics corresponds to {\tt 1991-05-17T00:00:00 UT}.}

Finally, the {\tt GCOMSelection} class also supports the specification of phase intervals, so that events 
can be selected according to the orbital phase of a gamma-ray binary or the rotational phase of a pulsar.
Owing to the different timescales involved, the handling of orbital phases differs from that of pulsar
phases.
For orbital phases, the phase as a function of event time is defined via an instance of the 
{\tt GModelTemporalPhaseCurve} class, and the phase selection is done at the level of superpackets, 
implying that only orbital periods that are significantly longer than the superpacket duration of 16.384 s 
will produce meaningful results.
We illustrate this capability below by a phase-resolved analysis of the gamma-ray binary LS~5039 
(cf.~Sect. \ref{sec:ls5039}).
For pulsar phases, the phase selection is done at the level of individual events, using pulsar 
ephemerides and a Solar System barycentre time correction that is applied to the onboard time of the 
events.
Details of the implementation are given in Appendix \ref{app:pulsar}, and we illustrate this capability
below by a phase-resolved analysis of the Crab pulsar and pulsar wind nebula 
(cf.~Sect. \ref{sec:crabpulsar}).

Only events for modules that are signalled as active are considered for the event binning.
The detector module status is provided by the {\tt GCOMStatus} class that relies on a database of daily 
status values that is provided with GammaLib.
During the COMPTEL operations a certain number of photomultiplier tubes (PMTs) of the D2 modules
failed, degrading the localisation precision of the event interactions in the respective modules.
One option is to exclude these modules from the analysis, reducing the number of available
events by about 25\%.
Alternatively, events from zones around the faulty PMTs can be excluded in the analysis, allowing
recovery of a large fraction of the events for the analysis.
Both analysis options are implemented in GammaLib (see Appendix \ref{app:fpmt}) and are
explored in the analysis of LS~5039 (cf.~Sect. \ref{sec:ls5039}).

\begin{figure*}[!th]
\centering
\includegraphics[width=18cm]{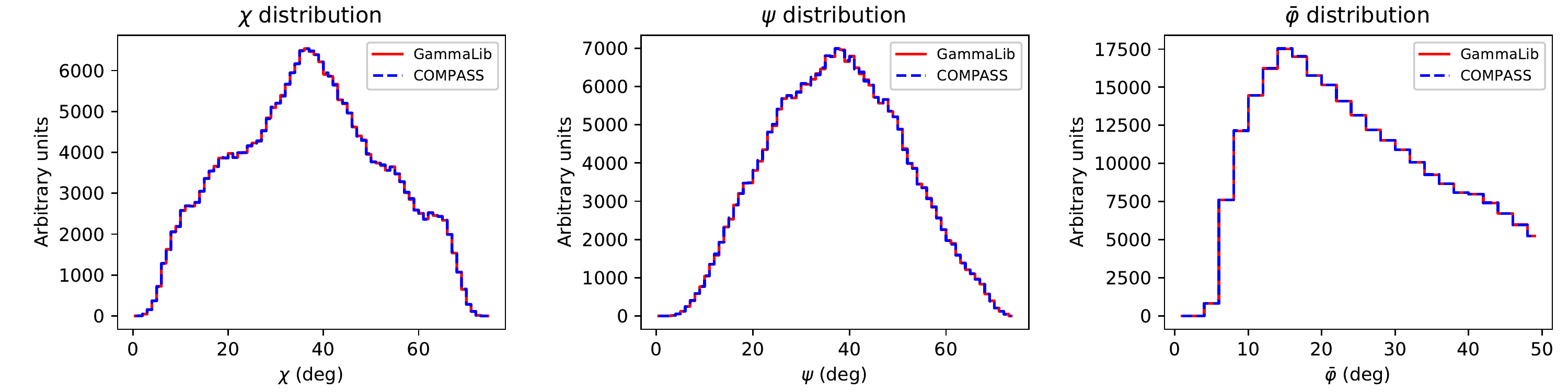}
\caption{
Comparison of an event cube generated with GammaLib for viewing period 2.0 and
the $1-3$ MeV energy band (red solid) and the equivalent event cube {\tt MPE-DRE-50607}
produced by the COMPASS software (blue dashed).
The left panel shows the distribution of $\chi$ values, obtained by summing over all $\psi$ and
$\bar{\varphi}$, the middle panel shows the distribution of $\psi$ values, obtained by summing over all
$\chi$ and $\bar{\varphi}$, and the right panel shows the distribution of $\bar{\varphi}$ values, obtained
by summing over all $\chi$ and $\psi$.
\label{fig:drecmp}
}
\end{figure*}

COMPTEL event cubes are stored as three-dimensional FITS images 
with the file type designation {\tt DRE}.
Projections of an event cube generated using GammaLib for viewing period 2.0\footnote{
  We have chosen the second viewing period in the HEASARC archive to illustrate the data 
  cube comparisons in our paper since for the first viewing period we encountered percent-level
  differences that are plausibly attributed to differences in our and the COMPASS processing 
  configuration that we were not able to track down.
}
and total energies in the interval $1-3$ MeV are shown in Fig.~\ref{fig:drecmp} where for comparison
the projections of an equivalent {\tt DRE} obtained using the COMPASS software are shown.
The projections of the datasets are nearly indistinguishable, illustrating that the event cubes 
generated by GammaLib are equivalent to those produced by the COMPASS software.

\subsection{Instrument response function}
\label{sec:irf}

\subsubsection{Factorisation}

The COMPTEL instrument response function, given in units of events cm$^2$ photons$^{-1}$, is 
factorised using
\begin{equation}
R(\chi, \psi, \bar{\varphi} | \alpha, \delta, E_{\gamma}) = \frac{L}{T}
\frac{
{\tt DRX}(\alpha, \delta) \,\,
{\tt DRG}(\chi, \psi, \bar{\varphi}) \,\,
{\tt IAQ}(\bar{\varphi} | \varphi_{\rm geo}, E_{\gamma})
}{T} ,
\label{eq:response}
\end{equation}
where ${\tt DRX}(\alpha, \delta)$ is the exposure in units of cm$^2$s towards a given celestial direction
         $(\alpha, \delta)$,
${\tt DRG}(\chi, \psi, \bar{\varphi})$ is a geometry function that specifies the probability that a photon 
         that was scattered in one of the D1 modules into direction $(\chi, \psi)$ will encounter one of 
         the active D2 modules (the term also accounts for the Earth horizon event selection, which introduces
         the $\bar{\varphi}$ dependence),
${\tt IAQ}(\bar{\varphi} | \varphi_{\rm geo}, E_{\gamma})$ is the probability that a 
         gamma-ray photon with energy $E_{\gamma}$ being Compton scattered in D1 interacts in D2, 
         with 
         \begin{equation}
         \varphi_{\rm geo} = \arccos \left( \sin \psi \sin \delta + \cos \psi \cos \delta \cos (\chi-\alpha) \right)
         \end{equation}
         being the angular separation between $(\chi, \psi)$ and $(\alpha, \delta)$,
$L$ is the livetime in units of s, and $T$ is the exposure time in units of s.

The fraction $L/T$ is the so-called deadtime correction factor, specifying the fraction of time
during which the telescope is able to register events.
For COMPTEL this fraction is rather constant and was determined by \citet{vandijk1996} to be
$L/T= 0.965$.
This value is hardcoded in GammaLib, and automatically applied to all analysis results.

\subsubsection{Exposure}

The exposure map ${\tt DRX}(\alpha, \delta)$ is computed in GammaLib by the method 
{\tt GCOMDri::compute\_drx} using
\begin{equation}
{\tt DRX}(\alpha, \delta) = T_{\rm sp} \sum_{i \in \{ S \}}
7 \pi r_1^2 \cos \theta_i \frac{1 - \exp \left( -\tau / \cos \theta_i \right)}{1 - \exp \left( -\tau \right)}
,\end{equation}
where
$T_{\rm sp}=16.384$ s is the duration of a superpacket and $r_1=13.8$ cm is the radius of a D1
module.
The $\cos \theta_i$ factor takes into account the reduction of the effective D1 surface when viewed
from an off-axis direction $\theta_i$, measured as the angle between the celestial direction $(\alpha, \delta)$
and the COMPTEL pointing direction for a given superpacket $i$.
The fraction accounts for the increased interaction length of off-axis photons within the D1 modules,
where $\tau=0.2$ is the typical thickness of a D1 module in radiation lengths.
The sum is taken over all superpackets $\{ S \}$ that satisfy the same selection criteria that were
applied to the event selection (cf.~Sect.~\ref{sec:cube}).

\begin{figure}[!th]
\centering
\includegraphics[width=8.8cm]{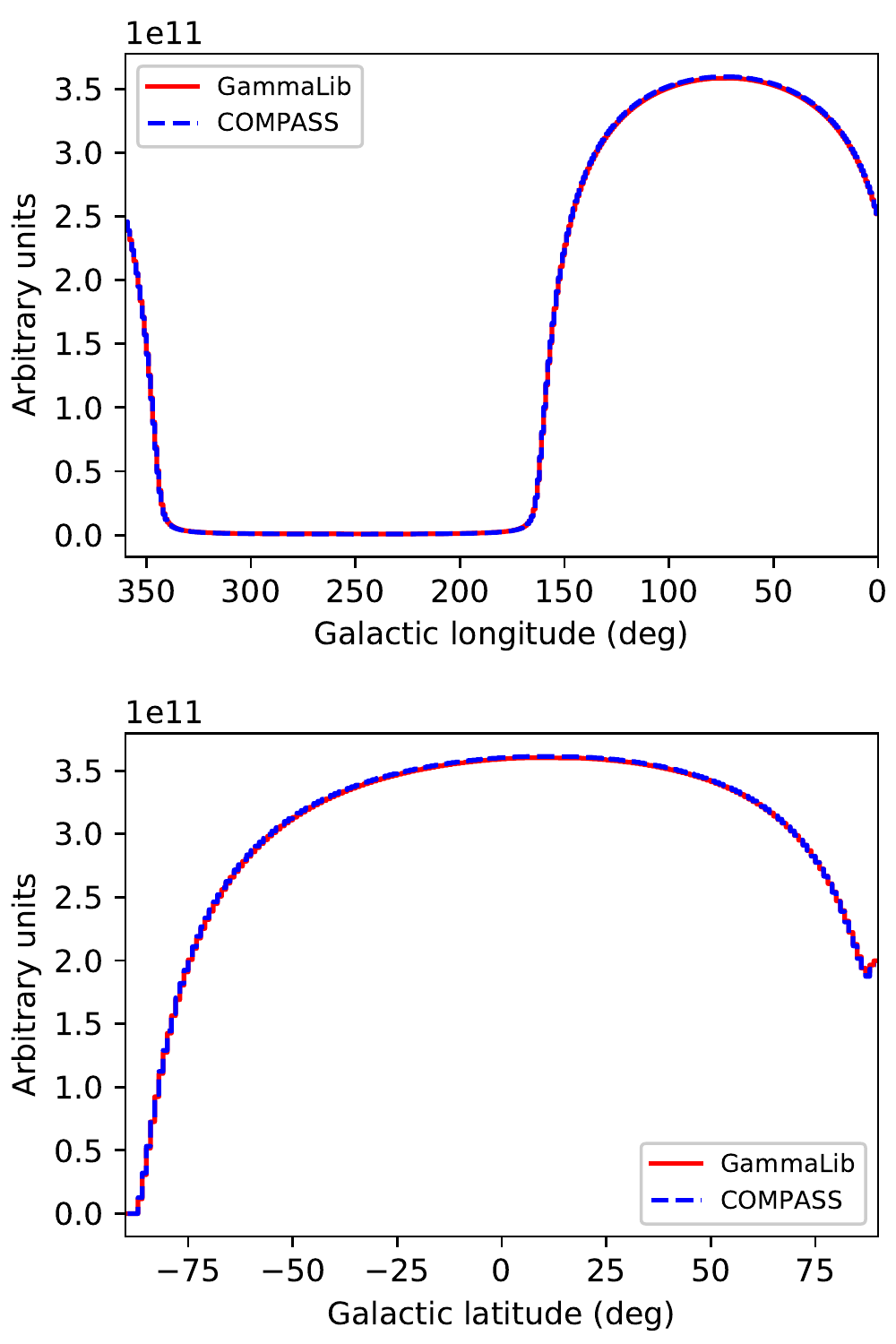}
\caption{
Comparison of an exposure map obtained with GammaLib for viewing period 2.0 (red solid) 
to the exposure map {\tt MPE-DRX-32302} obtained with COMPASS (blue dashed).
The upper panel shows the distribution of Galactic longitude values, obtained by summing over all 
Galactic latitudes, and the lower panel shows the distribution of Galactic latitude values, obtained
by summing over all Galactic longitudes.
\label{fig:drxcmp}
}
\end{figure}

The exposure map is given in units of cm$^2$s and is stored as a two-dimensional FITS image
with the file type designation {\tt DRX}.
For illustration, Fig.~\ref{fig:drxcmp} compares projections of the {\tt DRX} obtained by GammaLib
for viewing period 2.0 to the projections for an equivalent {\tt DRX} that was obtained by COMPASS.
The projections of the datasets are indistinguishable, illustrating that the exposure maps generated
by GammaLib are equivalent to those produced by the COMPASS software.

\subsubsection{Geometry function}

\begin{figure*}[!th]
\centering
\includegraphics[width=18cm]{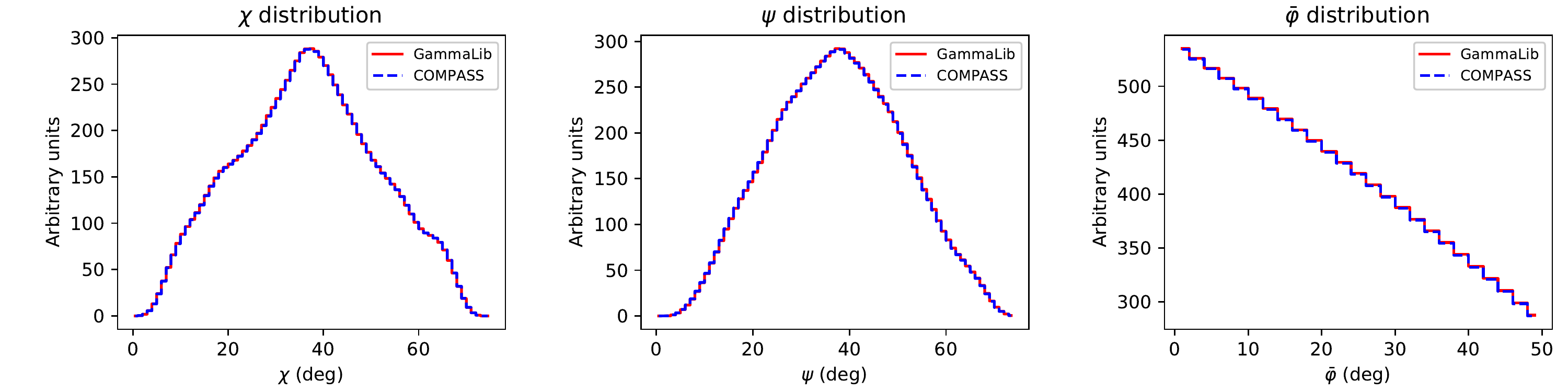}
\caption{
Comparison of a geometry function obtained with GammaLib for viewing period 2.0 (red solid) 
to the geometry function {\tt MPE-DRG-35128} obtained with COMPASS (blue dashed).
The left panel shows the distribution of $\chi$ values, obtained by summing over all $\psi$ and
$\bar{\varphi}$, the middle panel shows the distribution of $\psi$ values, obtained by summing over all
$\chi$ and $\bar{\varphi}$, and the right panel shows the distribution of $\bar{\varphi}$ values, obtained
by summing over all $\chi$ and $\psi$.
\label{fig:drgcmp}
}
\end{figure*}

The geometry function ${\tt DRG}(\chi, \psi, \bar{\varphi})$ is the probability that a photon that 
was scattered in one of the D1 modules towards the direction $(\chi, \psi)$ will encounter one 
of the active modules of D2.
It is computed by determining the geometrical area of the shadow of the D1 modules that is cast
on the D2 modules relative to the total area of all seven D1 modules as a function of the scatter 
direction.
Optionally, zones around failed D2 module PMTs can be excluded during this computation.
The geometry function also accounts for the Earth horizon event selection, which introduces a 
$\bar{\varphi}$ dependence in the probabilities.
As the position of the Earth with respect to the telescope changes continuously, the geometry 
function is recomputed for each superpacket, and the results are then averaged to
provide an effective geometry function that applies to a given event cube.
Similar to the computation of the exposure map, the superpackets $\{ S \}$ considered are those
that satisfy the same selection criteria that were applied to the event selection.
Details are provided in Appendix \ref{app:drg}.

The computation of the geometry function is implemented in the method 
{\tt GCOMDri::compute\_drg} and results are stored as three-dimensional FITS images 
with the file type designation {\tt DRG}.
Figure \ref{fig:drgcmp} illustrates the result of the geometry function computation, again in the form 
of projections for viewing period 2.0.
Equivalent projections for a geometry function obtained using the COMPASS system are
superimposed.
The projections of the datasets are indistinguishable, illustrating that the geometry functions 
generated by GammaLib are equivalent to those produced by the COMPASS software.

\subsubsection{Compton scattering probabilities}

\begin{figure*}[!th]
\centering
\includegraphics[width=18cm]{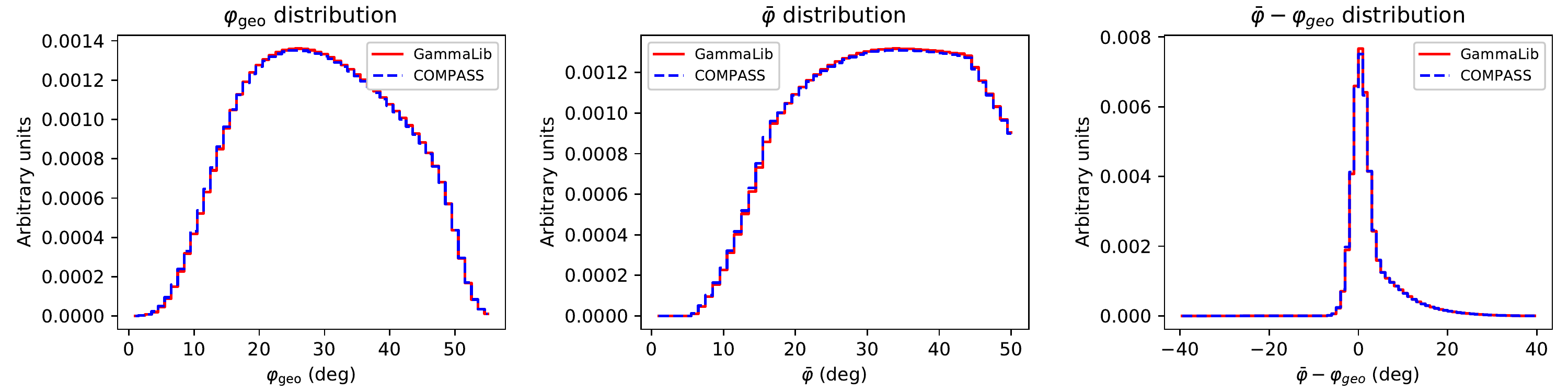}
\caption{
Comparison of projections of an {\tt IAQ} obtained with GammaLib for the energy band $1-3$~MeV
(red solid) to projections for {\tt ROL-IAQ-755} obtained with COMPASS (blue dashed).
The left panel shows the distribution of $\varphi_{\rm geo}$ values, obtained by summing over
all $\bar{\varphi}$, the middle panel shows the distribution of $\bar{\varphi}$ values, obtained by
summing over all $\varphi_{\rm geo}$, and the right panel shows angular resolution
measure $\bar{\varphi}-\varphi_{\rm geo}$.
\label{fig:iaqcmp}
}
\end{figure*}

The Compton scattering probabilities ${\tt IAQ}(\bar{\varphi} | \varphi_{\rm geo}, E_{\gamma})$
were computed employing the method {\tt GCOMIaq::set} using
\begin{equation}
\begin{split}
{\tt IAQ}(\bar{\varphi} | \varphi_{\rm geo}, E_{\gamma}) = &
\int_{\varphi_{\rm geo,min}}^{\varphi_{\rm geo,max}}
P_{\rm eff}(\varphi'_{\rm geo}, E_{\gamma}) \times
P_{\rm KN}(\varphi'_{\rm geo}, E_{\gamma}) \, \times \\
& \int_{E_1^{\rm min}}^{E_1^{\rm max}}
R_{\rm D1}(E_1|\hat{E_1}) \, R_{\rm D2}(E_2|\hat{E_2}) \, J(\bar{\varphi}|E_1) \, dE_1 \times \\
& K(\varphi'_{\rm geo}|\varphi_{\rm geo},\theta) \, d{\varphi'_{\rm geo}}
\label{eq:iaq}
\end{split}
,\end{equation}
where
$P_{\rm eff}(\varphi'_{\rm geo}, E_{\gamma})$ is an efficiency factor that is detailed in
Appendix~\ref{app:efficiency},
$P_{\rm KN}(\varphi'_{\rm geo}, E_{\gamma})$ is the contribution of the Klein-Nishina cross-section
to a given bin in $\varphi'_{\rm geo}$ that is detailed in Appendix~\ref{app:kleinnishina}, and
$R_{\rm D1}(E_1|\hat{E_1})$ and $R_{\rm D2}(E_2|\hat{E_2})$ are the energy response functions
of the D1 and D2 modules, respectively, that are detailed in Appendix~\ref{app:specrsp}.
The transformation from $(\varphi'_{\rm geo}, E_{\gamma})$ to $(\hat{E_1},\hat{E_2})$ is done
using
\begin{equation}
\hat{E_2} = \frac{E_{\gamma}}{(1 - \cos \varphi'_{\rm geo}) \frac{E_{\gamma}}{m_e c^2} +1}
\end{equation}
and $\hat{E_1} = E_{\gamma} - \hat{E_2}$;
\begin{equation}
E_2 = \frac{1}{2} \times \left(
\sqrt{\frac{4 m_e c^2 E_1}{1 - \cos \bar{\varphi}} + E_1^2} - E_1
 \right)
\end{equation}
gives the D2 energy deposit as a function of the D1 energy deposit ($E_1$) and the Compton
scattering angle ($\bar{\varphi}$),
and
\begin{equation}
J(\bar{\varphi}|E_1) = \frac{m_e c^2 \sqrt{E_1} \sin \bar{\varphi}}
{(1 - \cos \bar{\varphi})^2 \sqrt{\frac{4 m_e c^2}{1 - \cos \bar{\varphi}} + E_1}}
\end{equation}
is the Jacobian for the variable transformation.
Here $K(\varphi'_{\rm geo}|\varphi_{\rm geo},\theta)$ is a Gaussian kernel that provides some smearing 
of the response in $\varphi_{\rm geo}$ to take into account the event location uncertainties in the D1
and D2 modules.
The kernel formally depends on the zenith angle $\theta$ of the incoming gamma-ray photons.
However, the {\tt IAQ} is assumed independent of $\theta$, and hence the kernel will be computed for an
average zenith angle of $\theta=25\degrees$ (see Appendix \ref{app:smearing} for details).

The Compton scattering probabilities are stored as a two-dimensional FITS image 
with the file type designation {\tt IAQ}.
For illustration, Fig.~\ref{fig:iaqcmp} compares projections of the Compton scattering probabilities
obtained by GammaLib for the energy band $1-3$~MeV to the projections for an equivalent file 
that was obtained using COMPASS.
Both projections are indistinguishable, and the same is true for the other energy bands that were
investigated.
As demonstrated, the GammaLib implementation accurately reproduces the response computations 
that were implemented in COMPASS.
We note that the implementation of the COMPTEL response computation within GammaLib is
crucial for the preservation of COMPTEL data analysis capabilities, since response functions
are lacking in the HEASARC archive.
In addition to the internal response computation, GammaLib also includes several response 
functions that were derived using simulations \citep{stacy1996}, and that can be used as an 
alternative to the internally computed response functions.

\subsection{Background modelling}
\label{sec:background}

A considerable effort was undertaken by the COMPTEL collaboration to understand and model
the instrumental background in the three-dimensional COMPTEL data space 
\citep[e.g.][]{bloemen1994,knoedlseder1994,vandijk1996,weidenspointner2001}.
Satisfactory, although not perfect, results were obtained using the so-called SRCLIX
method that was developed by \citet{bloemen1994} and that has undergone several evolutions
\citep{vandijk1996}.
The SRCLIX method iteratively applies the BGDLIX algorithm, of which we implemented two
variants in GammaLib.
For reference, we also implemented the PHINOR algorithm (described below), and all 
background modelling methods are invoked in GammaLib via the
{\tt GCOMObservation::compute\_drb}
method.
The iterative SRCLIX method is implemented as a ctools script (cf.~Sect.~\ref{sec:scripts}).

\subsubsection{PHINOR}
\label{sec:phinor}

The motivation for the PHINOR algorithm is the observation that the ratio of {\tt DRE} and {\tt DRG}
multiplied by the solid angle $\Omega(\chi, \psi)$ of the event cube bins is to first order independent
of $\chi$ and $\psi$.
This leads to a background model given by  
\begin{equation}
\begin{split}
{\tt DRB}_{\rm phinor}(\chi,\psi,\bar{\varphi}) = \, &
{\tt DRG}(\chi, \psi, \bar{\varphi}) \, \Omega(\chi, \psi) \, \times \\
& \frac{\sum_{\chi',\psi'} {\tt DRE}(\chi', \psi', \bar{\varphi}) - {\tt DRM}(\chi', \psi', \bar{\varphi}))}
{\sum_{\chi',\psi'} {\tt DRG}(\chi', \psi', \bar{\varphi}) \, \Omega(\chi', \psi')}
\end{split}
\label{eq:phinor}
,\end{equation}
where
\begin{equation}
{\tt DRM}(\chi', \psi', \bar{\varphi}) =
\int_{\alpha,\delta, E_{\gamma}} I(\alpha, \delta, E_{\gamma}) \, R(\chi', \psi', \bar{\varphi} | \alpha, \delta, E_{\gamma})
\, d\alpha \, d\delta \, dE_{\gamma}
\label{eq:drm}
\end{equation}
is the convolution of a model of celestial sources $I(\alpha, \delta, E_{\gamma})$ with the
instrument response function used to subtract any source contribution from the data before
normalisation.
The evaluation of ${\tt DRM}(\chi', \psi', \bar{\varphi})$ is implemented by the method
{\tt GCOMObservation::drm}.

Despite its simplicity, the PHINOR algorithm yields generally empty residual maps
above 10 MeV. However, for lower energies significant large-scale residuals are frequently
observed \citep{vandijk1996}.

\begin{figure*}[!th]
\centering
\includegraphics[width=18cm]{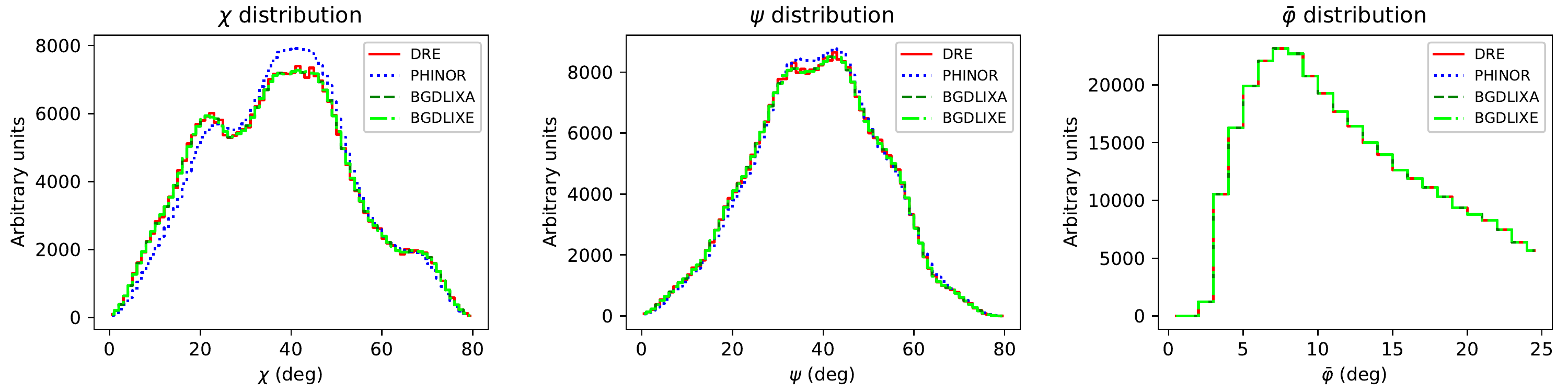}
\caption{
Comparison of PHINOR (blue dotted), BGDLIXA (dark green dashed), and BGDLIXE (light green
dashed-dotted) background model for viewing period 1.0 and the energy band $1-3$~MeV to the 
{\tt DRE} event distribution (red solid).
The left panel shows the distribution of $\chi$ values, obtained by summing over all $\psi$ and
$\bar{\varphi}$, the middle panel shows the distribution of $\psi$ values, obtained by summing over all
$\chi$ and $\bar{\varphi}$, and the right panel shows the distribution of $\bar{\varphi}$ values, obtained
by summing over all $\chi$ and $\psi$.
\label{fig:drbcmp}
}
\end{figure*}

\subsubsection{BGDLIXA}
\label{sec:bgdlixa}

The BGDLIXA algorithm applies a correction to the PHINOR background model by adjusting
$\bar{\varphi}$ templates ${\tt TPL}(\chi,\psi,\bar{\varphi})$ over a limited interval of
$\bar{\varphi}$ values to a normalisation function $n(\chi,\psi,\bar{\varphi})$ that reflects
the background event distribution ${\tt DRE}(\chi, \psi, \bar{\varphi}) - {\tt DRM}(\chi, \psi, \bar{\varphi})$
smoothed in $\chi$ and $\psi$ using a running average (see below).
The BGDLIXA algorithm evaluates the expression
\begin{equation}
{\tt DRB}'_{\rm bgdlixa}(\chi,\psi,\bar{\varphi}) = {\tt TPL}(\chi,\psi,\bar{\varphi}) \,
\frac{\sum_{\bar{\varphi}' \in \{ R_{\bar{\varphi}} \} } n(\chi,\psi,\bar{\varphi}')}
{\sum_{\bar{\varphi}' \in \{ R_{\bar{\varphi}} \} }  {\tt TPL}(\chi,\psi,\bar{\varphi}')}
\label{eq:bgdlixa}
,\end{equation}
where the summation is performed over the range
$R_{\bar{\varphi}} =
[\bar{\varphi} - \Delta \bar{\varphi}_{\rm incl} , \bar{\varphi} - \Delta \bar{\varphi}_{\rm excl}]
\cup
[\bar{\varphi} + \Delta \bar{\varphi}_{\rm excl}, \bar{\varphi} + \Delta \bar{\varphi}_{\rm incl}]$
with
$\Delta \bar{\varphi}_{\rm incl} = \Delta \bar{\varphi} \lfloor (N_{\rm incl}-1)/2 \rfloor$,
$\Delta \bar{\varphi}_{\rm excl} = \Delta \bar{\varphi} \lfloor (N_{\rm excl}-1)/2 \rfloor$,
$N_{\rm incl}$ and $N_{\rm excl}$ are either odd integers or zero, and
$\lfloor \cdot / \cdot \rfloor$ designates the integer division operator.
$\Delta \bar{\varphi}$ is the bin size in $\bar{\varphi}$ of the three-dimensional event cube,
which typically is $2\degrees$.
While in early days $N_{\rm excl}>0$ was used, later analyses set $N_{\rm excl}=0$, which
simplifies the summation range to
$R_{\bar{\varphi}} = [\bar{\varphi} - \Delta \bar{\varphi}_{\rm incl}, \bar{\varphi} + \Delta \bar{\varphi}_{\rm incl}]$.
Past COMPTEL analyses generally used $N_{\rm incl}=13$ resulting in
$R_{\bar{\varphi}} = [\bar{\varphi} - 12\degrees, \bar{\varphi} + 12\degrees]$.
As we will see later, $N_{\rm incl}$ is the most critical parameter of the BGDLIXA algorithm,
and specifies the number of $\bar{\varphi}$ layers of the three-dimensional event cube over
which the templates are adjusted.
If $N_{\rm incl}$ is large the fraction in Eq.~(\ref{eq:bgdlixa}) varies little with $\bar{\varphi}$,
leading to a background model that essentially follows the templates.
For small values of $N_{\rm incl}$ the fraction in Eq.~(\ref{eq:bgdlixa}) accommodates for 
differences between the templates and the normalisation function, leading to a background 
model that follows more closely the background event distribution at the expense of 
including also some of the source events.

The $\bar{\varphi}$ templates are computed using
\begin{equation}
\begin{split}
{\tt TPL}(&\chi,\psi,\bar{\varphi}) = \, {\tt DRB}_{\rm phinor}(\chi,\psi,\bar{\varphi}) \, \times \\
& \frac{\sum_{\chi' \in \{ R_{\chi} \}} \sum_{\psi' \in \{ R_{\psi} \}} \sum_{\bar{\varphi}'} 
{\tt DRE}(\chi', \psi', \bar{\varphi}') - {\tt DRM}(\chi', \psi', \bar{\varphi}') }
{\sum_{\chi' \in \{ R_{\chi} \}} \sum_{\psi' \in \{ R_{\psi} \}} \sum_{\bar{\varphi}'}
{\tt DRB}_{\rm phinor}(\chi',\psi',\bar{\varphi'})}
\end{split}
\label{eq:template}
,\end{equation}
which are an adjustment of the PHINOR background model to match the
$\bar{\varphi}$-integrated $\chi,\psi$ distribution of the data after subtracting the contributions
of celestial sources.
To reduce the impact of statistical fluctuations of the data on the model, the match is performed 
by a running-average summation over a small range in $\chi$ and $\psi$, with
$R_{\chi} = [\chi - \Delta \chi_{\rm runav}, \chi + \Delta \chi_{\rm runav}]$ and
$R_{\psi} = [\psi - \Delta \psi_{\rm runav}, \psi + \Delta \psi_{\rm runav}]$
where
$\Delta \chi_{\rm runav} = \Delta \chi \, N_{\rm runav}$ and
$\Delta \psi_{\rm runav} = \Delta \psi \, N_{\rm runav}$.
Here, $2 N_{\rm runav}+1$ specifies the number of $\chi$ and $\psi$ bins over which the 
running-average summation is performed, with $\Delta \chi$ and $\Delta \psi$ being
the bin size in $\chi$ and $\psi$ of the three-dimensional event cube, and $N_{\rm runav}$
is an integer number.
Usually, $\Delta \chi = \Delta \psi = 1\degrees$, and past analyses employed 
$N_{\rm runav}=3$ resulting in a running averaging of $\pm3\degrees$ around each 
$\chi,\psi$ pixel \citep{vandijk1996}.

The adjustment of the $\bar{\varphi}$ templates is done using a normalisation function
computed using
\begin{equation}
\begin{split}
n(\chi,\psi,\bar{\varphi}') & = \, 
{\tt DRG}(\chi, \psi, \bar{\varphi}') \, \Omega(\chi, \psi) \, \times \\
& \frac
{\sum_{\chi' \in \{ A_{\chi} \}} \sum_{\psi' \in \{ A_{\psi} \}} {\tt DRE}(\chi', \psi', \bar{\varphi}') - {\tt DRM}(\chi', \psi', \bar{\varphi}')}
{\sum_{\chi' \in \{ A_{\chi} \}} \sum_{\psi' \in \{ A_{\psi} \}} {\tt DRG}(\chi', \psi', \bar{\varphi}') \, \Omega(\chi', \psi')}
\end{split}
\label{eq:norm}
,\end{equation}
which is an analogue of Eq.~(\ref{eq:phinor}) that is limited to a small range of $\chi,\psi$ around the
pixel of interest.
Specifically,
$A_{\chi} = [\chi - \Delta \chi_{\rm avgr}, \chi + \Delta \chi_{\rm avgr}]$ and
$A_{\psi} = [\psi - \Delta \psi_{\rm avgr}, \psi + \Delta \psi_{\rm avgr}]$
where
$\Delta \chi_{\rm avgr} = \Delta \chi \, \lfloor (N_{\rm avgr}-1)/2 \rfloor$ and
$\Delta \psi_{\rm avgr} = \Delta \psi \, \lfloor (N_{\rm avgr}-1)/2 \rfloor$.
Here, $N_{\rm avgr}$ is an odd integer that specifies the number of $\chi$ and 
$\psi$ bins over which the adjustment is performed, with $\Delta \chi$ and $\Delta \psi$ 
being the bin size in $\chi$ and $\psi$ of the three-dimensional event cube.
Past analyses usually employed $N_{\rm avgr}=3$ resulting in an averaging
of
$A_{\chi} = [\chi - 1\degrees, \chi + 1\degrees]$ and
$A_{\psi} = [\psi - 1\degrees, \psi + 1\degrees]$ \citep{vandijk1996}.

We want to point out that the COMPASS analysis system implemented
\begin{equation}
\begin{split}
n(\chi,\psi,\bar{\varphi}') = & \, 
{\tt DRG}(\chi, \psi, \bar{\varphi}') \, \Omega(\chi, \psi) \, \times \\
& 
\sum_{\chi' \in \{ A_{\chi} \}} \sum_{\psi' \in \{ A_{\psi} \}}
\frac{{\tt DRE}(\chi', \psi', \bar{\varphi}')}{{\tt DRG}(\chi', \psi', \bar{\varphi}') \, \Omega(\chi', \psi')}
\end{split}
\end{equation}
instead of Eq.~(\ref{eq:norm}), which can lead to unreasonably large normalisations
at the edge of the three-dimensional event cube where ${\tt DRG}$ values are small.
Using Eq.~(\ref{eq:norm}) avoids such problems, and in fact leads to a simplified background modelling
algorithm that we implemented in GammaLib under the acronym BGDLIXE (see below).
Furthermore, we added a global $\bar{\varphi}$ normalisation step,
\begin{equation}
\begin{split}
{\tt DRB}_{\rm bgdlixa}(\chi,\psi,\bar{\varphi}) = \, &
{\tt DRB}'_{\rm bgdlixa}(\chi,\psi,\bar{\varphi}) \, \times \\
& \frac{\sum_{\chi',\psi'} {\tt DRE}(\chi', \psi', \bar{\varphi}) - {\tt DRM}(\chi', \psi', \bar{\varphi}))}
{\sum_{\chi',\psi'} {\tt DRB}'_{\rm bgdlixa}(\chi',\psi',\bar{\varphi})}
\end{split}
\label{eq:renorm}
,\end{equation}
at the end of the computation so that the model is normalised 
to the number of background events in each $\bar{\varphi}$ layer.

\subsubsection{BGDLIXE}
\label{sec:bgdlixe}

Recognising that Eqs.~(\ref{eq:template}) and (\ref{eq:norm}) both do the same thing (that is, they normalise
the solid-angle-weighted geometry function to the measured event distribution), the BGDLIX background
model can be simplified to
\begin{equation}
\begin{split}
{\tt DRB}'_{\rm bgdlixe}&(\chi,\psi,\bar{\varphi}) = \, {\tt DRB}_{\rm phinor}(\chi,\psi,\bar{\varphi}) \, \times \\
& \frac{\sum_{\chi' \in \{ A_{\chi} \}} \sum_{\psi' \in \{ A_{\psi} \}} \sum_{\bar{\varphi}' \in \{ R_{\bar{\varphi}} \} } {\tt DRE}(\chi', \psi', \bar{\varphi}') - {\tt DRM}(\chi', \psi', \bar{\varphi}')}
{\sum_{\chi' \in \{ A_{\chi} \}} \sum_{\psi' \in \{ A_{\psi} \}} \sum_{\bar{\varphi}' \in \{ R_{\bar{\varphi}} \} } {\tt DRB}_{\rm phinor}(\chi',\psi',\bar{\varphi}')}
\end{split}
,\end{equation}
which is a local fit of the PHINOR background model to the three-dimensional event distribution
after subtracting any celestial sources.
We note that $N_{\rm runav}$ is no longer a parameter of the model, and the number of pixels in $\chi$
and $\psi$ that are used for the local fit is determined by $N_{\rm avgr}$.
We also included a $\bar{\varphi}$ normalisation at the end of the computation equivalent to
Eq.~(\ref{eq:renorm}).

Figure \ref{fig:drbcmp} compares the event cube projections to the projections for
background models obtained using PHINOR, BGDLIXA and BGDLIXE using
the parameters $N_{\rm runav}=3$, $N_{\rm avgr}=3$, $N_{\rm incl}=13$ and $N_{\rm excl}=0$
proposed by \citet{vandijk1996}.
While the PHINOR model provides a first-order description of the event distribution, it shows significant
differences in the details.
The BGDLIXA and BGDLIXE models follow the event distribution very closely, and both models are 
in fact indistinguishable from each other, demonstrating that their results are equivalent for the
parameter values chosen.

\subsection{ctools scripts}
\label{sec:scripts}

To support the science analysis of COMPTEL data, we extended the ctools package with a number
of dedicated Python scripts, providing basic building blocks that each perform well-defined science
data analysis tasks.
This includes the generation of a database based on the data available in the HEASARC archive,
the selection of COMPTEL viewing periods from this database, event binning and response 
computation, data combination, background modelling, model fitting, generation of test statistic (TS)
maps, residual inspection, observation simulations, and source detection, as well as generation of 
pulsar pulse profiles.
These scripts complement the already existing generic science analysis tools in ctools that may be
used in combination with the new scripts.
Table \ref{tab:comscripts} summarises the COMPTEL-specific scripts that we added to
ctools.

\begin{table}[!t]
\caption{COMPTEL-specific ctools scripts.
\label{tab:comscripts}}
\centering
\begin{tabular}{l l}
\hline\hline
Script & Task \\
\hline
{\tt comgendb} & Generate database \\
{\tt comobsselect} & Select observations  \\
{\tt comobsbin} & Bin COMPTEL observations \\
{\tt comobsback} & Generate background model \\
{\tt comobsadd} & Combine observations \\
{\tt comobsmodel} & Generate XML model for analysis \\
{\tt comlixfit} & Fit model to data using SRCLIX method \\
{\tt comlixmap} & Create TS map using SRCLIX method \\
{\tt comsrcdetect} & Detect sources in TS map \\
{\tt comobsres} & Generate residuals \\
{\tt comobssim} & Simulate COMPTEL observations \\
{\tt compulbin} & Generate pulsar pulse profile \\
\hline
\end{tabular}
\end{table}

Before starting a COMPTEL data analysis, a database needs to be generated from the data
that are available in the HEASARC archive.
This analysis step, which only needs to be performed once on a given computer, is performed
by the {\tt comgendb} script.

Once the database is generated, a typical COMPTEL analysis starts with executing {\tt comobsselect} 
to select the relevant viewing periods for a given source or source region and observing time
interval from the database.
Results are provided as an observation definition file in XML format, which contains all the
relevant information for the subsequent analysis steps.
Following selection, the data are binned into three-dimensional event cubes and the corresponding
{\tt DRX} and {\tt DRG} cubes are generated using {\tt comobsbin}.
The binning can be done for an arbitrary number of energy bands, enabling a joint spectral-spatial
analysis that was not supported by the original COMPASS software.
{\tt comobsbin} also computes an initial background model {\tt DRB} using the PHINOR algorithm
(see Sect. \ref{sec:bgdmodel}), as well as the  {\tt IAQ} response matrices for the relevant energy 
bands.
All output files are stored as FITS files in a data store, avoiding the recomputation of identical data
files in subsequent analyses.
Alternative background models using the algorithms described in Sect. \ref{sec:background}
can be generated using {\tt comobsback}.

\begin{figure*}[!th]
\centering
\includegraphics[width=18cm]{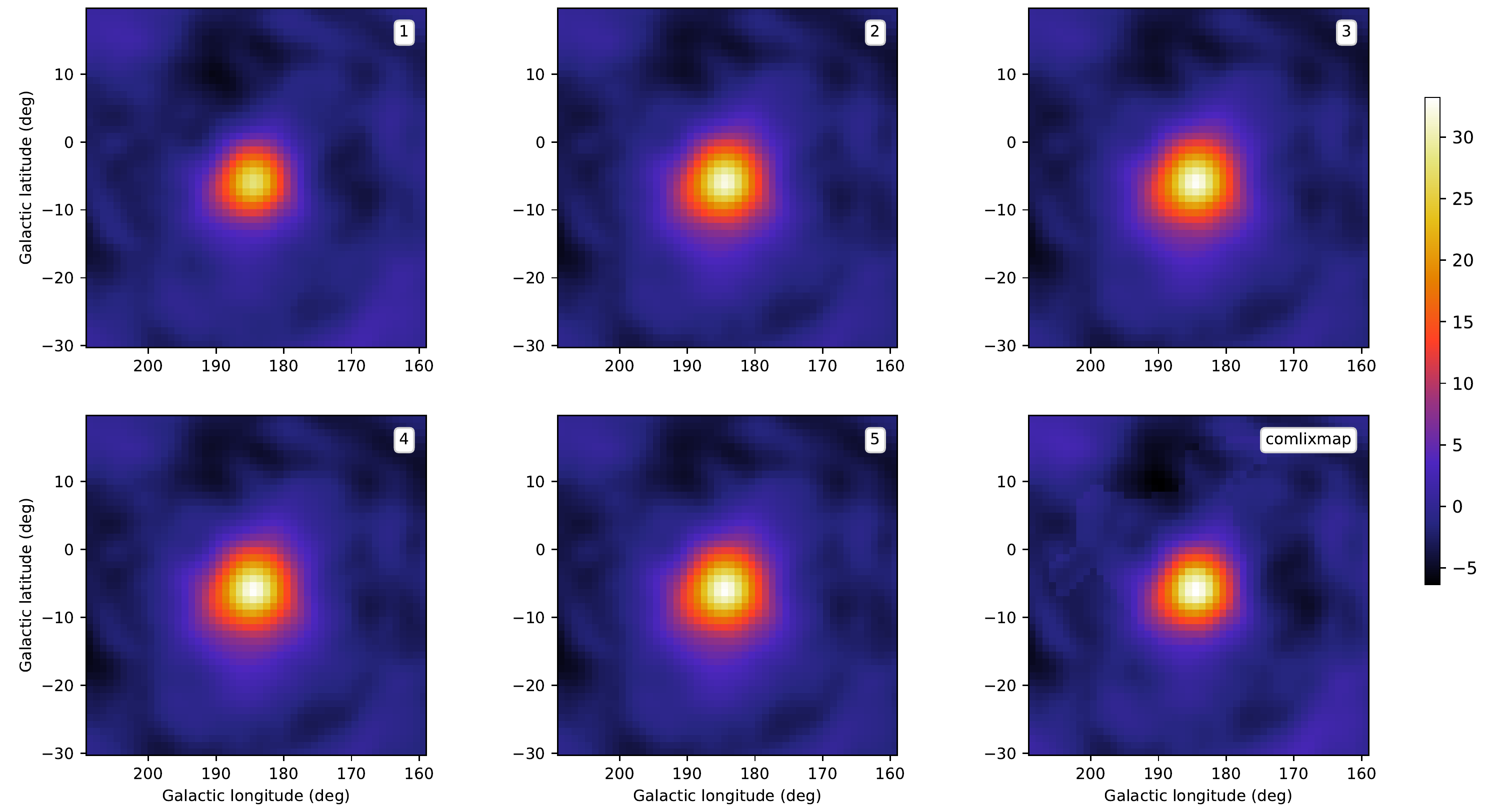}
\caption{
Test statistic maps of viewing period 1.0 (during which the Crab pulsar and pulsar wind nebula
was observed) for the combined analysis of the four standard COMPTEL energy bands as a function 
of the number of SRCLIX iterations, indicated in the upper-right corner of each panel.
The BGDLIXE algorithm with $N_{\rm avgr}=5$, $N_{\rm incl}=15,$ and $N_{\rm excl}=0$ was
employed.
The lower-right panel shows the TS map obtained using the {\tt comlixmap} script.
Colour maps are shown in units of $\sqrt{\rm TS}$.
Negative values correspond to negative fluxes of the test source.
\label{fig:srclix}
}
\end{figure*}

Individual viewing periods can be combined into single event cubes for each energy band
$\{E\}$ using the {\tt comobsadd} script, leading to a considerable speed-up of the data analysis.
The combination of viewing periods is however not required, and alternatively the data can be 
analysed using a joint maximum-likelihood analysis of the selected viewing periods, a possibility 
that was not supported by the original COMPASS software.
Combination of the viewing periods $i$ is done using
\begin{equation}
{\tt DRE}(\chi, \psi, \bar{\varphi}|\{E\}) = \sum_i {\tt DRE}_i(\chi, \psi, \bar{\varphi}|\{E\})
\end{equation}
\begin{equation}
{\tt DRB}(\chi, \psi, \bar{\varphi}|\{E\}) = \sum_i {\tt DRB}_i(\chi, \psi, \bar{\varphi}|\{E\})
\end{equation}
\begin{equation}
{\tt DRX}(\alpha, \delta) = \sum_i \max_{\alpha',\delta'} {\tt DRX}_i(\alpha', \delta')
\label{eq:drxadd}
\end{equation}
\begin{equation}
{\tt DRG}(\chi, \psi, \bar{\varphi}) = \frac{1}{T} \sum_i T_i \, {\tt DRG}_i(\chi, \psi, \bar{\varphi})
,\end{equation}
where $T_i$ is the exposure time of viewing period $i$ and $T=\sum_i T_i$ is the total
exposure time of the combined data.
We note that Eq.~(\ref{eq:drxadd}) leads to an exposure map that is independent of $\alpha, \delta$
and that is given by the maximum of the {\tt DRX} for each viewing period, an approximation
that is justified by the fact that the {\tt DRX} is rather flat (see Fig.~\ref{fig:drxcmp}) and
that the zenith angle variation in the response computation (cf.~Eq.~(\ref{eq:response})) is
dominated by the geometry function {\tt DRG} \citep[see also][]{knoedlseder1994}.

In order to perform a maximum-likelihood analysis of the data, a model definition file
in XML format needs to be generated.
We automatised this task with the {\tt comobsmodel} script, which in particular generates an adequate
model definition for fitting the background model {\tt DRB} to the data.
COMPTEL background model fitting is done using the {\tt DRBPhibarBins} model type, which introduces
one scaling factor for each $\bar{\varphi}$ layer for all viewing periods and energy bands.
In addition, {\tt comobsmodel} supports adding of a point source and diffuse model components
to the model definition XML file.

The main script for maximum-likelihood model fitting is {\tt comlixfit}, which implements the iterative
SRCLIX algorithm.
The algorithm starts from an input model definition XML file and computes a {\tt DRM} model
of the celestial sources using Eq.~(\ref{eq:drm}), which is then used to compute an initial
background model using one of the PHINOR, BGDLIXA, or BGDLIXE algorithms.
The script then uses the {\tt ctlike} tool to fit the source and background models to the binned data.
The celestial source model that results from the fit is then used to update the {\tt DRM} model
and to regenerate a background model using the selected algorithm.
A new {\tt ctlike} fit is then performed using the updated model, and the procedure is repeated
until the maximum log-likelihood change between subsequent iterations becomes negligible, typically
less than $0.05$.
Usually, the SRCLIX algorithm converges after a few iterations.

Figure \ref{fig:srclix} illustrates the algorithm by showing TS maps for subsequent 
SRCLIX iterations for viewing period 1.0 that were obtained using the {\tt cttsmap} tool.
The TS is defined as 
\begin{equation}
\mathrm{TS} = 2 \ln L(M_s+M_b) - 2 \ln L(M_b)
\label{eq:ts}
\end{equation}
\citep{mattox1996},
where $\ln L(M_s+M_b)$ is the log-likelihood value obtained when fitting the source and the background
models together to the data, and $\ln L(M_b)$ is the log-likelihood value obtained when fitting only the
background model to the data.
Under the hypothesis that the background model $M_b$ provides a satisfactory fit of the data, TS
follows a $\chi^2_n$ distribution with $n$ degrees of freedom, where $n$ is the number of 
free parameters in the source model component. Therefore,
\begin{equation}
p = \int_\mathrm{TS}^{+\infty} \chi^2_n(x) \:\: \mathrm{d}x
\end{equation}
gives the chance probability (p-value) that the log-likelihood improves by TS/2 when adding the source
model $M_s$ due to statistical fluctuations only \citep{cash1979}.

We note that viewing period 1.0 is an observation of the Crab pulsar and pulsar wind nebula, which is 
by far the brightest source of gamma rays in the COMPTEL energy range.
For the SRCLIX analysis, the data of the four standard energy bands were analysed jointly, and the 
Crab was modelled as a point source with power-law spectrum.
The source location, prefactor and spectral index were free parameters in the fit.
After each iteration of the SRCLIX algorithm, a TS map\footnote{
  The TS map comprises $50\times50$ pixels of size $1\degrees\times1\degrees$ situated 
  around the Crab position.}
was generated by replacing the Crab in the fitted model by a test source with fixed power-law 
spectral index of $\Gamma=2$.
The source visible at the centre of the TS maps is the Crab, which is already significantly detected
after the first SRCLIX iteration.
Subsequent iterations slightly increase the source significance, but overall the TS maps change little.
The SRCLIX algorithm converged after six iterations.

The TS maps show a halo of marginal significance around the Crab, which can be explained
by the fact that the maps were obtained using a background model that assumed the presence of a
point source at the location of the Crab.
The event cone of a test source placed a few degrees away from the Crab will pick up some of the
excess counts left by the background model, which explains the halo in maps.

An alternative way to generate TS maps is provided by the {\tt comlixmap} script, which applies the
SRCLIX algorithm to each test source position, and which is the algorithm that was implemented
by the COMPASS task SRCLIX.
Here each pixel in the TS map corresponds to a different background model, and when moving away
from the source location, no excess counts will remain in the data.
This reduces the halo around the sources, as illustrated in the last panel of Fig.~\ref{fig:srclix} that shows
the TS map that was obtained using {\tt comlixmap} for viewing period 1.0.
At the same time, negative residuals are amplified, which is explained by the fact that the events
of the Crab for test source positions offset from the source will be included as a smoothed event
cone in the background model.
We note, however, that this is only a feature of the TS maps, as ultimately, an adequate model of
celestial sources that describes the COMPTEL event distribution should be derived from the data.
Using that model as {\tt DRM} in the BGDLIX algorithm will exclude any source events from the
smoothing algorithm, and hence provides a reliable background model.
For this approach the {\tt comsrcdetect} script can be used, which extracts significant sources from
TS maps and adds them to a model definition XML file.
A subsequent run of the {\tt comlixmap} script will then show whether any additional celestial
sources remain in the data, building up iteratively an adequate model of celestial sources. 

Following model fitting an inspection of the fit residuals is crucial.
The {\tt comobsres} script enables such an inspection by projecting the residual between the event and
model cubes onto the sky by summing their content along the event cone using
\begin{equation}
N(\alpha,\delta) = \sum_{\chi,\psi,\bar{\varphi}} \left\{
  \begin{array}{l l}
  \displaystyle
  {\tt DRE}(\chi, \psi, \bar{\varphi}), & \mbox{if $ARM_{\rm min} \le \bar{\varphi}-\varphi_{\rm geo} \le ARM_{\rm max}$} \\
  \displaystyle
  0, & \mbox{otherwise} \\
  \end{array}
\right.
\label{eq:armdata}
\end{equation}
and
\begin{equation}
M(\alpha,\delta) = \sum_{\chi,\psi,\bar{\varphi}} \left\{
  \begin{array}{l l}
  \displaystyle
  {\tt DRM}(\chi, \psi, \bar{\varphi}), & \mbox{if $ARM_{\rm min} \le \bar{\varphi}-\varphi_{\rm geo} \le ARM_{\rm max}$} \\
  \displaystyle
  0, & \mbox{otherwise} \\
  \end{array}
\right.
\label{eq:armmodel}
,\end{equation}
where $\varphi_{\rm geo}$
is the angular distance between a sky position $(\alpha,\delta)$ and the Compton scatter direction $(\chi, \psi)$,
and $[ARM_{\rm min},ARM_{\rm max}]$ defines a selection window for the so-called angular resolution measure
that is typically taken to be $[-3\degrees,3\degrees]$.
We note that {\tt DRM} designates here the model cube that comprises both the source and background
components.
By default, {\tt comobsres} uses 
\begin{equation}
\begin{split}
R(\alpha,\delta) = & \sgn(N(\alpha,\delta)-M(\alpha,\delta)) \, \times \\
& \sqrt{ 2 \left( N(\alpha,\delta) \ln \frac{N(\alpha,\delta)}{M(\alpha,\delta)} + M(\alpha,\delta) - N(\alpha,\delta) \right)}
\end{split}
\end{equation}
to compute the significance of the residuals $R(\alpha,\delta)$ in Gaussian $\sigma$,
where the sign term indicates whether the measured number of counts is larger or smaller than
the number of counts predicted by the model.
Some special cases need to be treated separately.
Namely, if $N(\alpha,\delta) = 0$ the residual significance is
\begin{equation}
R(\alpha,\delta) = \sgn(N(\alpha,\delta)-M(\alpha,\delta)) \sqrt{2 M(\alpha,\delta)} ,
\end{equation}
while if $M(\alpha,\delta)=0$ the significance cannot be computed and we set $R(\alpha,\delta)=0$.

The {\tt comobssim} script enables the simulation of {\tt DRE} event cubes by sampling the events
according to a Poisson distribution using the expectation given by a model.
Specifically, simulated events for a given celestial source model can be added by {\tt comobssim}
to existing observations, allowing the study of celestial sources with known properties in real data,
a possibility that we use extensively in the next section.

Finally, {\tt compulbin} will generate pulsar phase profiles by applying the algorithms described in
Appendix \ref{app:pulsar} to individual events.
Only events that satisfy the ARM selection according to Eq.~(\ref{eq:armdata}) will be retained
in the phase profiles, with typical values for the ARM window being $[-3.5\degrees,3.5\degrees]$.

\section{Science validation}
\label{sec:validation}

Having verified that GammaLib and ctools reproduce data products that are identical to the ones
that were generated with the COMPASS system, we now verify that the use of our software for
an analysis of the data provided by HEASARC reproduces COMPTEL science results that were
published in the literature.
If not stated otherwise, for the analyses that follow we apply the event selection parameters
specified in Table \ref{tab:select}, a value of $\zeta_{\rm min}=5\degrees$, and we exclude D2
modules for which there were faulty photomultipliers.

\subsection{Background model validation}
\label{sec:bgdmodel}

\begin{figure*}[!th]
\centering
\includegraphics[width=18cm]{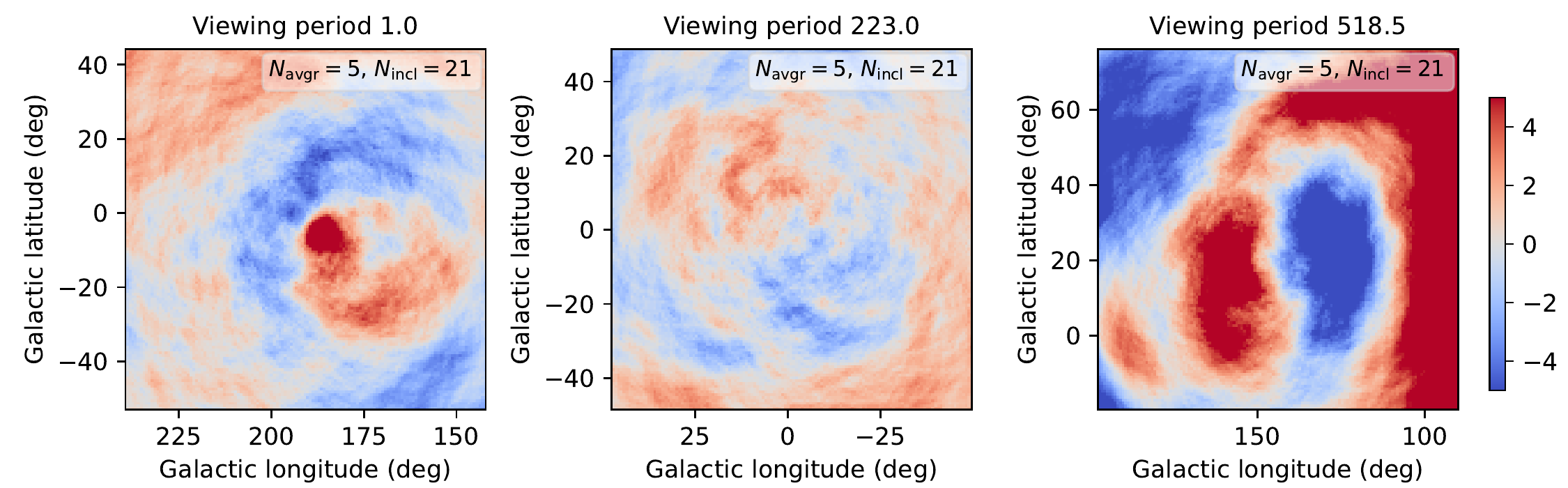}
\includegraphics[width=18cm]{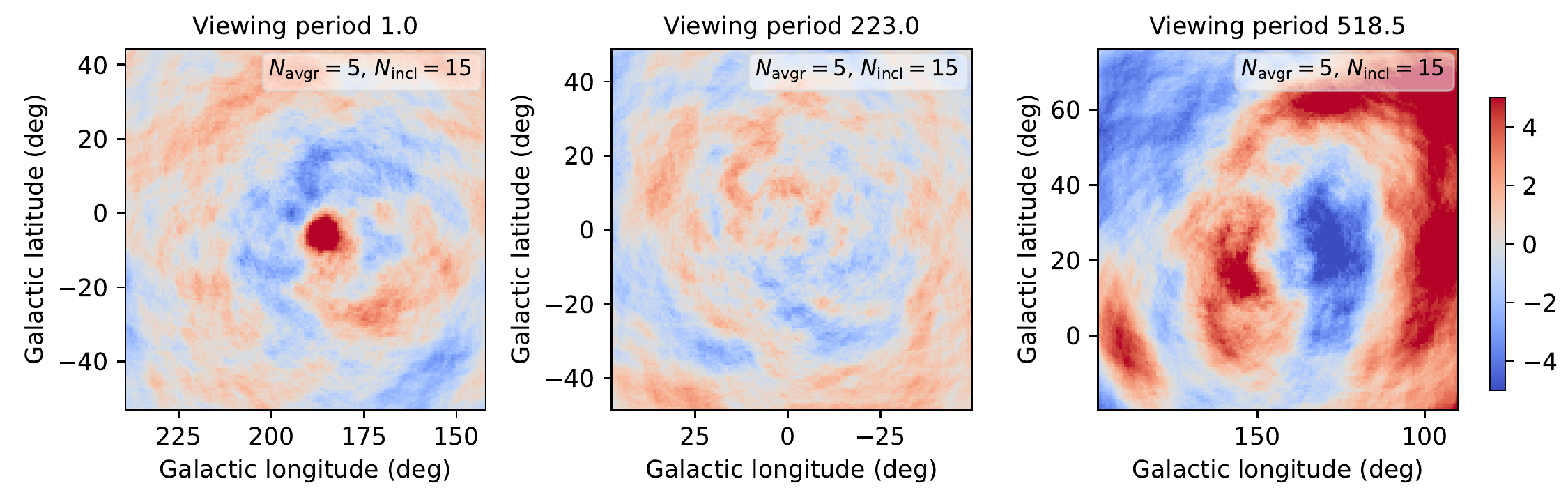}
\includegraphics[width=18cm]{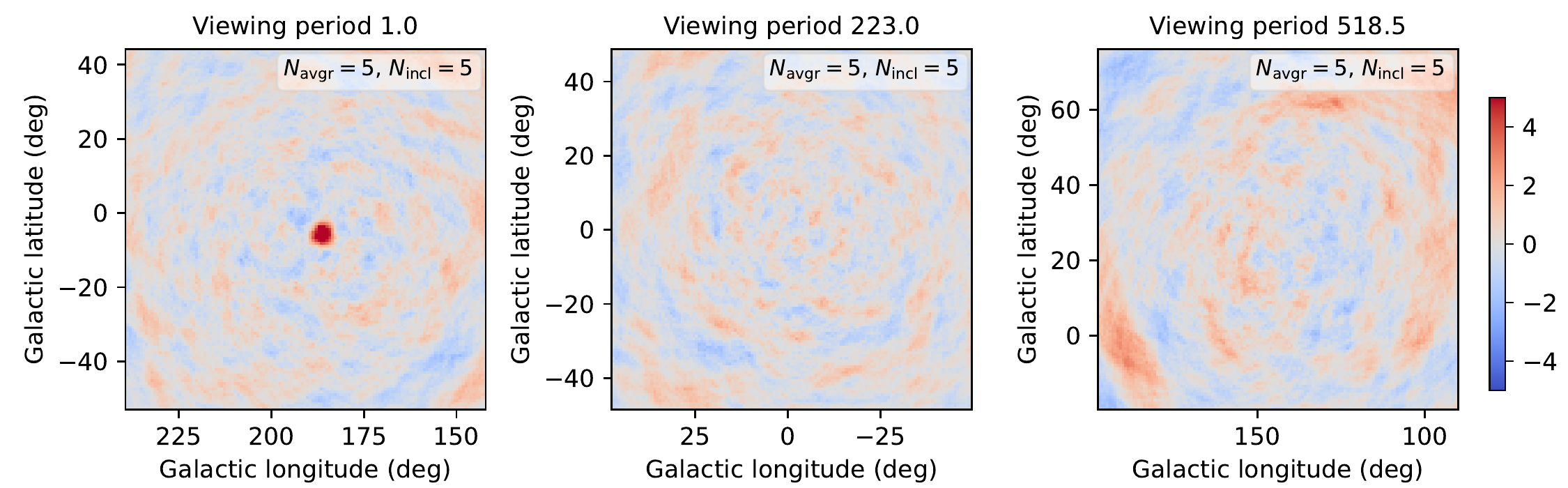}
\caption{
Residual maps for three viewing periods and the $1-3$~MeV energy band with background modelled
using the BGDLIXE algorithm for $N_{\rm avgr}=5$ and different values of $N_{\rm incl}$.
The colour scale is limited to the range $-5\sigma$ (blue) to $+5\sigma$ (red).
Viewing period 1.0 was selected because it contains the Crab, viewing period 223.0 because it
has a rather flat residual, and viewing period 518.5 because it is among the viewing periods with 
the worst residuals.
\label{fig:resmaps1}
}
\end{figure*}

An important step prior to any science analysis is the validation of the background model.
An accurate background model predicts the background event distribution within statistical fluctuations,
and allows for a reliable and accurate determination of the contribution of celestial gamma-ray
sources to the measured events.
As is obvious from Fig.~\ref{fig:drbcmp}, the PHINOR model certainly does not satisfy the
first criterion; hence, we no longer consider this model here.
On the other hand, the BGDLIXA and BGDLIXE models look promising, and were generally used
in the past for science analysis of COMPTEL data.
Since the BGDLIXA and BGDLIXE algorithms are equivalent as long as
$N_{\rm runav} \le N_{\rm avgr}$, we limit our study here to the simpler BGDLIXE algorithm.
Specifically, we investigate which values of $N_{\rm avgr}$ and $N_{\rm incl}$ provide
reliable background models without introducing a significant bias in the reconstruction of
celestial gamma-ray source characteristics.
In agreement with previous studies of the algorithms \citep[cf.][]{vandijk1996} we
always set $N_{\rm excl}=0$.

\begin{figure*}[!th]
\centering
\includegraphics[width=18cm]{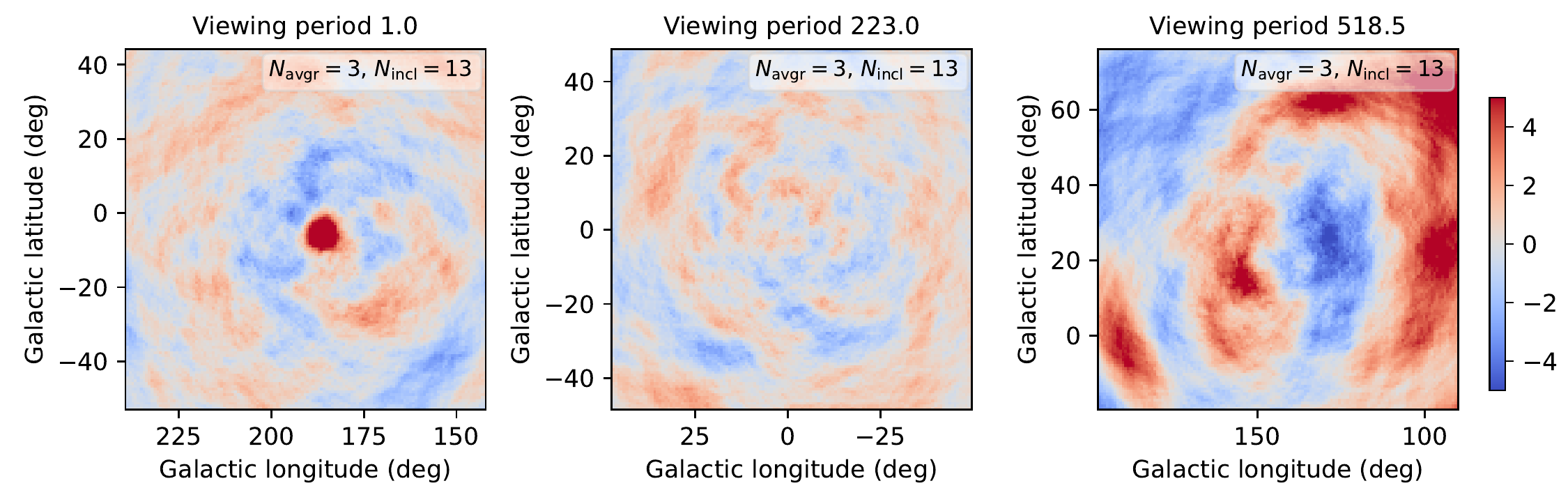}
\includegraphics[width=18cm]{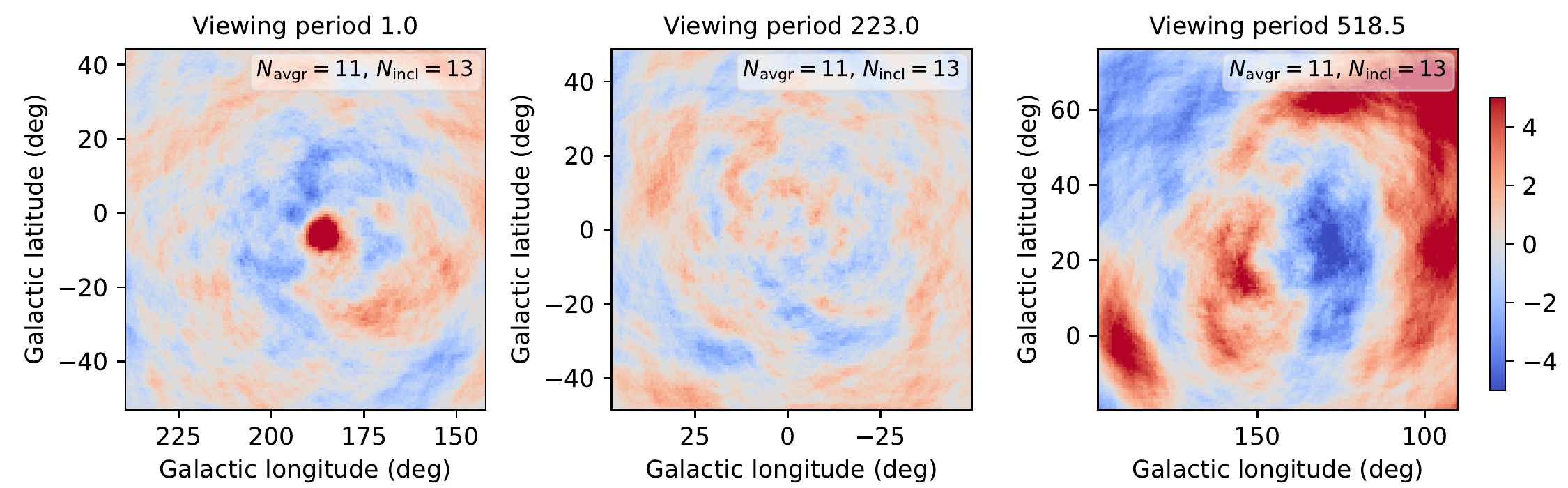}
\includegraphics[width=18cm]{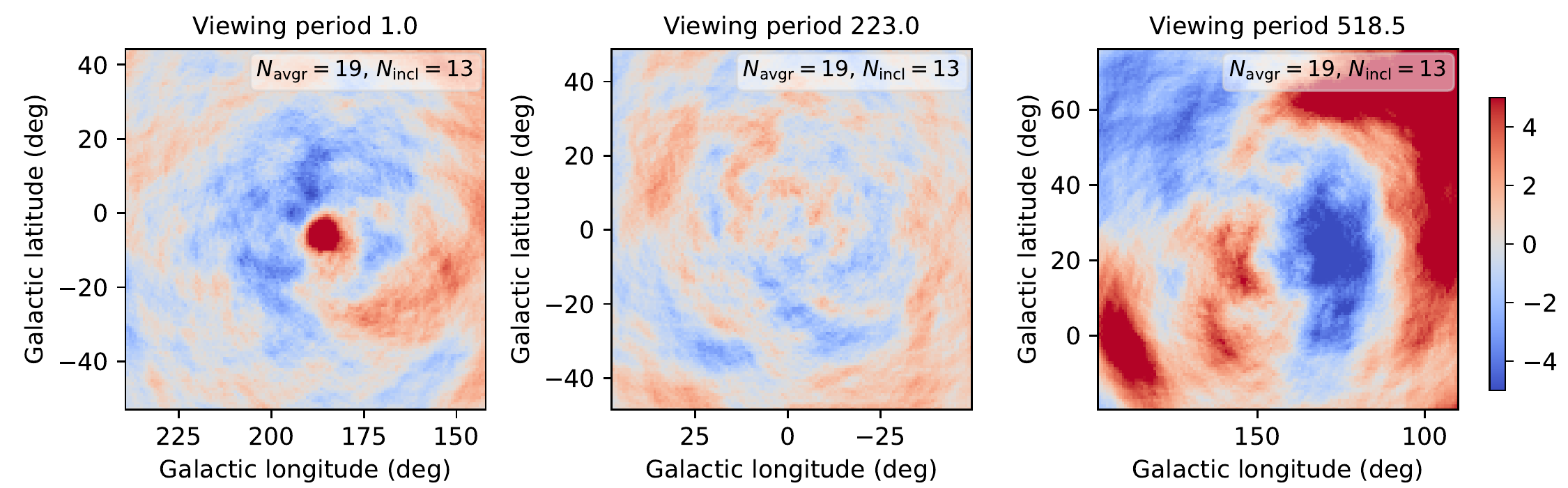}
\caption{
Same as Fig.~\ref{fig:resmaps1} but for $N_{\rm incl}=13$ and different values of $N_{\rm avgr}$.
\label{fig:resmaps2}
}
\end{figure*}

\subsubsection{Residual maps}
\label{sec:bgdres}

We started with modelling the background for all COMPTEL viewing periods and the four
standard COMPTEL energy bands using {\tt comobsback} and the BGDLIXE algorithm 
under variation of the parameters $N_{\rm avgr}$ and $N_{\rm incl}$.
For each viewing period and energy band we created residual maps and histograms using
{\tt comobsres} that we inspected visually.
We note that due to the dominance of the instrumental background in COMPTEL data, we
do not expect to see celestial gamma-ray emission in the residual maps of individual
viewing periods, with the exception of emission from the Crab pulsar and pulsar wind nebula,
which is the strongest gamma-ray source at MeV energies.
We find a general trend of stronger residuals at lower energies, with the largest residuals
observed in the $1-3$~MeV band.
Notably, residuals are stronger for viewing periods for which the Earth horizon selection
Eq.~(\ref{eq:eha}) introduces a strong $\bar{\varphi}$ dependence in the $\chi,\psi$ distribution
of the events.
The amplitude of the residuals changes strongly under variation of $N_{\rm incl}$, with larger
residuals for larger values of $N_{\rm incl}$, while variations of $N_{\rm avgr}$ impact the
residuals only moderately.

\begin{figure*}[!th]
\centering
\includegraphics[width=18cm]{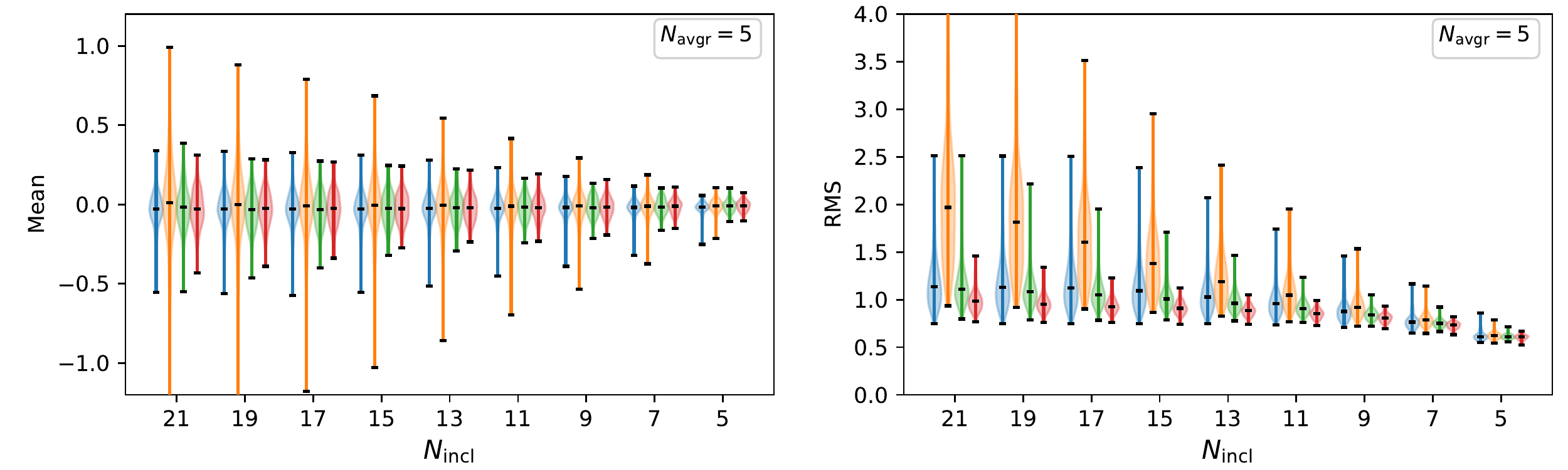}
\includegraphics[width=18cm]{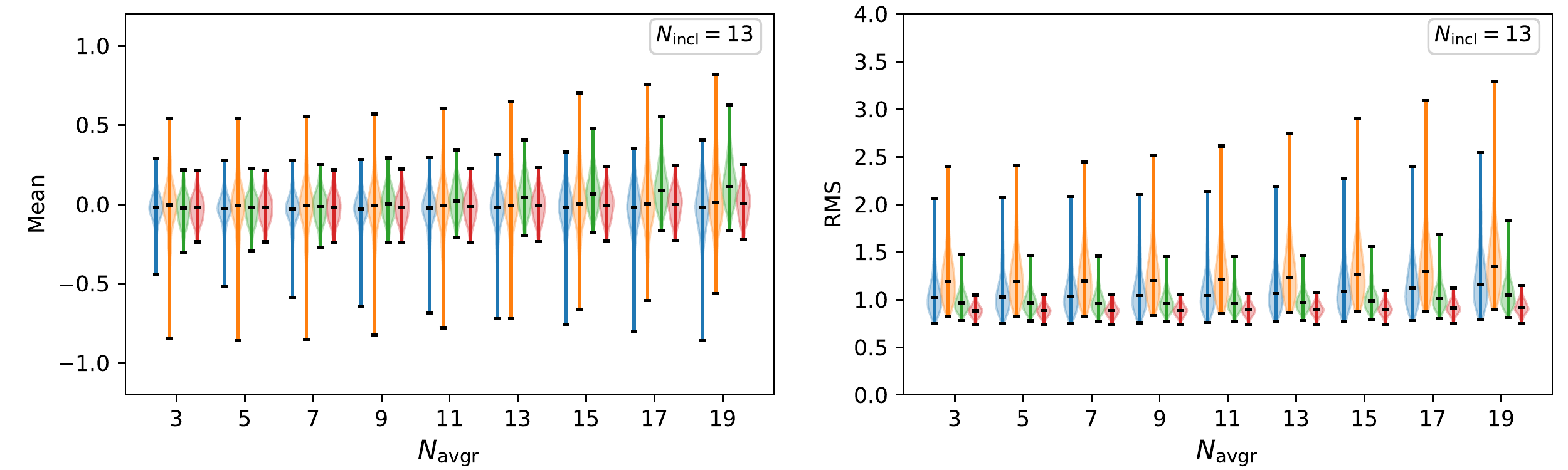}
\caption{
Violin plots of the mean and rms, in units of $\sigma$, of the residual maps with background modelled using the BGDLIXE
algorithm for all viewing periods except the ones that include the Crab.
The mean and rms are shown as a function of $N_{\rm incl}$ for $N_{\rm avgr}=5$ in the upper row
and as a function of $N_{\rm avgr}$ for $N_{\rm incl}=13$ in the lower row. 
The groups of four violins correspond to the four standard energy bands;
from left to right: $0.75-1$~MeV (blue), $1-3$~MeV (orange), $3-10$~MeV (green), and $10-30$~MeV
(red).
Horizontal black bars show the maximum, the median, and the minimum values, respectively.
\label{fig:resmapsviolos}
}
\end{figure*}

For illustration we show residual maps obtained using the BGDLIXE algorithm for different values 
of $N_{\rm avgr}$ and $N_{\rm incl}$ and for three representative viewing periods and the $1-3$~MeV 
energy band in Figs.~\ref{fig:resmaps1} and \ref{fig:resmaps2}.
Viewing period 1.0 is an observation of the Crab pulsar and pulsar wind nebula, which is clearly
visible in the residuals maps.
Viewing period 223.0 is an observation of 1E~1740.7--2942, a low-mass X-ray binary that is also known
as {the `Great Annihilator'} and that is situated near the Galactic centre.
The residuals in this viewing period are relatively modest. 
Viewing period 518.5 is an observation of the BL Lacertae object S5~0716+714, which is among
the viewing periods with the worst residuals, featuring large zones of significant excess counts and 
negative depressions.
The amplitude of the residuals clearly decreases with decreasing value of $N_{\rm incl}$, while at the
same time smaller value of $N_{\rm incl}$ also reduce the signal from the Crab.
As we will show later, some of this signal can be recovered using the iterative SRCLIX algorithm, yet
small values of $N_{\rm incl}$ tend to lead to an underestimation of source fluxes.
Therefore, the choice of $N_{\rm incl}$ is necessarily a trade-off between the amplitude of 
background residuals and the suppression of source flux.
On the other hand, the amplitude of the residuals changes only moderately with $N_{\rm avgr}$,
with a slight increase of the residual amplitudes for increasing values of $N_{\rm avgr}$.

To systematically quantify the residuals for a given choice of BGDLIXE parameters we computed
for all viewing periods and the four standard COMPTEL energy bands the mean and random mean
scatter (rms) of the residual maps.
The results for the BGDLIXE algorithm as a function of $N_{\rm incl}$ for $N_{\rm avgr}=5$ and as
a function of $N_{\rm avgr}$ for $N_{\rm incl}=13$ are shown in Fig.~\ref{fig:resmapsviolos}.
The violins represent the density distribution of the mean and rms of the residual maps for all
viewing periods.
Viewing periods including the Crab were excluded to avoid any bias due to the presence of a strong
source.

The plots confirm that the largest spread in the mean and rms are observed for lower energies,
with a particularly large spread for the $1-3$~MeV energy band.
A large spread indicates that for some viewing periods the background model results in important
residuals, while for other viewing periods the algorithm performs rather well, as illustrated in
Figs.~\ref{fig:resmaps1} and \ref{fig:resmaps2}.
Reducing $N_{\rm incl}$ considerably reduces the spread in the mean and rms values for all
energy bands, yet at some point the rms drops below the expected value of 1, indicating that the 
background model partially follows the statistical fluctuations of the data.
This is also the regime where the source fluxes start to get underestimated.

This overfitting can be slightly reduced by increasing $N_{\rm avgr}$ so that more events get
included in the $\chi,\psi$ averaging, as indicated in the lower row of Fig.~\ref{fig:resmapsviolos}.
On the other hand, increasing $N_{\rm avgr}$ leads to a slight increase of the mean and rms
distributions; hence, the selected value of $N_{\rm avgr}$ should also not be too large.

\begin{figure*}[!t]
\centering
\includegraphics[width=17.5cm]{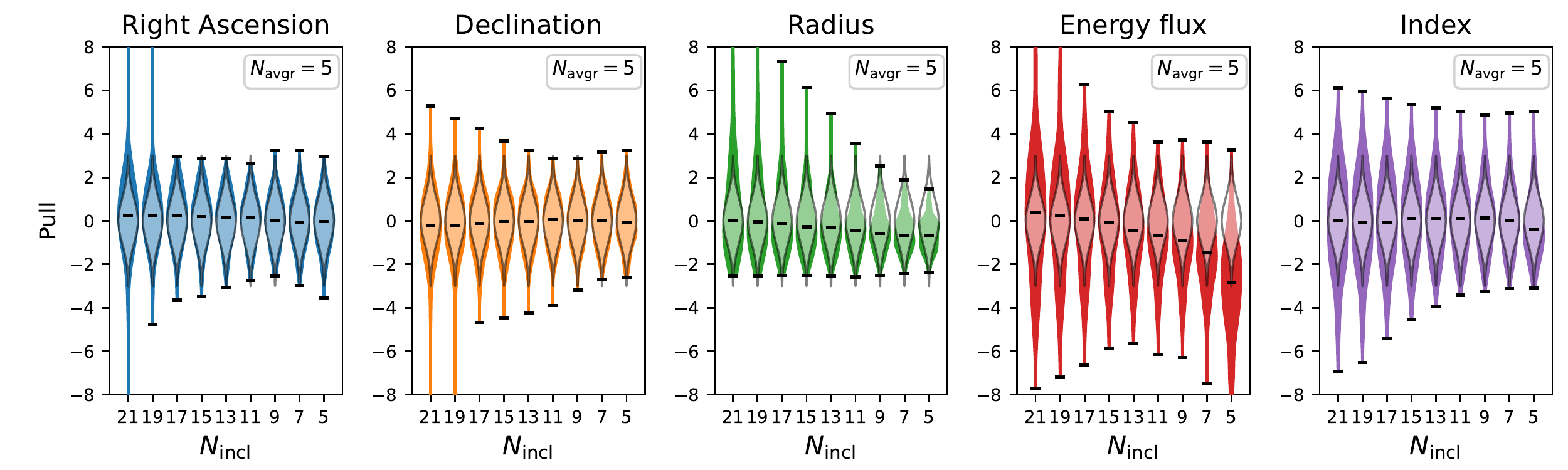}
\includegraphics[width=17.5cm]{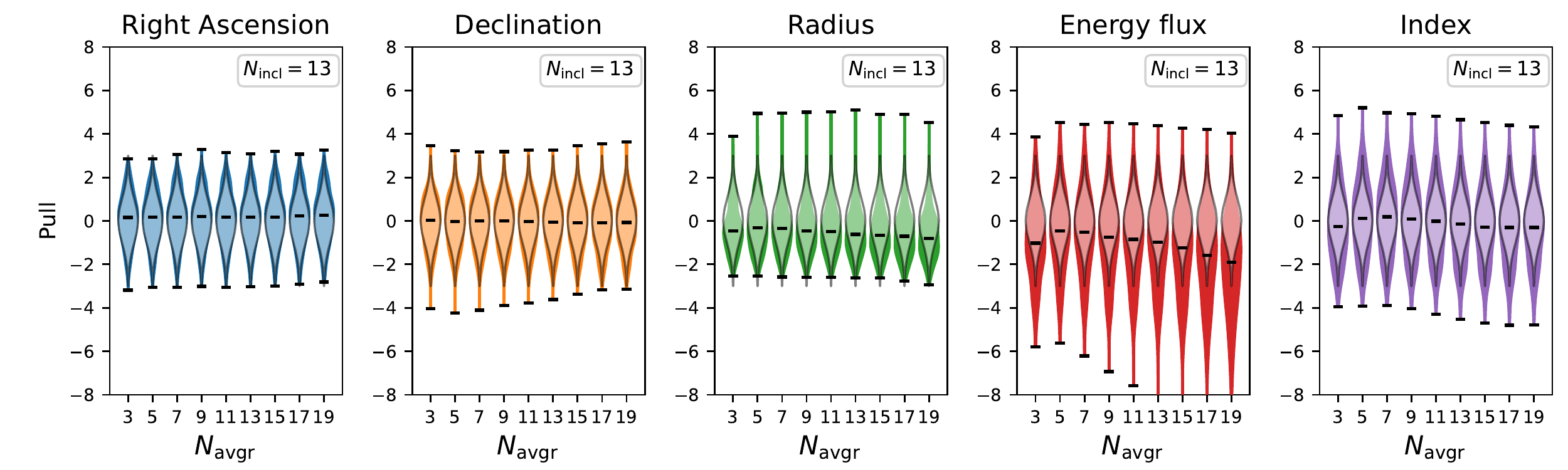}
\caption{
Violin plots of the fitted parameter pull distributions for a simulated source at a $20\degrees$ off-axis
angle for all viewing periods except those that include the Crab.
The upper row shows results as a function of $N_{\rm incl}$ for $N_{\rm avgr}=5$; the lower
row shows results as a function of $N_{\rm avgr}$ for $N_{\rm incl}=13$. 
For all simulations the source was simulated as a spatially extended disk with a $3\degrees$ radius
and with a power-law spectrum of index $\Gamma=2.1$.
Horizontal black bars show the maximum, the median, and the minimum values, respectively.
Semi-transparent violins with grey borders indicate the expected pull distributions from statistical
fluctuations only.
\label{fig:pullviolons}
}
\end{figure*}

\subsubsection{Flux reconstruction}

As the next step we studied the impact of the BGDLIXE parameters on the fitted values of celestial 
source parameters, such as source position, extent, flux and spectral index.
We do this by using {\tt comobssim} to add a simulated source at an offset angle of 
$20\degrees$ with respect to the pointing axis to the data of each viewing period for the four 
standard energy bands.
In that way, our study relies essentially on the observed event distribution, and hence is representative
for a real analysis situation, while the characteristics of the celestial source are known.
As simulated source model we use a spatially extended disk component with radius of $3\degrees$
combined with a spectral power-law component with an energy flux of
$1.068\times10^{-8}$ \eunit\ within $0.75-30$~MeV and a spectral index of $2.1$,
which roughly corresponds to the spectral parameters that are observed for the Crab.
The simulated data of each viewing period were then fit jointly for the four energy bands using
{\tt comlixfit}, determining the maximum likelihood right ascension, declination, disk radius,
energy flux, and spectral index of the source.
Initial values for the iterative fitting procedure were offset from the simulated values since in
a real data analysis the true source parameters are generally not known in advance.
Viewing periods including the Crab were excluded from the analysis to avoid any interference
with a known strong source of gamma rays.
We furthermore assume that no other source of gamma rays is significantly detected in any of
the remaining individual viewing periods, considering these viewing periods as empty fields for
the purpose of this analysis.

For each fitted source parameter, $i$, we determine the pull
\begin{equation}
{\rm Pull}(p_i) = \frac{p_i - v_i}{\sigma_i}
,\end{equation}
where $p_i$ is the fitted value, $v_i$ is the simulated value, and $\sigma_i$ is the statistical 
parameter uncertainty as determined by {\tt comlixfit}.
In the absence of systematic uncertainties, and under the assumption that the statistical
uncertainties are following a Gaussian distribution, ${\rm Pull}(p_i)$ follows a Gaussian
distribution with a mean of zero and a standard deviation of $\sigma=1$.

Figure \ref{fig:pullviolons} summarises the results of the analysis, showing the pull distributions 
for all considered viewing periods as violin plots, with violins for right ascension, declination, 
disk radius, energy flux, and spectral index.
The expectations for purely statistical parameter fluctuations are indicated by semi-transparent 
violins with grey borders, the horizontal black bars indicate the maximum, median and minimum 
pull of the distributions.
The upper row shows results as a function of $N_{\rm incl}$ for $N_{\rm avgr}=5$, the
lower row shows results as a function of $N_{\rm avgr}$ for $N_{\rm incl}=13$.

Figure \ref{fig:pullviolons} indicates that large values of $N_{\rm incl}$ lead to pull distributions
that are broader than expected from statistical fluctuations only, in particular for the energy flux, 
but also for the spectral index and to a lesser extent for the other source parameters.
The broadening is due to background residuals in some of the viewing periods for large $N_{\rm incl}$, 
as illustrated in Fig.~\ref{fig:resmaps1}, which impact the reconstructed source parameters.
Reducing $N_{\rm incl}$ brings the pull distributions more in line with the expectations, yet
for $N_{\rm incl}\le13$ the median pull of the fitted energy flux drops below zero,
indicating a systematic bias towards too low fluxes.
Specifically, for $N_{\rm incl}=5$, where background residuals are very small 
(cf.~Fig.~\ref{fig:resmaps1}), the flux reconstruction is strongly biassed, resulting is a significant
underestimation of gamma-ray fluxes.
The optimum value for $N_{\rm incl}$ is hence a trade-off between reduction of background
residuals and biassing the flux determination.

Flux biassing is also observed with increasing value of $N_{\rm avgr}$, with a minimum bias
that occurs for $N_{\rm avgr}=5$.
Using hence $N_{\rm avgr}=5$ and $N_{\rm incl}=15$ for the BGDLIXE background modelling
leads to results that are basically free from any systematic bias, although some significant 
background residuals may persist for this choice of values.
As illustrated in Fig.~\ref{fig:pullviolons}, these background residuals translate into an additional
uncertainty beyond the statistical fluctuations only.
In the present case, the standard deviation of the energy flux pull distribution for $N_{\rm avgr}=5$ 
and $N_{\rm incl}=15$ is about twice as large as expected from statistical fluctuations only.
In other words, when analysing individual viewing periods using BGDLIXE parameters 
$N_{\rm avgr}=5$ and $N_{\rm incl}=15$, the uncertainties in the energy flux related to the 
background modelling roughly doubles the uncertainties due to statistical fluctuations only.

In the following we use $N_{\rm avgr}=5$ and $N_{\rm incl}=15$ for the analysis in our
paper, and we generally recommend to use these parameters for COMPTEL data analysis with 
GammaLib and ctools.
We note that these values apply for a binning of $1\degrees$ in $\chi$ and $\psi$ and
$2\degrees$ in $\bar{\varphi}$, and that for a different binning the parameters need to be adjusted
accordingly.
Specifically, we used an equivalent value of $N_{\rm incl}=29$ for analyses in our paper for which 
a binning of $1\degrees$ is applied in $\bar{\varphi}$.

\subsection{Gamma-ray emission from the Crab}
\label{sec:crab}

We began the science validation of our software with a spectral analysis of the gamma-ray emission
from the Crab pulsar and pulsar wind nebula, which is the brightest source of gamma rays at MeV 
energies.
The MeV flux is dominated by emission from the pulsar wind nebula, yet using pulsar ephemerides
derived from radio observations the emission from the pulsar is also clearly detectable over the entire 
COMPTEL energy range.
The emission from the Crab pulsar and pulsar wind nebula was studied extensively by COMPTEL 
in the past
\citep[e.g.][]{much1995a,much1995b,much1996,vandermeulen1998,kuiper2001}.

\subsubsection{Total spectrum}

\begin{figure}[!t]
\centering
\includegraphics[width=8.8cm]{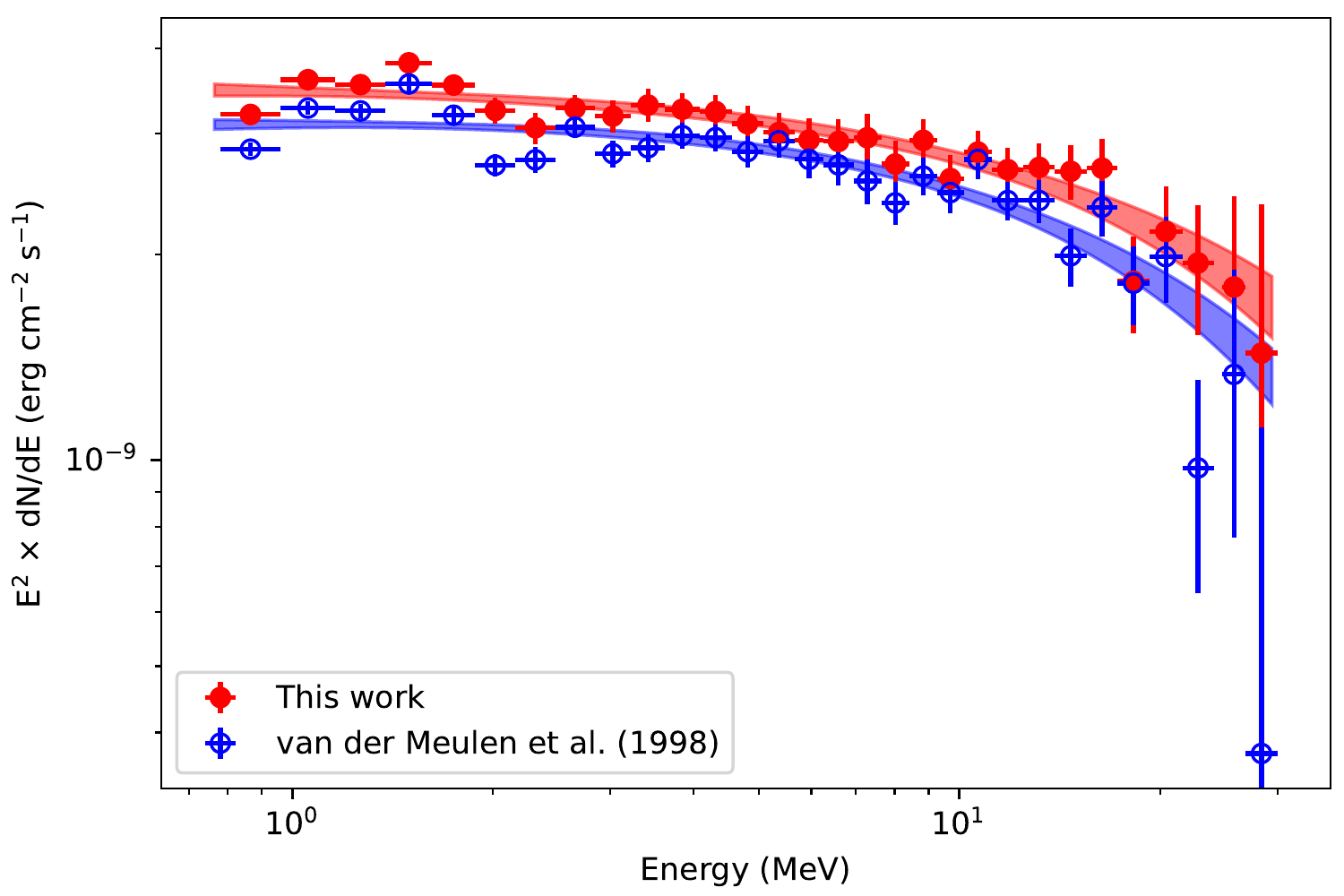}
\caption{
Spectral energy distribution of the total emission from the Crab pulsar and pulsar wind nebula as 
measured by COMPTEL in 30 bins covering the energy band $0.78-30$~MeV.
Filled red dots correspond to results obtained with ctools, and open blue dots
correspond to results obtained by \citet{vandermeulen1998} using COMPASS.
The shaded regions correspond to the $1\sigma$ uncertainty bands of the fitted
exponentially cut-off power-law models.
\label{fig:crab}
}
\end{figure}

We first considered the combined emission from the Crab pulsar and pulsar wind nebula.
In their analysis of five years of COMPTEL observations, \citet{vandermeulen1998} analysed the
spectrum of the total Crab emission within the energy range $0.78-30$ MeV in 30 narrow energy
bins.
Since this is the only work that quotes total flux values for the Crab (cf.~Table 2 of the publication),
we used this study as reference.

We analysed the same data that was used by \citet{vandermeulen1998} with GammaLib and ctools, 
except for viewing period 0 that is not available at HEASARC and viewing period 426.0 for which the EVP
file in the HEASARC archive has a corrupted content.
We binned the data according to the 30 energy bins defined by \citet{vandermeulen1998} and combined
the data for all viewing periods using {\tt comobsadd} using 80 bins in $\chi$ and $\psi$ that were
centred on the position of the Crab pulsar, taken here to be $83.6331\degrees$ in right ascension 
and $22.0145\degrees$ in declination (J2000).
Similar to \citet{vandermeulen1998} we used 50 bins in $\bar{\varphi}$, bin sizes of $1\degrees$ in all 
three data space dimensions, and instrument response functions derived by analytical modelling.

We modelled the Crab using a point source with fixed position, as given above, and using power-law, 
exponentially cut-off power-law or curved power-law spectral models.
We fitted the data jointly for the 30 energy bins using {\tt comlixfit} with BGDLIXE background model 
parameters $N_{\rm avgr}=5$ and $N_{\rm incl}=29$.
The best fit was obtained using the exponentially cut-off power law
\begin{equation}
I(E_{\gamma}) = k \left( \frac{E_{\gamma}}{3.5 MeV} \right)^{-\Gamma} \exp \left( -\frac{E_{\gamma}}{E_{\rm c}} \right)
\label{eq:eplaw}
,\end{equation}
with
$k=(1.81 \pm 0.06) \times 10^{-4}$ \fmev,
$\Gamma=2.00 \pm 0.03$, and
$E_{\rm c}=39.1 \pm 9.7$ MeV.
\citet{vandermeulen1998} do not quote the parameters of a fitted spectral model to the total Crab
emission data, so we used the GammaLib multi-wavelength interface to adjust the same spectral models
to the data of their Table 2, which also favoured the exponentially cut-off power law with best fitting
parameters
$k=(1.70 \pm 0.05) \times 10^{-4}$ \fmev,
$\Gamma=1.97 \pm 0.02$, and
$E_{\rm c}=29.8 \pm 5.0$ MeV.
While our prefactor $k$ is about $6$\% larger than the one obtained from fitting the spectrum of
\citet{vandermeulen1998}, the other spectral parameters are equivalent within statistical uncertainties.

We then used {\tt ctbutterfly} to determine uncertainty bands for the spectral models and {\tt csspec}
to derive flux points for the 30 energy bins.
The results of our analysis are compared to those of \citet{vandermeulen1998} in Fig.~\ref{fig:crab}.
The uncertainty bands for the data of \citet{vandermeulen1998} were determined using {\tt ctbutterfly}.
Overall the agreement between both analyses is quite good, yet as mentioned earlier, our flux points 
lie somewhat above the ones determined by \citet{vandermeulen1998}.
Possibly this discrepancy may be related to correction factors that were applied posterior to COMPASS
analyses at the time that were not automatically taken into account by the software.
These correction factors include an energy-dependent ToF correction factor (cf.~Table \ref{tab:tofcor}), an 
energy-independent deadtime correction factor of $0.965$ as well as an energy-independent flux 
correction factor that was eventually applied to SRCLIX analyses to correct for a flux suppression
that arose from the modification of the instrument response function \citep{vandijk1996}.
We recall that GammaLib automatically applies the ToF and deadtime correction factors to the results.
Whether or not such correction factors were applied by \citet{vandermeulen1998} is not known, yet they
may plausibly explain the $6\%$ discrepancy observed between the analyses.

\begin{figure}[!t]
\centering
\includegraphics[width=8.8cm]{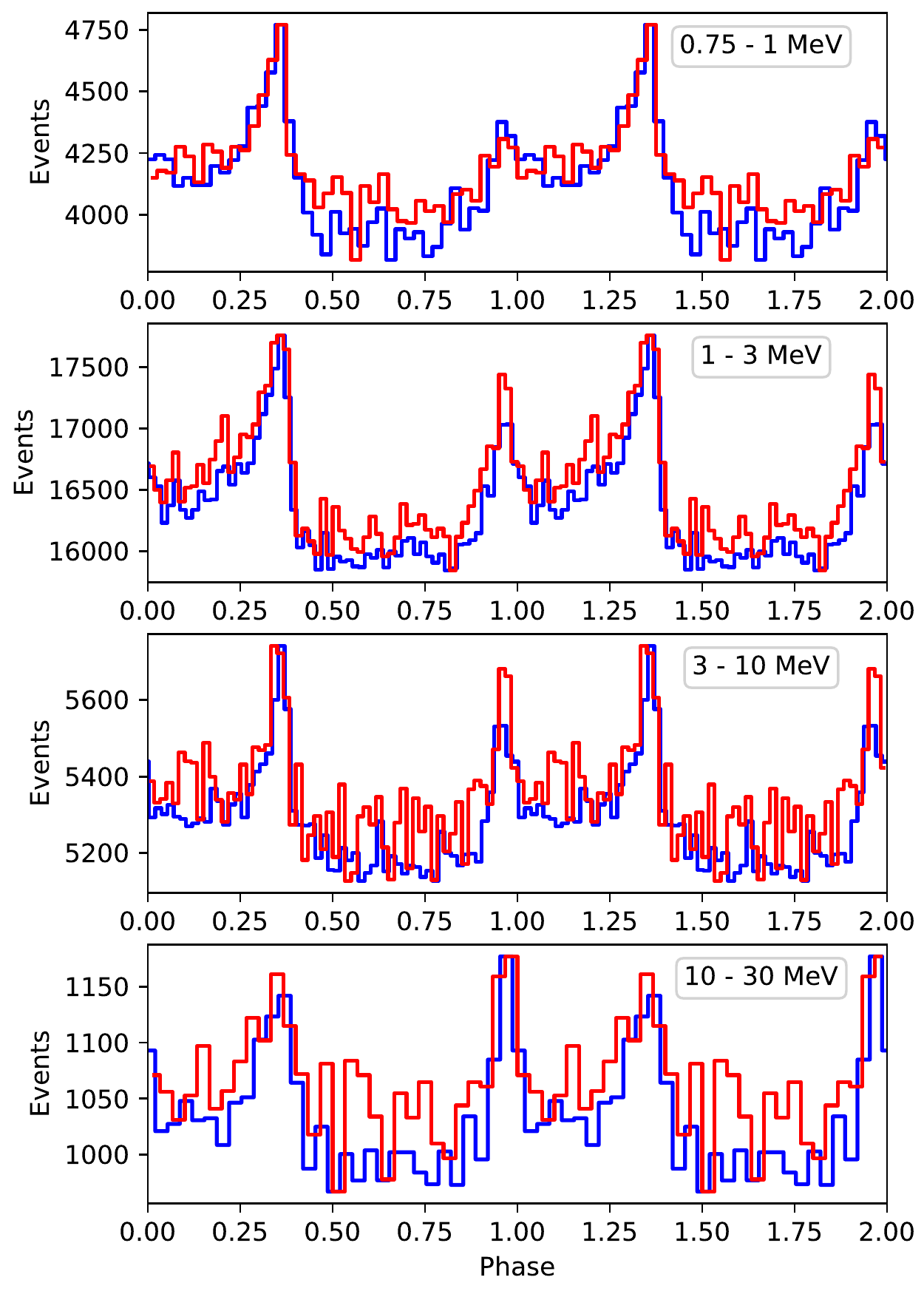}
\caption{
Pulse profiles of the Crab pulsar derived using ctools for the four standard energy bands
(red) compared to pulse profiles derived by \cite{kuiper2001} using the COMPASS software
(blue). The latter profiles were scaled to match our results in amplitude, and a phase shift of 
$-0.03$ was applied to match the profiles in phase.
\label{fig:crabpulsar}
}
\end{figure}

\subsubsection{Pulsar and nebula components}
\label{sec:crabpulsar}

We now turn to an analysis that separates the emission from the Crab pulsar and the pulsar wind
nebula to validate the implementation of the pulsar phase computations.
The most comprehensive analysis of the emission from the Crab pulsar and pulsar wind nebula
components using COMPTEL data was performed by \cite{kuiper2001} using data collected over
the nine years mission duration of CGRO.
While \cite{kuiper2001} used data from 33 viewing periods with pointing axis within $30\degrees$
of the Crab pulsar, we analysed 24 viewing periods that we found with the same selection criteria in
the HEASARC database, covering the viewing periods specified in Table 1 of \cite{kuiper2001} 
between viewing period 1.0 and viewing period 616.1.
As in the analysis above, viewing period 426.0 was excluded from the list since no usable EVP file 
exists for this observation in the HEASARC database.
Similarly to \cite{kuiper2001} we used ephemerides for the Crab pulsar from the Princeton radio pulsar
database, provided in the form of an ASCII file named {\tt psrtime.dat}\footnote{
  Details are provided at \url{https://heasarc.gsfc.nasa.gov/lheasoft/ftools/fhelp/fasebin.html}} 
that is part of the reference data of the X-ray Timing Explorer (XTE) module of the HEASoft software 
(version 6.29).\footnote{
  The HEASoft software can be downloaded from \url{https://heasarc.gsfc.nasa.gov/docs/software/lheasoft/}}

\begin{figure}[!t]
\centering
\includegraphics[width=8.8cm]{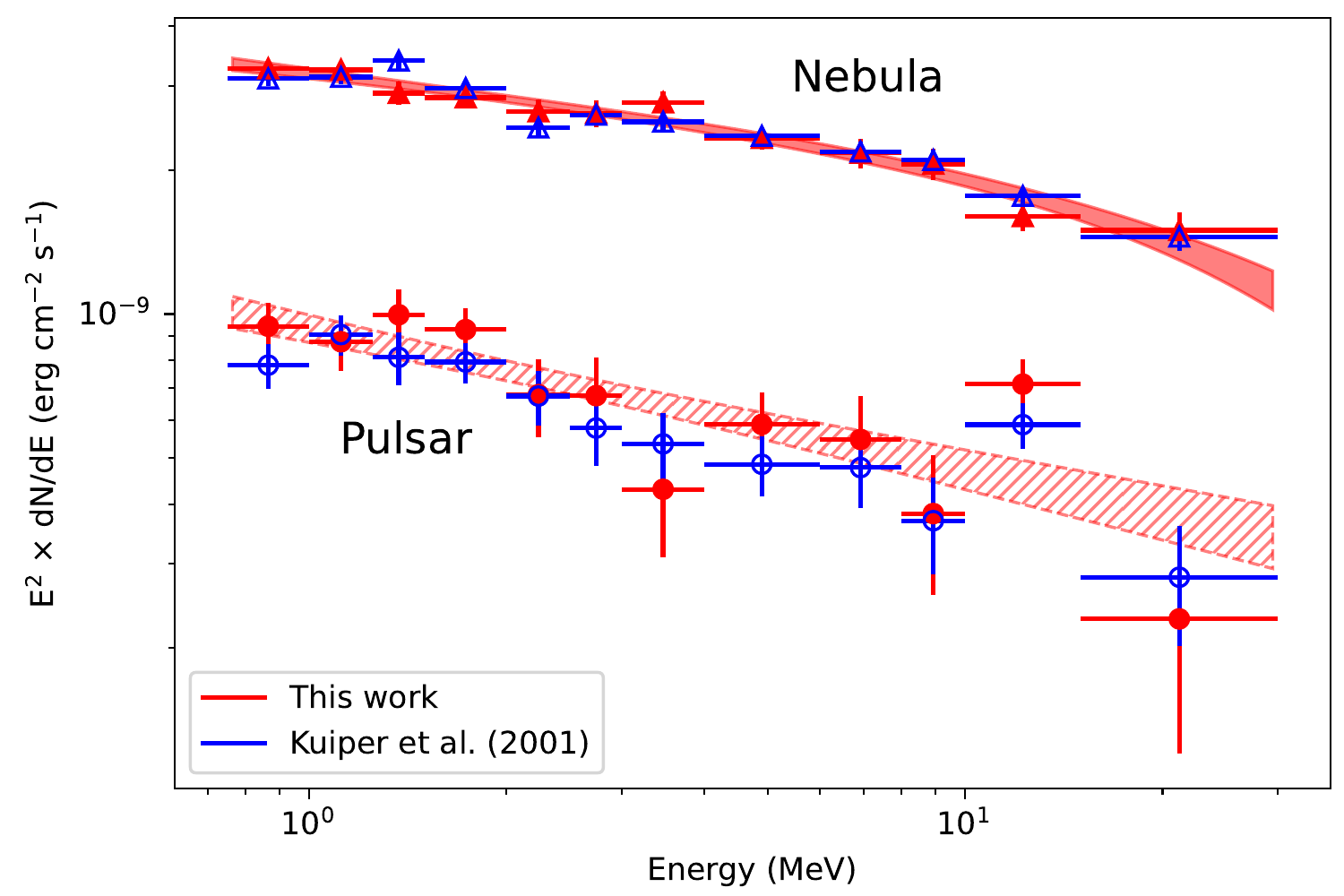}
\caption{
Spectral energy distributions of the Crab pulsar and pulsar wind nebula components as determined 
using GammaLib and ctools (red) and by \cite{kuiper2001} using COMPASS (blue).
Results for the Crab pulsar are shown as dots, and results for the Crab pulsar wind nebula are 
shown as triangles.
The figure also shows the $1\sigma$ uncertainty bands of the best fitting spectral models for
both components.
\label{fig:crab-pulsar-sed}
}
\end{figure}

We first used {\tt compulbin} with an angular resolution measure of $\pm3.5\degrees$ to produce pulse 
profiles for the four standard COMPTEL energy bands, as displayed in Fig.~2 of \cite{kuiper2001}.
The results of this analysis are shown in Fig.~\ref{fig:crabpulsar}, on which we overlay for comparison 
the pulse profiles obtained by \cite{kuiper2001}.
Since the \cite{kuiper2001}  profiles were obtained for a larger dataset and angular resolution measures that were not
specified in their paper, we scaled the profiles so that the minimum and maximum number of events in 
the profiles matches the numbers that we obtained in our analysis.
We also applied a phase shift of $-0.03$ to the pulse profiles of \cite{kuiper2001} to match them
to our profiles.
While we do not know the origin of this small discrepancy in the pulse phases, we note that a phase 
shift of $-0.03$ corresponds to a difference of about $0.5$ arcsec in the assumed right ascension 
of the Crab pulsar.
In GammaLib, the pulsar position is taken from the ephemerides file, which in the present case is
the radio position in the Princeton database, while \cite{kuiper2001} do not specify the position that 
they assumed for the Crab pulsar.
We note that the radio position in the Princeton database differs by about $0.5$ arcsec from the 
International Celestial Reference System (ICRS) position provided by the Set of Identifications, 
Measurements, and Bibliography for Astronomical Data (SIMBAD) service,\footnote{
  See \url{http://simbad.u-strasbg.fr/simbad/sim-basic?Ident=Crab+pulsar&submit=SIMBAD+search}}
which could be at the origin of the observed phase shift.

\begin{figure*}[!t]
\centering
\includegraphics[width=18cm]{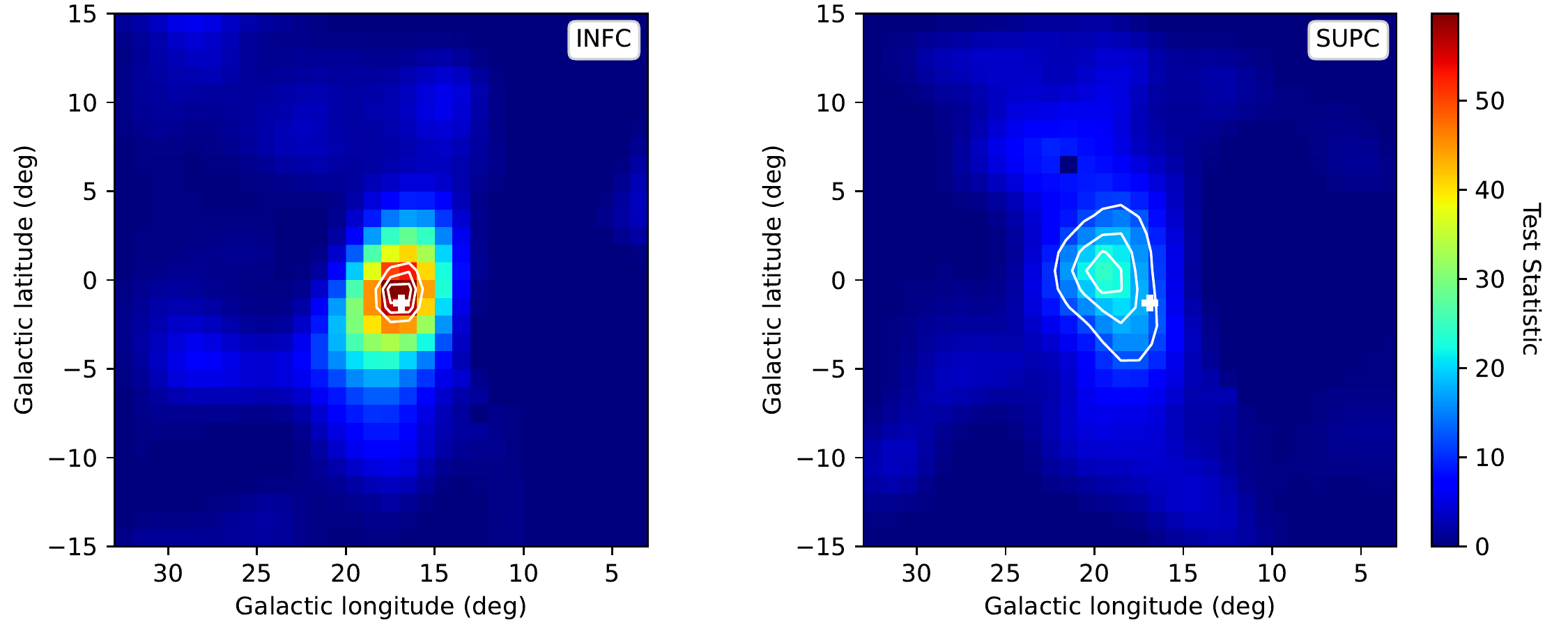}
\caption{
Test statistic maps of LS~5039 for inferior conjunction (left) and superior conjunction (right) phase
intervals derived by jointly fitting the four standard COMPTEL energy bands, covering
$0.75-30$ MeV.
The location of LS~5039 is shown by a white plus symbol, and contours show the $1\sigma$, $2\sigma$, 
and $3\sigma$ location uncertainties.
The quasar PKS~1830--210 was included in the source model and hence is not visible in the maps.
\label{fig:lsmaps}
}
\end{figure*}

As the next step we determined the spectra of the Crab pulsar and pulsar wind nebula to compare them
to those given in Table 3 of \cite{kuiper2001}.
For this purpose we binned the events using {\tt comobsbin} for the 12 energy bins specified in that
table.
The data were binned separately for the Off Pulse and Total Pulse phase intervals as defined in Table 2 
of \cite{kuiper2001}, shifted by $-0.03$ to accommodate for the observed phase shift.
Specifically, the Off Pulse interval comprises phases $0.49-0.85$ while the Total Pulse interval
comprises phases $0.85-1$ and $0-0.49$.
We fitted the data for both intervals jointly using {\tt comlixfit} with the BGDLIXE parameters 
$N_{\rm avgr}=5$ and $N_{\rm incl}=29$.
We used two point-source model components in our fit, one for the Crab nebula that was fitted to the
data of both phase intervals, and one for the Crab pulsar that was only fitted to the data of the Total 
Pulse interval.
Consequently, the Crab pulsar component modelled only events that were in excess of the pulsar 
wind nebula component.
Both point-source model components were located at the position of the Crab pulsar, as defined above, 
and had a spectral model with a free flux value for each of the 12 energy bins.
Similar to \cite{kuiper2001}, we used simulated instrument response functions and 50 bins in $\bar{\varphi}$ 
with bin sizes of $1\degrees$ for our analysis.
 
The spectra obtained with our analysis are shown in Fig.~\ref{fig:crab-pulsar-sed} together with 
the spectra obtained by \cite{kuiper2001}.
The agreement between the results is quite satisfactory and differences are generally well within
statistical uncertainties.
We note that there is no general flux offset between ours and the COMPASS analysis, as observed 
above for the total Crab spectrum, and that differences are plausibly explained by the use
of a different lists of viewing periods.
\cite{kuiper2001} noticed an enhanced emission in the $10-15$ MeV energy interval for the Crab
pulsar, and also in our analysis we found an equivalent feature.
We note, however, that by shifting the phase interval definition by $+0.03$ 
\citep[i.e.~using the original phase interval definition of][]{kuiper2001} 
this spectral enhancement is considerably reduced in our analysis, suggesting that the enhancement 
is probably a statistical fluctuation rather than a physical feature.

\begin{figure*}[!t]
\centering
\includegraphics[width=8.8cm]{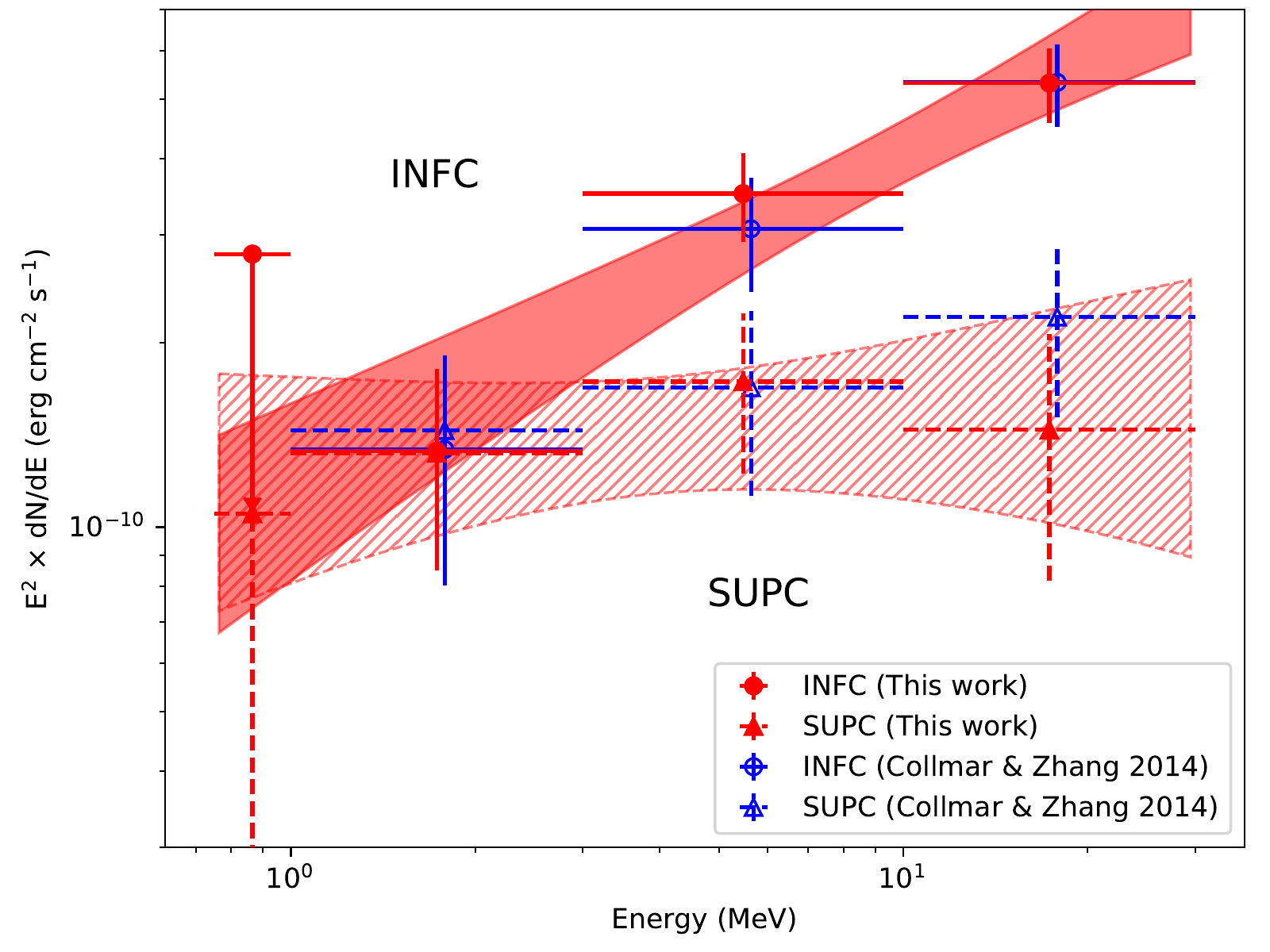}
\includegraphics[width=8.8cm]{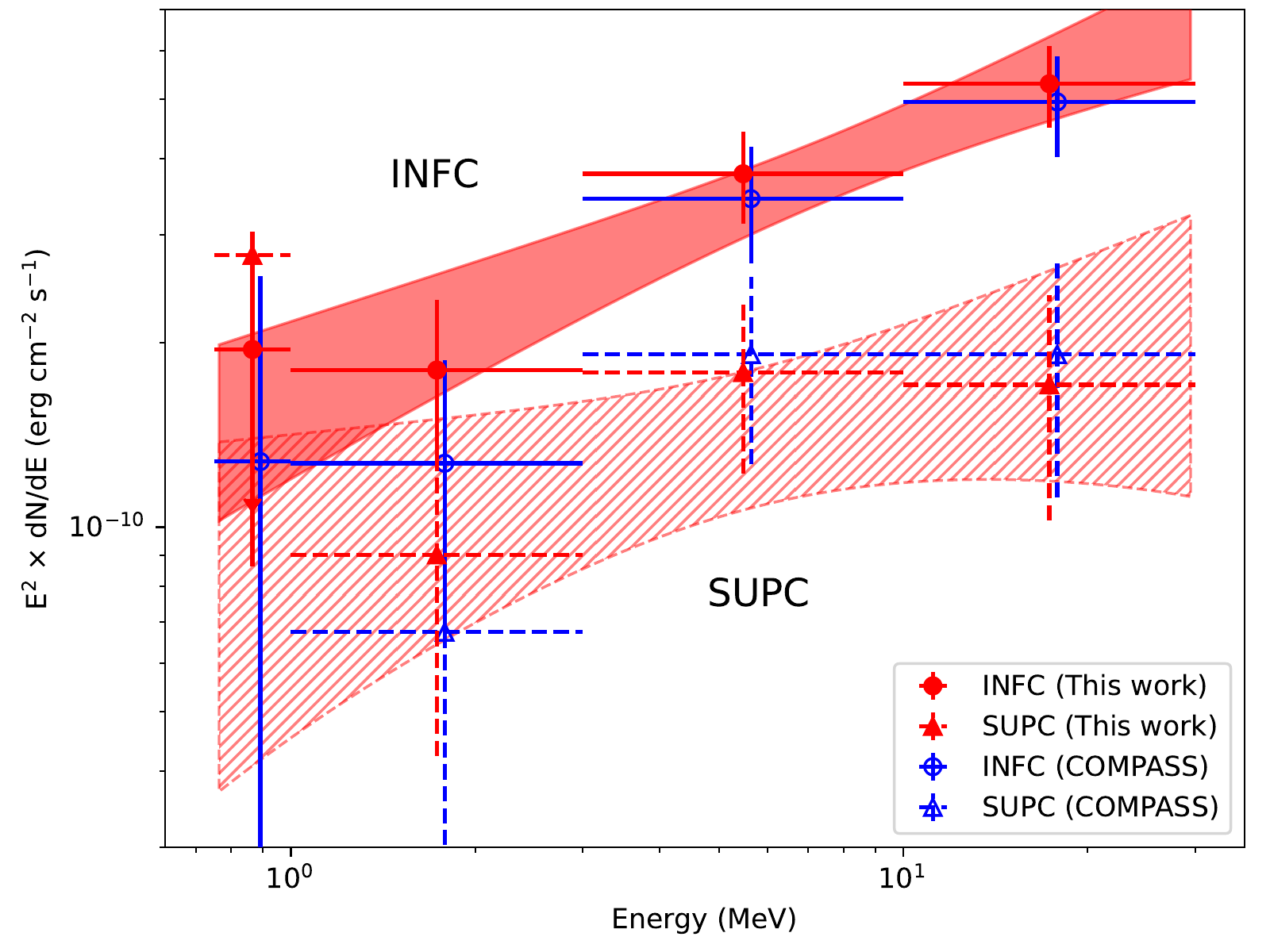}
\caption{
Spectral energy distributions and $1\sigma$ uncertainty bands of fitted power-law models for
LS~5039 for the INFC (solid lines and filled area) and the SUPC 
(dashed lines and hatched area). 
The left panel compares the ctools and GammaLib results (red) to the results obtained by 
\citet{collmar2014} for an identical event selection (blue).
The right panel compares the ctools and GammaLib results (red) to the results obtained using the
COMPASS software by excluding D2 modules with faulty PMTs (blue).
Data points from \citet{collmar2014} and derived using COMPASS were displaced by 3\% in energy 
for clarity.
\label{fig:ls5039}
}
\end{figure*}

We also fitted different spectral models to the data of the Crab pulsar and pulsar wind nebula,
including power laws, exponentially cut-off power laws and curved power laws.
We determined the corresponding uncertainty bands using {\tt ctbutterfly} and overlay
them on the spectral points in Fig.~\ref{fig:crab-pulsar-sed}.
Using an exponentially cut-off power law for the nebula component instead of a simple power
law improved the TS value of the nebula component by 6.5, corresponding to a detection
significance of $2.5\sigma$ for the spectral cutoff.
For the pulsar component no improvement was achieved when allowing for a cutoff or a curvature
in the power law.
For the Crab pulsar wind nebula, the best fitting parameters of Eq.~(\ref{eq:eplaw}) were
$k=(13.7 \pm 0.5) \times 10^{-5}$ \fmev,
$\Gamma=2.15 \pm 0.03$, and
$E_{\rm c}=53.3 \pm 15.5$ MeV.
For the Crab pulsar the best fitting power-law parameters were
$k=(5.1 \pm 0.3) \times 10^{-5}$ \fmev\ and
$\Gamma=2.29 \pm 0.06$.
This compares to the spectral indices of $2.227 \pm 0.013$ and $2.24 \pm 0.04$ determined by 
\cite{kuiper2001} for the nebula and pulsar components using power-law models, respectively.
While our index for the pulsar component is compatible with their result, our index for the 
nebula component is flatter, which can be explained by the spectral cutoff in our model.
Using a simple power law for the nebula component, as \cite{kuiper2001} did, we obtained a 
steeper index of $2.24 \pm 0.02$ that is compatible with their result.

\subsection{Phase-resolved analysis of LS~5039}
\label{sec:ls5039}

We now turn to an analysis of COMPTEL observations of the gamma-ray binary LS~5039 in
order to validate the ability to conduct phase-resolved analyses with GammaLib and ctools.
Using an orbit-resolved analysis, \citet{collmar2014} found strong evidence that the MeV flux
of GRO J1823-12, the strongest unidentified COMPTEL source in the Galactic plane, is modulated 
along the binary orbit of about 3.9 days of LS~5039.
Specifically, using maximum likelihood significance maps, the authors demonstrated that GRO J1823-12
shows a more significant signal for the inferior conjunction period of LS~5039 as compared to the
superior conjunction period.
The same trend was also observed in their spectral analysis.

We repeated the analysis of \citet{collmar2014} by choosing all viewing periods with pointing
within $35\degrees$ of $(l,b)=(17.5\degrees,-0.5\degrees)$ from the HEASARC database.
In total this resulted in a list of 41 viewing periods, starting with viewing period 5.0 and ending with
viewing period 712.0.
Up to viewing period 712.0 our list is identical to Table~1 of \citet{collmar2014}, yet the authors
included 12 more viewing periods in their analysis that are not available in the HEASARC
archive.
For our analysis we combined the data in a data space of  $140\times123\times25$ bins of 
$1\degrees\times1\degrees\times2\degrees$ in size and centred on 
$(l,b)=(15.0\degrees,-4.5\degrees)$, which corresponds to the same dimensions that were used
by \citet{collmar2014} in some of their analyses.
We used the four standard COMPTEL energy bands for our analysis.
Similar to \citet{collmar2014} we used the binary ephemeris of \citet{casares2005}, which is
an orbital period of 3.90603 days with periastron passage (corresponding to phase 0) at
JD $2451943.09$, and we define phases $0.45 \le \Phi < 0.9$ as the inferior conjunction 
interval (INFC) and phases $\Phi \ge 0.9$ and $\Phi < 0.45$ as the superior conjunction interval
(SUPC).

As the first step we created TS maps of the region around LS~5039 using {\tt comlixmap} 
for the INFC and SUPC phase intervals.
The data for the four standard energy bands were analysed jointly, using a model composed
of a test point source and an additional point source at the location of the quasar PKS~1830--210
that is spatially close to LS~5039.
The spectra of both components were modelled using power laws.
In addition, the source model included components for Galactic diffuse emission based on template maps 
for bremsstrahlung and inverse Compton emission with free scaling factors for each energy bin.\footnote{
  We used the bremsstrahlung emission map with the COMPASS identifier {\tt MPE-MIS-13006} and
  the inverse Compton emission map with the COMPASS identifier {\tt ROL-MIS-163} that were also 
  used by \citet{collmar2014}.}
Furthermore, an isotropic component was included to account for the cosmic gamma-ray background 
emission, with intensity fixed according to 
$I(E_{\gamma}) = 1.12 \times 10^{-4} ( E_{\gamma}/{5 \, {\rm MeV}} )^{-2.2}$ \fmevster\, as suggested 
by \citet{weidenspointner1999}.
The background was modelled using the BGDLIXE algorithm with parameters $N_{\rm avgr}=5$ and 
$N_{\rm incl}=15$.

The resulting TS maps are shown in Fig.~\ref{fig:lsmaps} for the INFC (left) and SUPC (right) phase
intervals.
The maps can be compared to those shown in Fig.~6 of \citet{collmar2014}, which were determined 
separately for the $3-10$ MeV and $10-30$ MeV energy bands.
In both analyses, LS~5039 is considerably more significant in the INFC phase interval but only weakly 
detected in the SUPC phase interval.
In the latter interval, the emission maximum seems to be displaced towards the north-east with 
respect to the position of LS~5039 in ours and the analysis of \citet{collmar2014}, yet the emission 
location is still compatible within the $3\sigma$ uncertainty contour with the position of LS~5039.

As the next step we derived the spectral energy distribution of LS~5039 for the four standard energy 
bands using {\tt csspec} for both phase intervals to reproduce Fig.~7 of \citet{collmar2014}.
In addition, we also derived for both phase intervals the uncertainty band of the fitted 
power-law model using {\tt ctbutterfly}.
The results are shown in Fig.~\ref{fig:ls5039}.
For the left panel we used exactly the same event selection that was used by \citet{collmar2014}
which is a minimum distance from the Earth horizon of $\zeta_{\rm min}=0\degrees$ and the
use of circular exclusion regions to handle D2 modules with failed PMTs
(i.e. {\tt fpmtflag = 2}; cf.~Appendix \ref{app:fpmt}).
Only with this event selection we were able to reproduce the $1-3$~MeV flux point of 
\citet{collmar2014}, while using our standard setting of $\zeta_{\rm min}=5\degrees$ and 
{\tt fpmtflag = 0} that excludes D2 modules with failed PMTs produced
a notable variation of the $1-3$~MeV flux point between inferior and superior conjunction.
To verify that this variation can indeed be attributed to differences in the event selection, 
we also did an equivalent analysis with COMPASS using $\zeta_{\rm min}=5\degrees$ and 
{\tt fpmtflag = 0}.
This resulted in a $1-3$~MeV flux variation between inferior and superior conjunction that
was similar to that observed in our analysis, confirming that the spectral differences are
due to differences in the event selection.
The corresponding results are summarised in the right panel of Fig.~\ref{fig:ls5039}.

As illustrated by the uncertainty bands of the fitted power-law model, the use of circular exclusion 
regions for D2 modules with faulty PMTs leads to a flatter (or softer) spectrum in 
superior conjunction, yet the variation seems still to remain within statistical uncertainties, given
the broadness of the uncertainty band.
After all, LS~5039 is a very faint source in SUPC, and hence its
spectral properties are only poorly constrained in this phase interval.

\begin{figure}[!t]
\centering
\includegraphics[width=8.8cm]{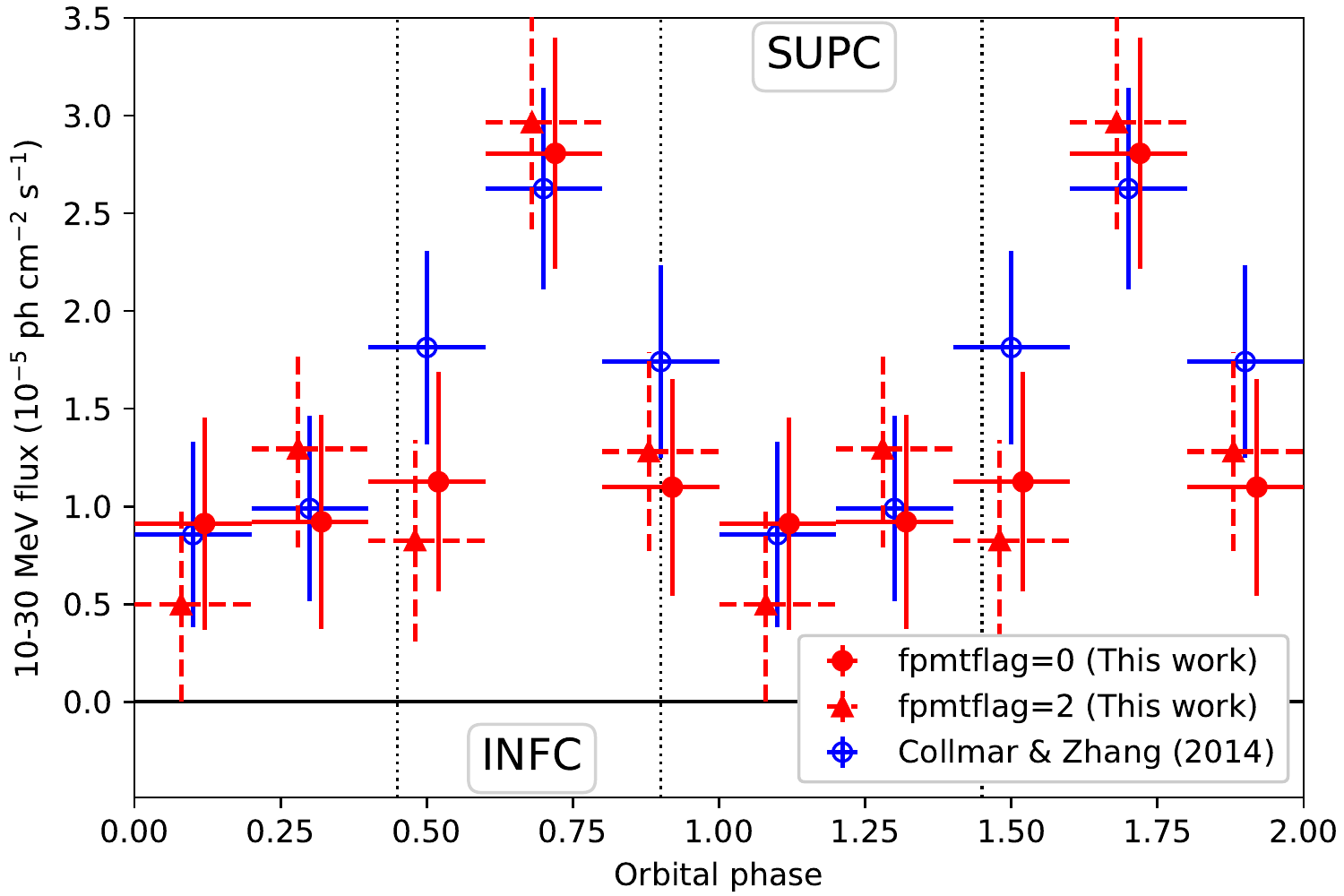}
\caption{
Orbital light curve of LS~5039 in the $10-30$ MeV energy band.
Data points for the ctools analysis were displaced by $\pm0.02$ in phase for clarity.
Analysis results obtained under analysis conditions identical to those in \citet{collmar2014}
($\zeta_{\rm min}=0\degrees$ and {\tt fpmtflag = 2}) are shown as triangles and
dashed error bars, and results obtained using $\zeta_{\rm min}=5\degrees$ and {\tt fpmtflag = 0}
are shown as dots and solid error bars.
Vertical lines indicate the definitions of the INFC and SUPC phase intervals.
\label{fig:orbit}
}
\end{figure}

Finally, we derived the orbital flux variation in the $10-30$ MeV energy band for the two different
event selections to reproduce Fig.~8 of \citet{collmar2014}.
For this purpose we split the $10-30$ MeV data into five phase intervals of equal length and fitted 
the data using {\tt comlixfit} with a source model comprising components for LS~5039, 
PKS~1830--210, bremsstrahlung emission, inverse Compton emission and cosmic background 
emission (see above for details).
The background was modelled using the BGDLIXE algorithm with parameters $N_{\rm avgr}=5$ 
and $N_{\rm incl}=15$.
The results are shown in Fig.~\ref{fig:orbit}.

Our analysis reproduces the orbital light curve of \citet{collmar2014} within statistical uncertainties.
Differences between their flux points and ours can be explained by differences in the event selection, 
as \citet{collmar2014} use a larger number of viewing periods compared to our analysis.
Variations of the same size are also observed in our analysis for the two different event selections,
which, however, are well within statistical uncertainties, as expected.

\subsection{\al26\ line emission from Carina}
\label{sec:line}

As the next step we validated the capacities of GammaLib and ctools for gamma-ray line emission 
analysis together with its abilities to assess the spatial morphology of the emission.
As reference, we chose the COMPTEL detection of point-like 1.8 MeV line emission from the Carina
region for this validation, as reported by \citet{knoedlseder1994} and \citet{knoedlseder1996a},
which are to our knowledge the only published studies where an analysis using a parametric 
spatial model was performed with COMPTEL.
\begin{table}[!b]
\caption{Energy bins for 1.8 MeV gamma-ray line analysis.
\label{tab:ebins18}}
\centering
\begin{tabular}{c c}
\hline\hline
Number & Energies (MeV) \\
\hline
1 & $1.000-1.584$ \\
2 & $1.584-1.634$ \\
3 & $1.634-1.684$ \\
4 & $1.684-1.734$ \\
5 & $1.734-1.784$ \\
6 & $1.784-1.834$ \\
7 & $1.834-1.884$ \\
8 & $1.884-1.934$ \\
9 & $1.934-1.984$ \\
10 & $1.984-2.034$ \\
11 & $2.034-3.000$ \\
\hline
\end{tabular}
\tablefoot{Bins 1 and 11 serve primarily to constrain continuum gamma-ray emission, while bins $2-10$
serve to trace the shape of the 1.8 MeV gamma-ray line.}
\end{table}

In their studies the authors analysed COMPTEL data from viewing periods 1 to 301 combined in
a data space of $100\times100\times25$ bins of $1\degrees\times1\degrees\times2\degrees$
in size, centred on $(l,b)=(286.5\degrees,0.5\degrees)$, 
which corresponds to the peak position of the observed 1.8 MeV line emission feature.
The emission feature was found to be compatible with a point-like source with an 1.8 MeV
flux of $(3.1\pm0.8) \times 10^{-5}$ \funit\ that was determined through model fitting.
Using fits of models with uniform intensity within a circular region centred on 
$(l,b)=(286.5\degrees,0.5\degrees)$,
\citet{knoedlseder1996a} determined a $2\sigma$ upper limit of $5.6\degrees$ for the diameter
of the 1.8 MeV emission region.
The analysis was done in a single energy bin covering $1.7-1.9$ MeV, and the instrumental
background was modelled using measurements in adjacent energy intervals that were 
corrected for the energy dependence of the Compton scatter angle $\bar{\varphi}$.
This method suppresses to first order emission from continuum gamma-ray sources 
\citep{knoedlseder1996b}.

\begin{figure}[!t]
\centering
\includegraphics[width=8.8cm]{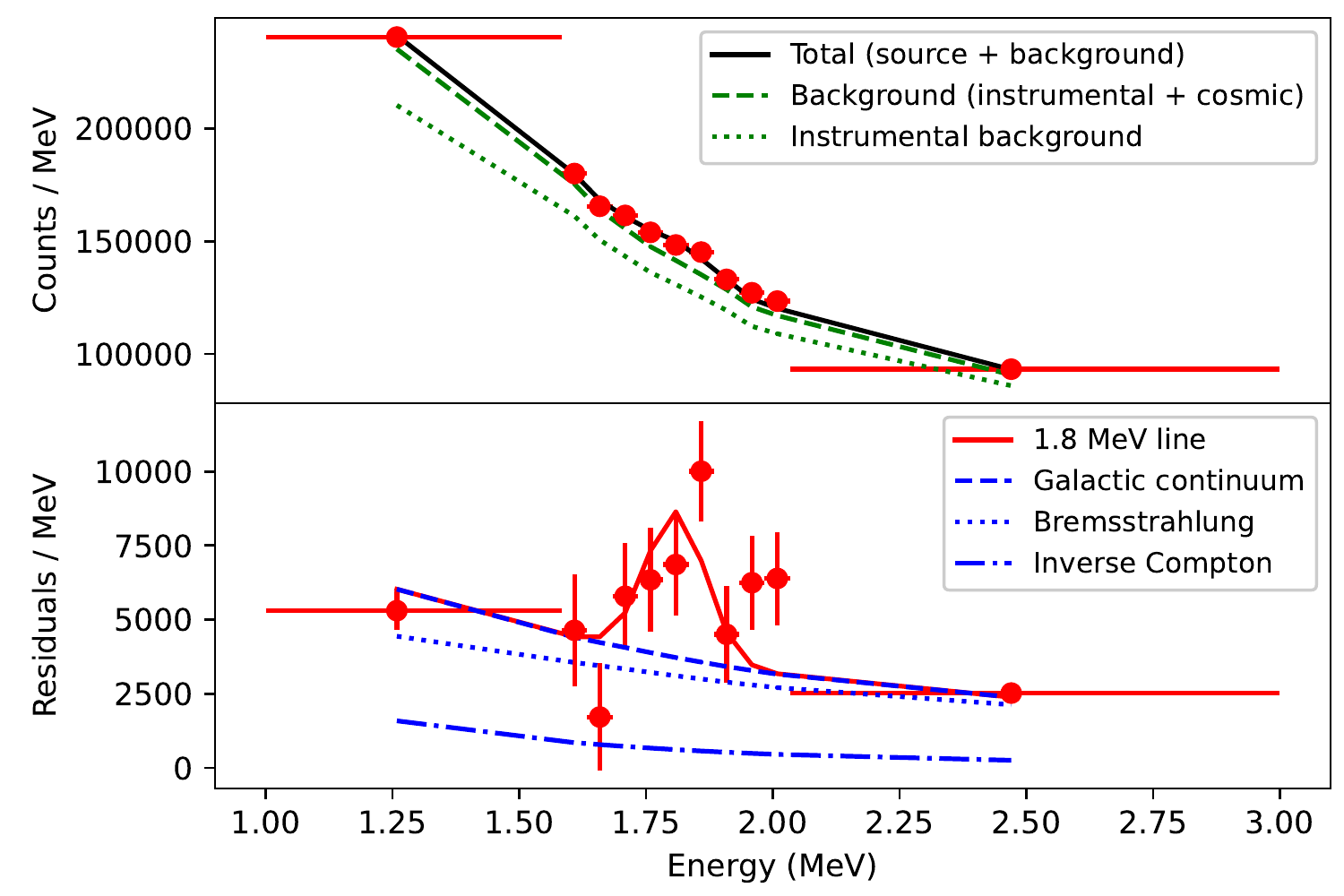}
\caption{
Count spectrum of the Carina region determined for 11 bins covering the energy band
$1-3$ MeV.
The top panel shows the measured number of counts per MeV together with the best fitting
background model (green dashed), composed of instrumental background (green dotted)
and cosmic gamma-ray background.
The combined source and background model is shown as a black solid line.
The bottom panel shows the background model-subtracted count spectrum together
with the best fitting 1.8 MeV line model (red solid) on top of the combined Galactic continuum 
components (blue dashed), composed of bremsstrahlung (blue dotted) and inverse Compton 
emission (blue dash-dotted).
\label{fig:carina-spectrum}
}
\end{figure}

\begin{figure*}[!t]
\centering
\includegraphics[width=18cm]{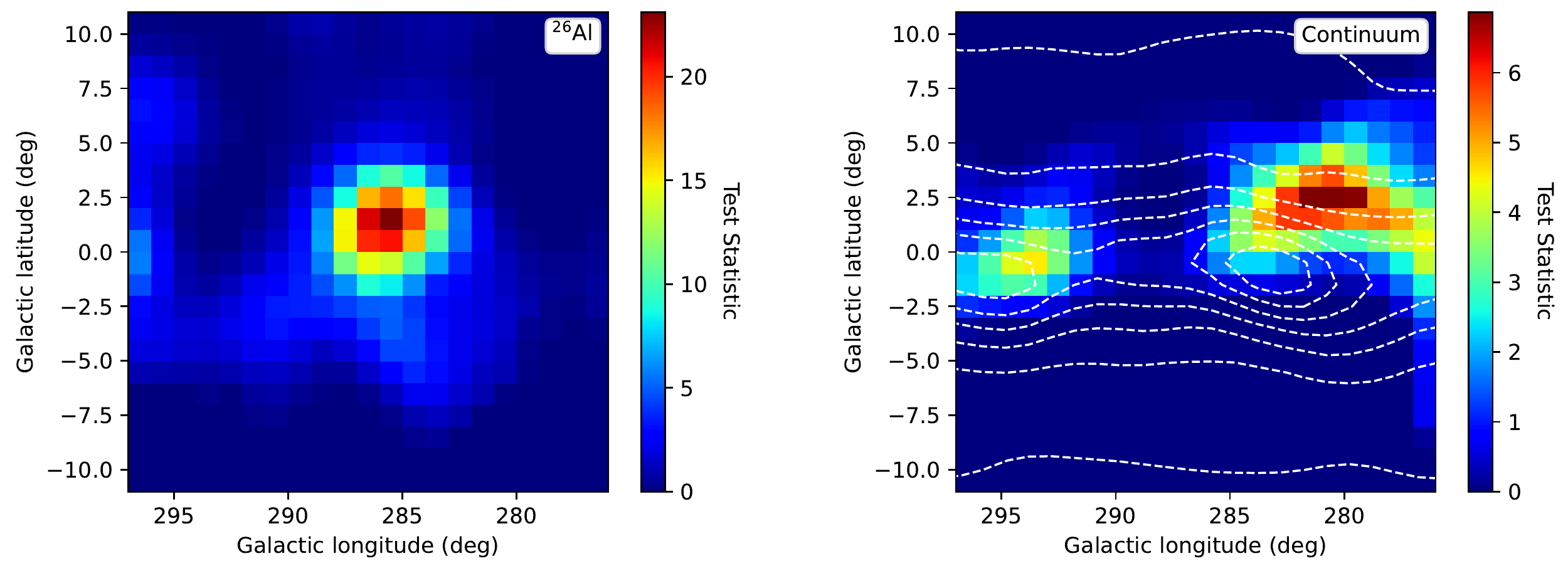}
\caption{
Test statistic  maps of 1.8 MeV line emission (left) and $1-3$ MeV continuum emission (right)
obtained using {\tt comlixmap} for the Carina region. The map on the left is equivalent to Fig.~2
of \citet{knoedlseder1996a}. Dashed contours in the map on the right reflect the intensity of a
combination of Galactic bremsstrahlung and inverse Compton emission maps as fitted in 
an independent analysis where a model of a 1.8 MeV line point source was
fitted together with a combination of bremsstrahlung and inverse Compton spatial maps
to the data (see text).
\label{fig:carmaps}
}
\end{figure*}

We analysed the same data and adopted the same data space definition that was used by 
\citet{knoedlseder1996a}, yet we tested an alternative analysis method that jointly handles
the \al26\ line signal and any underlying continuum emission.
This is more in line with the GammaLib and ctools philosophy of explicitly modelling all emission 
components and provides a more accurate handling of underlying continuum emission.
Specifically, we split the data within the $1-3$ MeV energy interval into 11 energy bins, specified in 
Table \ref{tab:ebins18}, and analysed them jointly using a combination of model components describing 
the 1.8 MeV line signal, any underlying continuum gamma-ray emission and the instrumental background.
To model the 1.8 MeV line emission spectrum we used a Gaussian spectral component with a fixed 
mean of 1.809 MeV and a standard deviation of $\sigma=58.9$ keV that corresponds to COMPTEL's 
instrumental energy resolution at that energy.
Similarly to our analysis of LS~5039 we modelled continuum emission using template maps for 
Galactic bremsstrahlung and inverse Compton emission and an isotropic component for 
the cosmic gamma-ray background.
The spectra of the three continuum components were model using power laws, where
the prefactors and indices were free for the Galactic components, while the prefactor and
index were fixed to 
$I(E_{\gamma}) = 1.12 \times 10^{-4} ( E_{\gamma}/{5 \, {\rm MeV}} )^{-2.2}$ \fmevster\ for the
cosmic gamma-ray background component, as suggested by \citet{weidenspointner1999}.
The TS map was generated using {\tt comlixmap} and model fitting was done using {\tt comlixfit}
with the standard BGDLIX parameters $N_{\rm avgr}=5$ and $N_{\rm incl}=15$.

We illustrate our analysis procedure in Fig.~\ref{fig:carina-spectrum}, which shows the counts spectrum
determined using Eq.~(\ref{eq:armdata}) for the position $(l,b)=(286.5\degrees,0.5\degrees)$ at which
\citet{knoedlseder1994} found the 1.8 MeV line emission maximum and an ARM window of 
$\pm3\degrees$.
We also show the model components that were fitted using {\tt comlixfit} and that we extracted using 
Eq.~(\ref{eq:armmodel}).
We used a point source located at the fixed position $(l,b)=(286.5\degrees,0.5\degrees)$ as spatial
model for the 1.8 MeV line component.
Figure \ref{fig:carina-spectrum} illustrates that the data are dominated by instrumental background,
followed by cosmic gamma-ray background.
The bottom panel illustrates that, once these components are subtracted, a clear line signal becomes 
apparent that is compatible with the expected signature of the \al26\ decay line.
In addition, a continuum signal is detected that is dominated by Galactic bremsstrahlung emission.

In their Fig.~2, \citet{knoedlseder1996a} present a 1.8 MeV line emission maximum likelihood 
map of the Carina region, and we produced an equivalent TS map with the same spatial binning 
using {\tt comlixmap}.
Specifically, we fitted point-source models for the 1.8 MeV line emission and the $1-3$ MeV 
continuum emission for a grid of source positions, producing hence TS maps 
for both emission components.
In the fitting the fluxes of both point sources were constrained to non-negative values.
The resulting maps are shown in Fig.~\ref{fig:carmaps}, where the left map can be
compared to the maximum likelihood map presented in Fig.~2 of \citet{knoedlseder1996a}.
Both maps show qualitatively comparable features, yet we obtained a maximum TS
value of $23.1$ at $(l,b)=(285.5\degrees,1.5\degrees)$ while \citet{knoedlseder1996a} 
found a lower maximum TS value of $14.7$ at $(l,b)=(286.5\degrees,0.5\degrees)$.
Eventually, these differences may be explained by the background modelling techniques
and analysis methods that differ significantly between the studies.
We recall that \citet{knoedlseder1996a} analyse the 1.8 MeV line data in a single energy
bin covering $1.7-1.9$~MeV and used a background model derived from adjacent energy
bands, which to first order includes also continuum emission, but which does not properly
account for spectral differences between instrumental background and diffuse emission
components as well as differences in their $\bar{\varphi}$ distributions \citep{bloemen1999}.

It is actually rather reassuring that both approaches produce qualitatively comparable maps, 
as already suggested by \citet{bloemen1999} who implemented a comparable analysis
method to ours.
The continuum TS map indicates that the continuum emission is located towards the
Galactic plane, suggesting that it originates from our Galaxy.
We emphasise, however, that the statistical significance of the emission features is modest, and 
consequently the appearance of the map is notably affected by the statistical fluctuations of the data.
Nevertheless, we note that observed TS maxima are not too far from maxima in the 
Galactic bremsstrahlung emission, which eventually may be the dominant MeV emission component
near the Galactic plane \citep{strong1996}.

As the next step we fitted the data with a source model composed of a 1.8 MeV line component
modelled using a point source and a $1-3$ MeV continuum component modelled using a 
combination of bremsstrahlung and inverse Compton spatial maps as well as an isotropic
component for the cosmic gamma-ray background.
The position of the point-source model as well as the spectral parameters for the Galactic 
continuum power-law components were free parameters in the fit.
Fitting the data using {\tt comlixfit} gave 
a best fitting position of $(l,b)=(285.4\degrees \pm 0.8\degrees, 1.3\degrees \pm 0.7\degrees)$,
a flux of $(3.1\pm0.6) \times 10^{-5}$ \funit\ and a TS value of $22.3$ for the 1.8 MeV line 
component.
Our 1.8 MeV line flux is consistent with the value of $(3.1\pm0.8) \times 10^{-5}$ \funit\ found by 
\citet{knoedlseder1996a}, yet our best fitting position is offset by about $1.4\degrees$ from the 
one found in their analysis.
Replacing the point source for the 1.8 MeV line component by a radial disk model did not
improve the fit and led to a best fitting disk radius of $0.003\degrees$, which is near the minimum 
value of $0.001\degrees$ that we allowed in the analysis.
Using the {\tt ctulimit} tool we derived a $2\sigma$ upper limit of $5.1\degrees$ for the diameter
of the disk, a bit smaller than the value of $5.6\degrees$ determined by 
\citet{knoedlseder1996a}.
This difference is plausibly explained by the larger detection significance of the 1.8 MeV line
emission signal in our analysis, allowing us to put a stronger constraint on the extent of
the emission region.

\begin{figure}[!t]
\centering
\includegraphics[width=8.8cm]{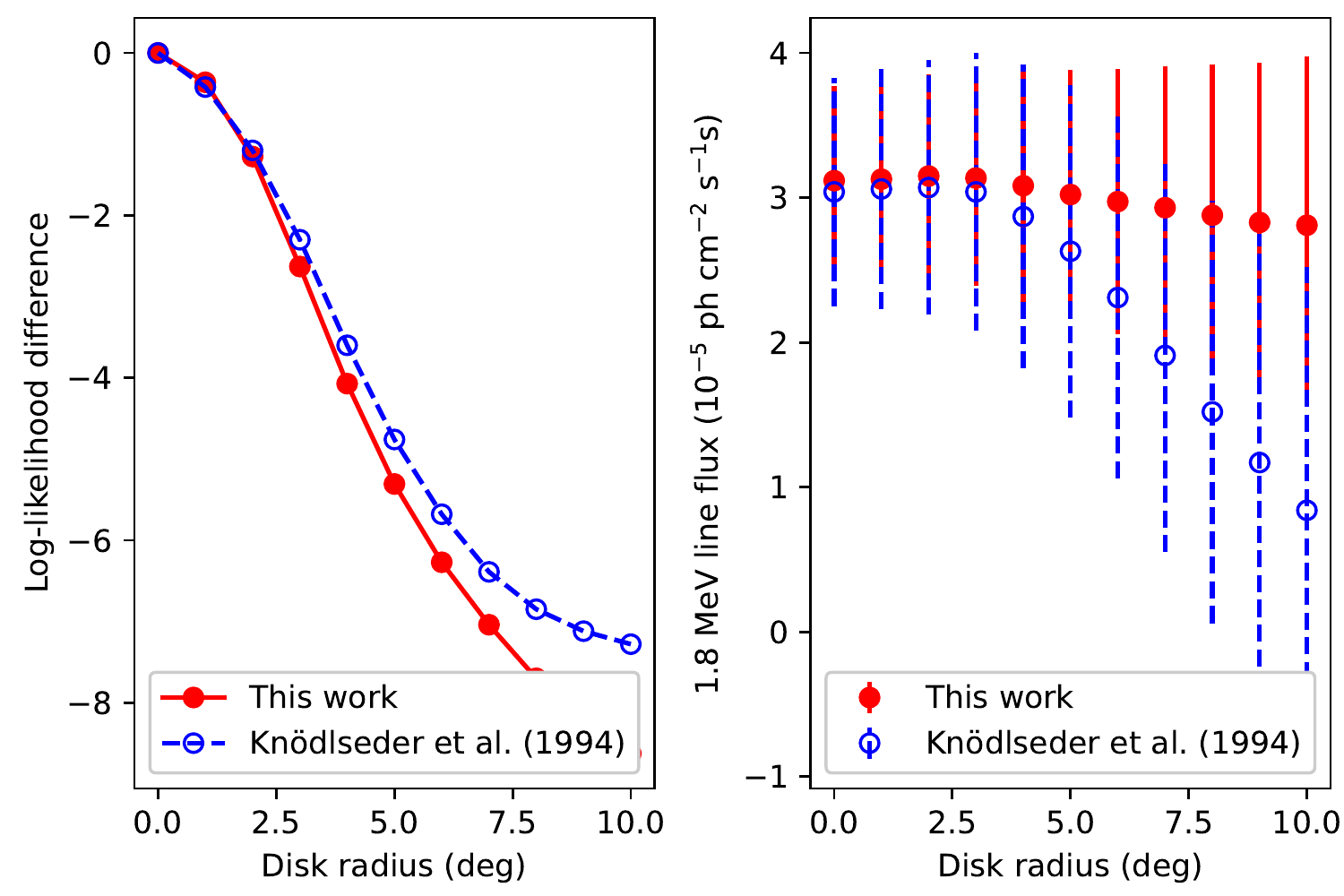}
\caption{
Variation in maximum log-likelihood with increasing radius of the disk model (left) and fitted 1.8 MeV 
line flux (right).
Data shown as red filled dots and solid lines are those derived in this work, and data shown as blue open 
dots and dashed lines are taken from \citet{knoedlseder1994}.
\label{fig:carina-radial}
}
\end{figure}

Finally, we tried to reproduce Fig.~5.5 of \citet{knoedlseder1994} that shows the variation of the 
maximum log-likelihood value and the fitted 1.8 MeV line flux as a function of spatial disk radius.
For that purpose, we determined the maximum likelihood solution for a disk model with 
position fixed at our maximum likelihood solution $(l,b)=(285.4\degrees, 1.3\degrees)$ for
a set of disk radius values starting with $0.001\degrees$, and followed by $1\degrees$
up to $10\degrees$ with a step size of $1\degrees$.
The results of our analysis are shown in Fig.~\ref{fig:carina-radial}, which also includes the
results shown in Fig.~5.5 of \citet{knoedlseder1994} for comparison.
The log-likelihood profile between both analyses is very similar, yet our analysis turns out to be
a bit more constraining, probably owing to the more significant detection of the 1.8 MeV
line emission feature with respect to the analysis of \citet{knoedlseder1994}.
The flux attributed to the 1.8 MeV line component changes actually very little with disk
radius in our analysis, while in \citet{knoedlseder1994} the flux decreases with increasing
disk radius.
This difference is probably partly due to fact that our analysis detects the 1.8 MeV line signal
more significantly, but may also be related to our analysis method that gives freedom to the
continuum emission model components to adjust as a function of the 1.8 MeV disk model radius, 
while in \citet{knoedlseder1994} the continuum emission was implicitly subtracted by the 
background model, resulting in a much more constrained overall model that leaves little freedom for
the model to adjust.

\section{Conclusion}
\label{sec:conclusion}

We have implemented a comprehensive science analysis framework for COMPTEL gamma-ray
data in the GammaLib and ctools software packages, and we have demonstrated that our
software reliably reproduces published analysis results that were derived using the COMPASS
software.
Having public, free, and validated software for COMPTEL science data analysis now opens 
the HEASARC COMPTEL archive to the community for scientific exploration.
In the medium-energy gamma-ray band, from $1-30$~MeV, the COMPTEL archive still contains 
the most sensitive observations ever performed, and a unique dataset for exploring the non-thermal 
Universe and nuclear transition lines.
Since the 1990s when the COMPTEL data were taken, the field of gamma-ray astronomy has
made impressive progress thanks to satellites such as the International Gamma-Ray Astrophysics 
Laboratory (INTEGRAL) and Fermi and ground-based observatories such as 
the High Energy Stereoscopic System (H.E.S.S.),
the Major Atmospheric Gamma-ray Imaging Cherenkov Telescope (MAGIC), and
the Very Energetic Radiation Imaging Telescope Array System (VERITAS).
Many source classes that are known today were not established as gamma-ray emitters during
the COMPTEL era, and the COMPTEL data were never comprehensively analysed with the
current knowledge in the field.
An exception to this is the post-mission discovery of the orbital modulation of MeV gamma-ray
emission of LS~5039 in the COMPTEL archival data by \citet{collmar2014}, a source that was
not an established gamma-ray emitter in the 1990s.
Thanks to GammaLib and ctools, such discoveries are now achievable by the community at
large.

We want to stress that our work was also motivated by the goal of reducing the
carbon footprint of astronomical research.
As recently pointed out by \citet{knoedlseder2022}, the current deployment rate of new astronomical 
observatories is not compatible with the imperative of reducing the carbon footprint across all 
activity sectors of modern societies, and this calls for fundamental changes in astronomical practices in 
the future.
Among the many possible alternatives to the building of ever more and ever bigger new
astronomical facilities is the exploitation of archival data from past missions that may 
have scientific treasures yet to be discovered.
Since version 2.0.0, ctools estimates the carbon footprint of its use based on the assessment of 
\citet{berthoud2020} for the GRICAD computing centre, and, using this feature, we estimate that 
the work presented in this publication resulted in the generation of $600 \pm 300$ kgCO$_2$e 
of greenhouse gases (GHGs) due to the computing related to the analysis of archival data.
This is about 40 times less than the median per-publication emissions associated with the 
analysis of data from an active astronomical observatory \citep{knoedlseder2022}.
Taking all sources of emissions related to this work into account, we estimate the carbon
footprint of this research to be $1.9 \pm 0.3$ tCO$_2$e
(see Appendix \ref{app:footprint}).

This assessment illustrates that the exploitation of archival data instead of the development of new
astronomical observatories has the potential to dramatically reduce the carbon footprint of
astronomical research, which would help realise the reductions that are needed to limit global 
warming and reach the goals of the Paris Climate Agreement.
Past missions can be seen as our scientific carbon legacy since the GHGs that were emitted during 
their construction and operations are to a large extent still present in the Earth's atmosphere.
To assure that these GHGs were at least not emitted in vain, conservation of the
archival data and the development and maintenance of software for their exploitation should 
therefore be of the utmost importance.

\begin{acknowledgements}
We would like to thank the anonymous referee for the very careful reading of the manuscript 
and the many constructive suggestions.
This research made use of ctools, a community-developed gamma-ray astronomy science analysis 
software \citep{ctools2016}. 
ctools is based on GammaLib, a community-developed toolbox for the scientific analysis of astronomical
gamma-ray data \citep{gammalib2011}.
This work has made use of the Python 2D plotting library matplotlib \citep{hunter2007}.
This research has made use of data and/or software provided by the High Energy Astrophysics 
Science Archive Research Center (HEASARC), which is a service of the Astrophysics Science Division 
at NASA/GSFC.
This research has made use of the SIMBAD database, operated at CDS, Strasbourg, France.
This research is part of the \#LowCarbonScience initiative that aims in reducing the carbon footprint
of scientific research, and has benefitted from discussions held in the context of the GDR Labos 1point5
and the Astro4Earth initiative.
\end{acknowledgements}

\bibliographystyle{aa} 
\bibliography{references}

\begin{appendix}

\section{HEASARC archive}
\label{app:heasarc}

COMPTEL data in the HEASARC archive are grouped by so-called viewing periods with typical
durations of two weeks during which the CGRO satellite had a stable pointing.
In total, there exist 359 viewing periods, of which 278 were archived at HEASARC,
covering viewing periods 1.0 to 719.0, where the first digit indicates the mission year or phase, 
with viewing periods smaller than 100 corresponding to mission year one.
Some of the viewing periods that are available in the HEASARC archive cannot be exploited due to 
corrupted or missing data files, or spacecraft operations that prevented data taking. 
The 23 viewing periods in the range 1.0 to 719.0 that are not exploitable are listed in Table \ref{tab:heasarc}.
This leaves 255 exploitable viewing periods in the HEASARC archive, which is about $71\%$ of the 
total number of 359 viewing periods that were executed during the CGRO mission.

\begin{table}[!th]
\caption{Inexploitable viewing periods in the range 1.0 to 719.0 in the COMPTEL HEASARC 
archive.
\label{tab:heasarc}}
\centering
\begin{tabular}{l l l}
\hline\hline
VP & Target & Reason \\
\hline
22.0 & Mrk 279 & Missing {\tt EVP} file \\
29.0 & G $224-40$ & {\tt EVP} file unreadable \\
225.0 & CGRO reboost & Missing {\tt EVP} file \\
229.3 & Perseids shower & Missing {\tt EVP} file \\
229.5 & Gal $5+5$ & Missing {\tt EVP} file \\
303.4 & Not executed & Missing {\tt EVP} file \\
303.5 & Reboost & Missing {\tt EVP} file \\
303.7 & Nova Cygnus 92 & Missing {\tt EVP} file \\
308.3 & Reboost & Missing {\tt EVP} file \\
308.6 & Virgo $283+75$ & Missing {\tt EVP} file \\
311.3 & Reboost & Missing {\tt EVP} file \\
311.6 & Virgo $284+75$ & Missing {\tt EVP} file \\
426.0 & Anticenter & Invalid event times in {\tt EVP} \\
617.2 & CGRO reboost test & Missing {\tt EVP} file \\
617.4 & CGRO reboost test & Missing {\tt EVP} file \\
617.6 & CGRO reboost & Missing {\tt EVP} file \\
619.0 & Cir X--1 & Missing {\tt EVP} file \\
619.5 & CGRO reboost & Missing {\tt EVP} file \\
627.0 & PSR $1055-52$ & Missing {\tt EVP} file \\
632.0 & Not executed & Missing {\tt EVP} file \\
632.1 & PKS $0235+164$ & Missing {\tt EVP} file \\
711.0 & 2EG J$1835+5919$ & Missing {\tt OAD} files \\
718.0 & Cen X--3 & Missing {\tt EVP} file \\
\hline
\end{tabular}
\end{table}

The HEASARC archive was created from the original COMPTEL data using dedicated file conversion 
software that generated FITS files following recognised standards \citep{pence2010}.
While interpreting these FITS files using GammaLib we found that the units in some of the table 
columns were not correct.
We summarise the inconsistencies that we encountered in Table \ref{tab:headerunits}.
We furthermore found that the good time 
interval dataset for viewing period 8.0 with the
identifier {\tt MPE-TIM-11481} only contains about half of the good time 
intervals that were 
defined in COMPASS, probably as the result of a file truncation error in the input ASCII file that was 
used for the generation of the FITS file in the HEASARC archive.

\begin{table}[!th]
\caption{Incorrect header units in FITS files in the COMPTEL HEASARC archive.
\label{tab:headerunits}}
\centering
\begin{tabular}{l l l l}
\hline\hline
File & Column & Written unit & True unit \\
\hline
{\tt EVP} & {\tt TOF} & ns & channels \\
{\tt EVP} & {\tt PHIBAR} & deg & rad \\
{\tt EVP} & {\tt GLON\_SCAT} & deg & rad \\
{\tt EVP} & {\tt GLAT\_SCAT} & deg & rad \\
{\tt EVP} & {\tt AZIMUTH\_SCAT} & deg & rad \\
{\tt EVP} & {\tt ZENITH\_SCAT} & deg & rad \\
{\tt EVP} & {\tt EARTH\_HORIZON} & deg & rad \\
{\tt BVC} & {\tt SSB\_X} & km & milli light seconds \\
{\tt BVC} & {\tt SSB\_Y} & km & milli light seconds \\
{\tt BVC} & {\tt SSB\_Z} & km & milli light seconds \\
\hline
\end{tabular}
\end{table}

The HEASARC archive also comprises binned data products for standard energy bands that can be 
used for analysis using GammaLib, yet we note that the world coordinate system information 
attached to these data products is incorrect.
While the FITS headers suggest that the data cubes are provided in Mercator projection, the event 
cubes are presented in a cartesian projection with a reference latitude value of 0.
For all event cubes given in Mercator projection GammaLib assumes that they originate from the 
HEASARC archive and corrects the world coordinate system information upon reading of the event cubes.

The HEASARC archive comprises further data products such as telescope housekeeping 
data, gamma-ray burst detector data and sky maps that are not used by GammaLib.

\section{Time of flight conversion}
\label{app:tof}

The HEASARC archive mixes different versions of {\tt EVP} files that have different levels of
processing for the ToF values.
The versions can be distinguished by the header keyword {\tt DSD\_REP} in the {\tt EVP} file,
specifying either 2 for ToF$_{\rm II}$ or 3 for ToF$_{\rm III}$, where the latter corrects for energy
dependent effects, aligning the forward peak at channel number 120
(see \citealt{vandijk1996} and \citealt{weidenspointner1999} for an explanation of the ToF corrections).
If a version 2 {\tt EVP} is encountered, the {\tt GCOMEventList} class will automatically convert
ToF$_{\rm II}$ values into ToF$_{\rm III}$ using
\begin{equation}
{\rm ToF}_{\rm III} = {\rm ToF}_{\rm II} + 120 - \left( \sum_{i=0}^{6} a_i E_1^i + \sum_{i=0}^{4} b_i E_2^i - 118.3 \right)
\end{equation}
with the coefficients given by Tables \ref{tab:tof3a} and \ref{tab:tof3b} and the D1 and D2 energy
deposits $E_1$ and $E_2$ given in MeV.
For the HEASARC archive, ToF values accessed through GammaLib are therefore always
ToF$_{\rm III}$ values.

\begin{table}[!th]
\caption{Coefficients, $a_i$, for the conversion from ToF$_{\rm II}$ to ToF$_{\rm III}$.
\label{tab:tof3a}}
\centering
\begin{tabular}{c c c}
\hline\hline
$E_1$ & $\le 2.25$ MeV & $> 2.25$ MeV \\
\hline
$a_0$ & $111.74858$ & $116.25374$ \\
$a_1$ & $28.280247$ & $0.500407092$ \\
$a_2$ & $-45.024305$ & $0.3818272$ \\
$a_3$ & $35.18321$ & $-0.080145513$ \\
$a_4$ & $-14.639463$ & $0.006556979$ \\
$a_5$ & $3.1342536$ & $-0.00024650067$ \\
$a_6$ & $-0.2711735$ & $3.507724 \times 10^{-6}$ \\
\hline
\end{tabular}
\end{table}

\begin{table}[!th]
\caption{Coefficients, $b_i$, for the conversion from ToF$_{\rm II}$ to ToF$_{\rm III}$.
\label{tab:tof3b}}
\centering
\begin{tabular}{c c c c}
\hline\hline
$E_2$ & $\le 1.4$ MeV & $1.4 - 5.5$ MeV & $> 5.5$ MeV \\
\hline
$b_0$ & $181.77024$ & $120.91608$ & $119.24278$ \\
$b_1$ & $-252.4107$ & $-0.1504849$ & $-0.43134699$ \\
$b_2$ & $371.09898$ & $-0.45526025$ & $0.06018308$ \\
$b_3$ & $-232.83985$ & $0.11710009$ & $-0.002684779$ \\
$b_4$ & $52.918785$ & $ -0.0082172427$ & $3.7720986 \times 10^{5}$ \\
\hline
\end{tabular}
\end{table}

\section{Flux correction due to time of flight selection}
\label{app:tofcor}

To correct for the photon rejection by the ToF selection an energy-dependent correction factor needs
to be applied to the instrument response.
Since this correction factor depends on the ToF selection interval, it is computed in GammaLib 
during the event binning and stored using the header keyword {\tt TOFCOR} in the {\tt DRE} FITS files.
The correction factor is determined from a linear interpolation of the values given in Table \ref{tab:tofcor},
evaluated at the geometric mean energy $\sqrt{E_{\rm min} E_{\rm max}}$ of the energy interval
$[E_{\rm min}, E_{\rm max}]$ for which the event cube is generated.
Only corrections for ToF$_{\rm max}=130$ are supported.

\begin{table}[!th]
\caption{Time of flight correction factors for ToF$_{\rm max}=130$.
\label{tab:tofcor}}
\centering
\begin{tabular}{l c c c c}
\hline\hline
& \multicolumn{4}{c}{Geometric mean energy (MeV)} \\
ToF$_{\rm min}$ & 0.8660 & 1.7321 & 5.4772 & 17.3205 \\
\hline
110 & 1.14 & 1.07 & 1.02 & 1.01 \\
111 & 1.17 & 1.09 & 1.03 & 1.01 \\
112 & 1.21 & 1.11 & 1.05 & 1.02 \\
113 & 1.26 & 1.15 & 1.07 & 1.04 \\
114 & 1.32 & 1.20 & 1.11 & 1.06 \\
115 & 1.40 & 1.27 & 1.17 & 1.11 \\
116 & 1.50 & 1.36 & 1.24 & 1.17 \\
117 & 1.63 & 1.47 & 1.35 & 1.28 \\
118 & 1.79 & 1.63 & 1.51 & 1.43 \\
119 & 2.01 & 1.85 & 1.73 & 1.67 \\
\hline
\end{tabular}
\end{table}

\section{Pulsar timing}
\label{app:pulsar}

GammaLib supports generation of event cubes, geometry functions and exposure maps for 
phase-resolved pulsar analysis.
For this purpose a specific processing is implemented in the methods
{\tt GCOMDri::compute\_dre},
{\tt GCOMDri::compute\_drg}, and
{\tt GCOMDri::compute\_drx} that is used if the specified {\tt GCOMSelection} instance
includes pulsar ephemerides data and the specification of pulsar phase intervals.
All three methods will first trim the good time 
intervals of the observation so that they
only cover periods for which the specified pulsar ephemerides are valid.
In that way, GammaLib assures that only data will be used for an analysis that cover intervals 
with valid pulsar ephemeris information.

The remaining pulsar-specific code is implemented in {\tt GCOMDri::compute\_dre}.
First, the method converts arrival times $t_{\rm CGRO}$ of COMPTEL events at the CGRO satellite, 
specified in the Coordinated Universal Time (UTC) time system, into arrival times $t_{\rm SSB}$ 
at the Solar System barycentre, specified in the barycentric dynamical time (TDB) system.

As a side note, before 1992-06-25T01:00:00 UTC the CGRO onboard clock was early by 2.042144
seconds, and this time difference needs to be subtracted from the measured onboard time to get
the true arrival time in UTC.
In GammaLib, this subtraction is automatically performed when converting onboard times, given in
truncated Julian days (TJDs) and tics, into {\tt GTime} objects using the {\tt gammalib::com\_time} 
function, and the conversion is undone when using the inverse functions {\tt gammalib::com\_tjd} 
and {\tt gammalib::com\_tics}.

After applying the clock correction, event times are converted using
\begin{equation}
t_{\rm SSB} = t_{\rm CGRO} + \Delta t_{\rm CGRO \to SSB} - \Delta t_{\rm Shapiro} + 
\Delta t_{\rm UTC \to TT} + \Delta t_{\rm TT \to TBD}
\label{eq:timecor}
,\end{equation}
where
$\Delta t_{\rm CGRO \to SSB}$ corrects for the light travel time from CGRO to the Solar System 
barycentre,
$\Delta t_{\rm Shapiro}$ corrects for the gravitational time delay near the Sun, also known as Shapiro 
delay,
$\Delta t_{\rm UTC \to TT}$ converts from UTC to terrestrial time (TT), and
$\Delta t_{\rm TT \to TBD}$ converts from TT to the TDB time system.
We note that all time correction terms are themselves time dependent, yet we ignore this time dependence
in Eq.~(\ref{eq:timecor}) to simplify the notation.
All terms are given in units of seconds.

The computation of the three terms 
$\Delta t_{\rm CGRO \to SSB} - \Delta t_{\rm Shapiro} + \Delta t_{\rm TT \to TBD}$ 
is implemented by the {\tt GEphemerides::geo2ssb} method, which
regroups all time corrections that depend on planetary ephemerides.
GammaLib deals with planetary ephemerides through the {\tt GEphemerides} class, and the software
includes the Jet Propulsion Laboratory (JPL) Development Ephemeris 200 including
information about the Sun position, the Earth position and its first three time derivatives, and the time 
difference between TT and TDB in seconds on a daily basis between Julian day (JD) 2436913 
(10 December 1959) and 2469807 (31 December 2049).
Specifically, the method computes
\begin{equation}
\Delta t_{\rm CGRO \to SSB} = \mathbf{\hat{n}} \cdot (\mathbf{e} + \mathbf{r})
\end{equation}
and
\begin{equation}
\Delta t_{\rm Shapiro} = -2 t_{\odot} \ln \left( 1 + \frac{\mathbf{\hat{n}} \cdot (\mathbf{e} + \mathbf{r} - \mathbf{s})}
{|\mathbf{e} + \mathbf{r} - \mathbf{s}|} \right)
\label{eq:shapiro}
,\end{equation}
where
\begin{equation}
\mathbf{\hat{n}} = 
\begin{bmatrix}
  \cos \delta \cos \alpha \\
  \cos \delta \sin \alpha \\
  \sin \delta
\end{bmatrix}
\end{equation}
is the unit vector in celestial coordinates of a pulsar with right ascension $\alpha$ and declination
$\delta$,
$\mathbf{e}$ is the vector from the Solar System barycentre to the centre of the Earth,
$\mathbf{s}$ is the vector from the Solar System barycentre to the centre of the Sun,
$\mathbf{r}$ is the vector from the centre of the Earth to the CGRO spacecraft, and
\begin{equation}
t_{\odot} = 4.92549 \times 10^{-6}
\end{equation}
is half the Schwarzschild radius of the Sun divided by the speed of light.
All vectors are given in units of light seconds and are specified in the celestial coordinate system.

The Earth vector $\mathbf{e}$ at a given time $t_{\rm CGRO}$ (specified in the TT time system) 
is computed using the Taylor expansion
\begin{equation}
\mathbf{e} = 
  \mathbf{e}(i) + 
  \mathbf{\dot{e}}(i) \Delta t + 
  \frac{1}{2} \mathbf{\ddot{e}}(i) \Delta t^2 + 
  \frac{1}{6} \mathbf{\dddot{e}}(i) \Delta t^3
,\end{equation}
where $i$ is the index of the nearest entry in time in the JPL database, 
$\mathbf{e}(i)$, $\mathbf{\dot{e}}(i)$, $\mathbf{\ddot{e}}(i)$, and $\mathbf{\dddot{e}}(i)$
is the Earth position and its first three time derivatives for this entry, and
\begin{equation}
\Delta t = t_{\rm CGRO} - t(i)
\end{equation}
is the time difference between the event time and the nearest entry in the JPL database in units of
days, which by definition is in the range $-0.5$ and $+0.5$.
Since the Sun moves only a little around the Solar System barycentre, it is sufficient to take for the Sun
vector the nearest entry in the JPL database, which is $\mathbf{s}  = \mathbf{s}(i)$.
Finally, the conversion from the TT to the TDB time system at a given time $t_{\rm CGRO}$ is computed
using 
\begin{equation}
\Delta t_{\rm TT \to TDB} = \Delta t_{\rm TT \to TDB}(i) + (\Delta t_{\rm TT \to TDB}(i+1) - \Delta t_{\rm TT \to TDB}(i)) \Delta t
,\end{equation}
where
$\Delta t_{\rm TT \to TDB}(i)$ is the nearest entry in the JPL database.

The last term in the time correction, $\Delta t_{\rm UTC \to TT}$, does not depend on planetary
ephemerides and is given by
\begin{equation}
\Delta t_{\rm UTC \to TT} = 32.184 + n_{\rm leap}
,\end{equation}
where $n_{\rm leap}$ is the number of leap seconds.
The computation of $\Delta t_{\rm UTC \to TT}$ is implemented by the {\tt GTime::utc2tt} method.

The HEASARC archive includes for most of the viewing periods files that provide for each superpacket 
the vectors $\mathbf{e} + \mathbf{r}$ as well as the correction terms
$\Delta t_{\rm UTC \to TT} + \Delta t_{\rm TT \to TBD}$, avoiding the need for planetary ephemerides.
These so-called {\tt BVC} data can be handled by GammaLib through the classes {\tt GCOMBvc} and
{\tt GCOMBvcs} that manage individual data records as well as entire files.
Specifically, the method {\tt GCOMBvcs::tdelta} computes
$\Delta t_{\rm CGRO \to SSB} - \Delta t_{\rm Shapiro} + \Delta t_{\rm UTC \to TT} + \Delta t_{\rm TT \to TBD}$,
and if the observation that should be binned includes {\tt BVC} information, {\tt GCOMDri::compute\_dre}
will use this method instead of the algorithm described above for the barycentric time correction.
The formulae used by {\tt GCOMBvcs::tdelta} are identical to those described above, except for the
Shapiro time delay for which the displacement of the Sun from the Solar System barycentre is
neglected:
\begin{equation}
\Delta t_{\rm Shapiro} = -2 t_{\odot} \ln \left( 1 + \frac{\mathbf{\hat{n}} \cdot (\mathbf{e} + \mathbf{r})}
{|\mathbf{e} + \mathbf{r}|} \right)
.\end{equation}

Following the time correction, the pulsar phase $\Phi$ is computed using the {\tt GPulsarEphemeris::phase}
method that implements
\begin{equation}
\Phi = \Phi_0 + \nu \Delta t + \frac{1}{2} \dot{\nu} \Delta t^2 + \frac{1}{6} \ddot{\nu} \Delta t^3
,\end{equation}
where
$\nu$, $\dot{\nu}$ and $\ddot{\nu}$ is the pulsar frequency and its first two time derivatives,
$\Delta t = t_{\rm SSB} - t_0$ is the elapsed time since the reference time of the ephemeris,
and $\Phi_0$ is the pulsar phase at the reference time.
Only the fractional part of the pulsar phase is retained so that its value is within the
interval $[0,1)$.

Pulsar information, including specifically the ephemerides of a pulsar, is handled by the
{\tt GPulsar} class that supports reading of ephemeris information from various file formats.
The format most relevant to COMPTEL is the {\tt psrtime} format, which is an ASCII file
format containing the radio pulsar database as maintained during the CGRO mission by the
pulsar group at Princeton and nowadays by the Jodrell Bank Centre for Astrophysics.\footnote{\url{https://www.jodrellbank.manchester.ac.uk}}
Upon loading of a {\tt psrtime} file, the {\tt GPulsar::load\_psrtime} method computes from
the data the pulsar phase at the reference time using
\begin{equation}
\Phi_0 = \nu \Delta t + \frac{1}{2} \dot{\nu} \Delta t^2 + \frac{1}{6} \ddot{\nu} \Delta t^3
,\end{equation}
where
\begin{equation}
\begin{split}
\Delta t = & ({\rm MJD}_0 - |{\rm MJD}_0|) \times 86400 + \Delta t_{\rm CGRO \to SSB} - \Delta t_{\rm Shapiro} + \\ 
& \Delta t_{\rm UTC \to TT} + \Delta t_{\rm TT \to TBD}
\end{split}
\label{eq:timeforphase}
\end{equation}
and ${\rm MJD}_0$ is the modified Julian day of the ephemeris reference (the 
{\tt GEphemerides::geo2ssb} method is used for this computation, and hence the Shapiro delay
in Eq.~(\ref{eq:timeforphase}) includes the displacement of the Sun around the Solar System 
barycentre, cf.~Eq.~(\ref{eq:shapiro})).

\section{Handling of failed photomultiplier tubes}
\label{app:fpmt}

COMPTEL comprised two detector layers, composed of 7 and 14 circular modules for D1 and D2, 
respectively.
Modules of the first layer were composed of the liquid scintillator NE~213A while modules of the
second layer were made of NaI(Tl) scintillator crystals.
Each D1 module was viewed by eight PMTs while each D2 module was 
viewed by seven PMTs.
The relative amplitudes of the PMT signals for a given module allowed localisation of the interactions
within the module with an average $1\sigma$ spatial resolution of 2.3 cm for D1 modules and 
1.5 cm for D2 modules \citep{schoenfelder1993}.

During the operations of COMPTEL a certain number of D2 module PMTs failed 
(cf.~Table \ref{tab:fpmt}), degrading the interaction localisation capabilities within the concerned
modules and hence increasing the uncertainties in the determination of the event scatter directions
$(\chi,\psi)$.
GammaLib implements several options for handling data from D2 modules with failed PMTs,
controlled through the {\tt GCOMSelection::fpmtflag} method that takes an integer value of
{\tt 0}, {\tt 1}, or {\tt 2}.
This integer value is a user parameter of the ctools scripts {\tt comobsbin} and {\tt compulbin} 
that, by default, is assumed to be {\tt 0}.

For {\tt fpmtflag = 0}, events registered in D2 modules with failed PMTs are ignored, and the 
corresponding modules are also excluded in the computation of the geometry function 
(cf.~Appendix \ref{app:drg}).
Conversely, for {\tt fpmtflag = 1} the failure of the PMTs is ignored, and events from D2 modules 
are treated as if the failed PMTs were still operating.
Finally, for {\tt fpmtflag = 2}, circular exclusion regions are applied around the zones of the failed
PMTs for dates after their failure, as defined in Table \ref{tab:fpmt}.
Events localised within these regions are ignored and the regions are removed in the computation 
of the geometry function (cf.~Appendix \ref{app:drg}).
Using circular exclusion regions was the default for most of the COMPTEL analysis published in
the literature in the past.
The impact of the {\tt fpmtflag} value on the analysis results is illustrated for the case of the
spectral energy distribution of LS~5039 in Sect. \ref{sec:ls5039}.

\begin{table}[!th]
\caption{D2 module exclusion circles applied following the failure of PMTs
at the specified dates.
\label{tab:fpmt}}
\centering
\begin{tabular}{l c c c c c c}
\hline\hline
TJD & VP & Date & Module & $x_e$ & $y_e$ & $r_e$ \\
\hline
8718     & 24.0 & 6-4-1992 & 13 & 9.0 & 41.2 & 9.0 \\
8737     & 26.0 & 25-4-1992     & 11 & -51.7 & 8.7 & 9.0 \\
8756 & 28.0 & 14-5-1992 & 14 & -34.7 & 49.0 & 9.0 \\
8981    & 204.0 & 25-12-1992 & 2         & \multicolumn{3}{c}{entire module} \\
\hline
\end{tabular}
\tablefoot{Module numbers are counted from 1. The circle centre locations $x$ and $y$ and the circle radii 
$r$ are given in units of cm.}
\end{table}

\section{Computation of the geometry function}
\label{app:drg}

The geometry function {\tt DRG} is computed in GammaLib by the method {\tt GCOMDri::compute\_drg}
using
\begin{equation}
{\tt DRG}(\chi, \psi, \bar{\varphi}) = \frac{1}{N} \sum_{i \in  \{ S \}} \tilde{G}_i(\chi, \psi, \bar{\varphi})
\end{equation}
with
\begin{equation}
\tilde{G}_i(\chi, \psi, \bar{\varphi}) =
\left\{
\begin{array}{l l}
\displaystyle
G_i(\chi, \psi), & \mbox{if ${\rm EHA}(\chi, \psi) \ge {\rm EHA}_{\rm min}(\bar{\varphi})$} \\
\displaystyle
0, & \mbox{otherwise}
\end{array}
\right.
,\end{equation}
where
${\rm EHA}(\chi, \psi)$ is the distance between scatter direction $(\chi, \psi)$ and the Earth
horizon,
${\rm EHA}_{\rm min}(\bar{\varphi})$ is given by Eq.~(\ref{eq:eha}), and
$N$ is the number of selected superpackets.

The $G_i(\chi, \psi)$ is the geometry factor for a given superpacket and corresponds to the area of
the shadow that is cast by all active D1 modules on all active D2 modules for a given scatter
angle $(\chi, \psi)$ divided by the total area of all D1 modules.
The computation of the geometry factor is implemented in {\tt GCOMDri::compute\_geometry} 
and calculated using
\begin{equation}
G_i(\chi, \psi) = \frac{1}{7} \sum_{k=0}^{6} \delta^{{\rm D1}}_k \sum_{l=0}^{13} \delta^{{\rm D2}}_l \tilde{o}_{kl}(\theta,\phi)
,\end{equation}
where
\begin{equation}
\delta^{{\rm D1}}_k = 
\left\{
\begin{array}{l l}
\displaystyle
1, & \mbox{D1 module $k$ active} \\
\displaystyle
0, & \mbox{otherwise} \\
\end{array}
\right.
\end{equation}
and
\begin{equation}
\delta^{{\rm D2}}_l = 
\left\{
\begin{array}{l l}
\displaystyle
1, & \mbox{D2 module $l$ active} \\
\displaystyle
0, & \mbox{otherwise} \\
\end{array}
\right.
\end{equation}
are determined using the {\tt GCOMStatus} class.
In case of {\tt fpmtflag = 0}, failed modules according to Table \ref{tab:fpmt} are also considered as
inactive for superpacket dates after the dates of PMT failure.
Furthermore,
\begin{equation}
\tilde{o}_{kl}(\theta,\phi) = 
\left\{
\begin{array}{l l}
\displaystyle
0, & \mbox{if $d_{kl} \ge r_1 + r_2$} \\
\displaystyle
1 - f_{kle}(\theta,\phi), & \mbox{if $d_{kl} \le r_2 - r_1 + 0.1$} \\
\displaystyle
o_{kl}(\theta,\phi) - f_{kle}(\theta,\phi), & \mbox{otherwise} \\
\end{array}
\right.
\label{eq:overlap}
,\end{equation}
with
$r_1=13.8$ cm being the radius of a D1 module and
$r_2=14.085$ cm being the radius of a D2 module.
We note the margin of $0.1$ cm in Eq.~(\ref{eq:overlap}), which assures numerical stability with 
respect to rounding errors. The 
$d_{kl}$ is the projected distance between the centres of the D1 and D2 modules, given
by
\begin{equation}
d_{kl} = \sqrt{(x_l - x_k + h \tan \theta \cos \phi)^2 + (y_l - y_k + h \tan \theta \sin \phi)^2}
,\end{equation}
where $x_k$ and $y_k$ are the geometric positions of the D1 modules and
$x_l$ and $y_l$ the positions of the D2 modules with respect to the optical axis in cm, 
$h=158$ cm being the vertical separation between D1 and D2 modules,
$(\theta,\phi)$ being the zenith and azimuth angles of the Compton scatter direction
$(\chi, \psi)$ with respect to the COMPTEL pointing direction, and
\begin{equation}
o_{kl}(\theta,\phi) = \frac{
r_1^2 (\alpha_{kl} - \sin \alpha_{kl} \cos \alpha_{kl}) +
r_2^2 (\beta_{kl} - \sin \beta_{kl} \cos \beta_{kl})
}{\pi r_1^2}
\end{equation}
being the projected overlap of a D1 module and a D2 module with
\begin{equation}
\cos \alpha_{kl} = \frac{d_{kl}^2 + (r_1^2 - r_2^2)}{2 d_{kl} r_1}
\end{equation}
and
\begin{equation}
\cos \beta_{kl} = \frac{d_{kl}^2 - (r_1^2 - r_2^2)}{2 d_{kl} r_2} .
\end{equation}

The term $f_{kle}(\theta,\phi)$ accounts for the exclusion of circular regions around failed
PMTs and differs from zero only for {\tt fpmtflag = 2}.
It quantifies the fractional overlap between the projected D1 module $k$ and the part of 
the exclusion region for failed PMTs that is contained within D2 module $l$.
The exclusion region is circular, and specified by a geometric centre position $x_e$ and 
$y_e$ and a radius $r_e$ as given in Table \ref{tab:fpmt}.
Specifically, for {\tt fpmtflag = 2} 
\begin{equation}
f_{kle}(\theta,\phi) = 
\left\{
\begin{array}{l l}
\displaystyle
0, & \mbox{if $d_{kl} > r_1 + r_2$ or $d_{ke} < r_1 + r_e$} \\
\displaystyle
f^{\rm contained}_{le}(\theta,\phi), & \mbox{if $d_{ke}+ r_e < r_1$} \\
\displaystyle
f^{\rm partial}_{kle}(\theta,\phi), & \mbox{otherwise} \\
\end{array}
\right.
,\end{equation}
with
\begin{equation}
d_{ke} = \sqrt{(x_e - x_k + h \tan \theta \cos \phi)^2 + (y_e - y_k + h \tan \theta \sin \phi)^2}
\end{equation}
being the distance between the projected D1 module centres $x_k$ and $y_k$ and the 
centre $x_e$ and $y_e$ of the exclusion region.

The $f^{\rm contained}_{le}(\theta,\phi)$ is the overlap for the case that the exclusion circle is
fully contained in the projected D1 module circumference.
In this case, the relevant overlap to take into account is the overlap between the exclusion 
circle and the D2 module, given by
\begin{equation}
f^{\rm contained}_{le}(\theta,\phi) = 
\left\{
\begin{array}{l l}
\displaystyle
r_2^2 / r_1^2, & \mbox{if $r_1 > d_{le} + r_2$} \\
\displaystyle
1, & \mbox{if $r_2 > d_{le} + r_1$} \\
\displaystyle
\tilde{f}^{\rm contained}_{le}(\theta,\phi), & \mbox{otherwise} \\
\end{array}
\right.
\label{eq:fcontained}
,\end{equation}
with
\begin{equation}
d_{le} = \sqrt{(x_e-x_l)^2 + (y_e-y_l)^2}
\end{equation}
being the distance between the centres of the exclusion circle and the D2 module, and
\begin{equation}
\tilde{f}^{\rm contained}_{le}(\theta,\phi) = \frac{
r_1^2 (\alpha_{le} - \sin \alpha_{le} \cos \alpha_{le}) +
r_2^2 (\beta_{le} - \sin \beta_{le} \cos \beta_{le})
}{\pi r_1^2}
\end{equation}
being the overlap of the exclusion region and the D2 module with
\begin{equation}
\cos \alpha_{le} = \frac{d_{le}^2 + (r_1^2 - r_2^2)}{2 d_{le} r_1}
\end{equation}
and
\begin{equation}
\cos \beta_{le} = \frac{d_{le}^2 - (r_1^2 - r_2^2)}{2 d_{le} r_2} .
\end{equation}
The evaluation of Eq.~(\ref{eq:fcontained})  is implemented by the method 
{\tt GCOMDri::compute\_surface}.

Finally, $f^{\rm partial}_{kle}(\theta,\phi)$ specifies the fractional overlap between the projected
D1 module, the D2 module and the exclusion circle.
This quantity is evaluated numerically by testing a grid of $25 \times 25$ $x$ and $y$ positions
around the exclusion circle.
The numerical evaluation is implemented by the method {\tt GCOMDri::compute\_overlap}.

\section{Response computation}
\label{app:response}

The following sections describe some details of the response computations implemented in
GammaLib.

\subsection{Efficiency factors}
\label{app:efficiency}

The efficiency factor $P_{\rm eff}(\varphi'_{\rm geo}, E_{\gamma})$ in Eq.~(\ref{eq:iaq}) is factorised
according to
\begin{equation}
\begin{split}
P_{\rm eff}(\varphi'_{\rm geo}, E_{\gamma}) = &
P_{\rm A1}(E_{\gamma}) \times 
P_{\rm V1}(E_{\gamma}) \times 
P_{\rm D1}(E_{\gamma}) \times 
P_{\rm C}(E_{\gamma}) \, \times \\ 
& P_{\rm MH}(E_{\gamma}) \times 
P_{\rm SV}(E_{\gamma}) \, \times \\ 
& P_{\rm PSD}(\hat{E_1}) \times 
P_{\rm A2}(\hat{E_2}) \times 
P_{\rm V23}(\hat{E_2}) \times 
P_{\rm D2}(\hat{E_2}) \, \times \\ 
& P_{\rm MS}(\varphi'_{\rm geo}, E_{\gamma}) 
\end{split}
,\end{equation}
with
\begin{equation}
\hat{E_2} = \frac{E_{\gamma}}{\left( 1 - \cos \varphi'_{\rm geo} \right) \frac{E_{\gamma}}{m_e c^2} +1} 
\end{equation}
being the energy of the photon entering the D2 module for a true scatter angle $\varphi'_{\rm geo}$
and an incident photon energy of $E_{\gamma}$, and $\hat{E_1} = E_{\gamma} - \hat{E_2}$ being 
the true energy deposit in D1 for a single Compton scattering by an angle of $\varphi'_{\rm geo}$.
The GammaLib methods that implement the computation of the efficiency factors are summarised in
Table~\ref{tab:efficiency}.

\begin{table}[!h]
\caption{GammaLib methods that implement the computation of the efficiency factors.
\label{tab:efficiency}}
\centering
\begin{tabular}{l l}
\hline\hline
Factor & Method \\
\hline
$P_{\rm eff}(\varphi'_{\rm geo}, E_{\gamma})$ & {\tt GCOMIaq::weight\_iaq} \\
$P_{\rm A1}(E_{\gamma})$ & {\tt GCOMInstChars::trans\_D1} \\
$P_{\rm V1}(E_{\gamma})$ & {\tt GCOMInstChars::trans\_V1} \\
$P_{\rm D1}(E_{\gamma})$ & {\tt GCOMInstChars::prob\_D1inter} \\
$P_{\rm C}(E_{\gamma})$ & {\tt GCOMIaq::weight\_iaq} \\
$P_{\rm MH}(E_{\gamma})$ & {\tt GCOMInstChars::prob\_no\_multihit} \\
$P_{\rm SV}(E_{\gamma})$ & {\tt GCOMInstChars::prob\_no\_selfveto} \\
$P_{\rm PSD}(\hat{E_1})$ & {\tt GCOMInstChars::psd\_correction} \\
$P_{\rm A2}(\hat{E_2})$ & {\tt GCOMInstChars::trans\_D2} \\
$P_{\rm V23}(\hat{E_2})$ & {\tt GCOMInstChars::trans\_V23} \\
$P_{\rm D2}(\hat{E_2})$ & {\tt GCOMInstChars::prob\_D2inter} \\
$P_{\rm MS}(\varphi'_{\rm geo}, E_{\gamma})$ & {\tt GCOMInstChars::multi\_scatter} \\
\hline
\end{tabular}
\end{table}

\begin{table*}[!th]
\caption{Efficiency parameters as a function of energy.
\label{tab:coefficients}}
\centering
\begin{tabular}{c c c c c c c}
\hline\hline
$E$ (MeV) &
$\mu_{\rm Veto}(E)$ & 
$\mu_{\rm Al}(E)$ & 
$\mu_{\rm D1}(E)$ & 
$\mu_{\rm D2}(E)$ & 
$P_{\rm MS}(E)$ &
$P_{\rm SV}(E)$ \\
\hline
0.05 & 0.205944 & & 0.178262 & & & \\
0.1 & 0.169298 & 0.432 & 0.146770 & 5.9821 & & \\
0.15 & 0.150704 & 0.3618 & 0.130690 & 2.15796 & & \\
0.2 & 0.137444 & 0.324 & 0.119207 & 1.14871 & 1.0 & \\
0.25 & 0.127823 & & 0.110860 & & & \\
0.3 & 0.119531 & 0.2781 & 0.103667 & 0.57986 & 1.0 & \\
0.35 & 0.112835 & & 0.097861 & & & \\
0.4 & 0.106932 & 0.24894 & 0.092743 & 0.41471 & 1.0 & \\
0.45 & 0.101949 & & 0.088426 & & & \\
0.5 & 0.097504 & 0.22707 & 0.084574 & 0.336906 & 1.0 & \\
0.55 & 0.093717 & & 0.081289 & & & \\
0.6 & 0.090226 & 0.21006 & 0.078260 & 0.293233 & 1.0 & \\
0.65 & 0.087195 & & 0.075629 & & & \\
0.7 & 0.084350 & & 0.073160 & & & \\
0.75 & 0.081692 & & 0.070853 & & & \\
0.8 & 0.079220 & 0.18468 & 0.068708 & 0.242954 & 1.0 & \\
0.85 & 0.077065 & & 0.066840 & & & \\
0.9 & 0.075009 & & 0.065057 & & & \\
0.95 & 0.073052 & & 0.063360 & & & \\
1.0 & 0.071194 & 0.16578 & 0.061749 & 0.212493 & 0.993 & \\
1.25 & 0.064013 & & 0.055524 & & & \\
1.5 & 0.058035 & & 0.050340 & 0.170655 & 0.991 & \\
1.75 & 0.053580 & & 0.046471 & & & \\
2.0 & 0.049689 & 0.1166 & 0.043092 & 0.150837 & 0.985 & \\
2.5 & 0.044222 & & 0.038345 & & & \\
3.0 & 0.039767 & 0.0953 & 0.034476 & 0.134322 & 0.976 & \\
3.5 & 0.036580 & & 0.031708 & & & \\
4.0 & 0.033893 & 0.0837 & 0.029373 & 0.128817 & 0.975 & \\
4.4 & & & & & & 0.940 \\
4.5 & 0.031814 & & 0.027565 & & & \\
5.0 & 0.030016 & 0.07641 & 0.026002 & 0.127349 & 0.973 & \\
5.5 & 0.028562 & & 0.024737 & & & \\
6.0 & 0.027268 & 0.07155 & 0.023611 & 0.127716 & 0.971 & \\
6.1 & & & & & & 0.912 \\
6.5 & 0.026254 & & 0.022727 & & & \\
7.0 & 0.025319 & & 0.021912 & & &  \\
7.5 & 0.024462 & & 0.021166 & & & \\
8.0 & 0.023684 & 0.06534 & 0.020488 & 0.130285 & 0.950 & \\
8.5 & 0.023043 & & 0.019929 & & & \\
9.0 & 0.022442 & & 0.019405 & & & \\
9.5 & 0.021880 & & 0.018915 & & & \\
10.0 & 0.021357 & 0.0621 & 0.018459 & 0.134689 & 0.929 & \\
12.0 & & & & & & 0.850 \\
12.5 & 0.019618 & & 0.016940 & & & \\
15.0 & 0.018280 & 0.05859 & 0.015770 & 0.147534 & 0.912 & \\
17.5 & 0.017477 & & 0.015066 & & & \\
20.0 & 0.016805 & 0.05805 & 0.014475 & 0.158544 & 0.913 & \\
20.5 & & & & & & 0.718 \\
25.0 & 0.015988 & & 0.013756 & & 0.915 & \\
30.0 & 0.015407 & 0.05859 & 0.013242 & 0.176894 & 0.917 & \\
35.0 & 0.015158 & & 0.013019 & & & \\
40.0 & 0.014948 & 0.06048 & 0.012830 & 0.190840 & & \\
45.0 & 0.014756 & & 0.012659 & & & \\
50.0 & 0.014646 & 0.06210 & 0.012558 & 0.201116 & & \\
\hline
\end{tabular}
\tablefoot{Interaction coefficients, $\mu$, are given in units of cm$^{-1}$.}
\end{table*}

The $P_{\rm A1}(E_{\gamma})$ is the transmission probability for photons for the material 
above D1, which is composed essentially of aluminium, and is computed using
\begin{equation}
P_{\rm A1}(E_{\gamma}) = e^{-\mu_{\rm Al}(E_{\gamma}) \, l_{\rm above}}
,\end{equation}
where
$\mu_{\rm Al}(E_{\gamma})$ is the energy-dependent interaction coefficient of aluminium
in units of cm$^{-1}$ that is interpolated using a log-log interpolation of the values given in
Table \ref{tab:coefficients} and $l_{\rm above}=0.147$ cm is the thickness of the material above D1.

The $P_{\rm V1}(E_{\gamma})$ is the transmission probability for photons for the first Veto dome
and is computed using
\begin{equation}
P_{\rm V1}(E_{\gamma}) = e^{-\mu_{\rm Veto}(E_{\gamma}) \, l_{\rm V1}}
,\end{equation}
where
$\mu_{\rm Veto}(E_{\gamma})$ is the energy-dependent interaction coefficient for the Veto dome
in units of cm$^{-1}$ that is interpolated using a log-log interpolation of the values given in
Table \ref{tab:coefficients} and $l_{\rm V1}=1.721$ cm is the thickness of the first Veto dome.

$P_{\rm D1}(E_{\gamma})$ is the interaction probability in D1 and is computed using
\begin{equation}
P_{\rm D1}(E_{\gamma}) = 1 - e^{-\mu_{\rm D1}(E_{\gamma}) \, l_{\rm D1}}
,\end{equation}
where
$\mu_{\rm D1}(E_{\gamma})$ is the energy-dependent D1 attenuation coefficient in units of
cm$^{-1}$ that is interpolated using a log-log interpolation of the values given in
Table \ref{tab:coefficients} and $l_{\rm D1}=8.5$ cm is the thickness of the D1 modules.

$P_{\rm C}(E_{\gamma})$ is the fraction of Compton scatter interactions among all possible
photon interactions within D1 and is given by
\begin{equation}
P_{\rm Compton}(E_{\gamma}) = \left\{
  \begin{array}{l l}
  \displaystyle
  1.067 - 0.0295 E_{\gamma} + 3.4 \, 10^{-4} E^2_{\gamma}, & \mbox{if $E_{\gamma} \ge 2$ MeV} \\
  \displaystyle
  1, & \mbox{otherwise} \\
  \end{array}
\right.
,\end{equation}
with the additional constraint of $0 \le P_{\rm Compton}(E_{\gamma}) \le 1$.

The $P_{\rm MH}(E_{\gamma})$ is the probability that there is no multi-hit.
This probability is computed using a log-log interpolation of the values of Table \ref{tab:coefficients}.
$P_{\rm SV}(E_{\gamma})$ is the probability that the photon was not self-vetoed.
This probability is computed using a linear interpolation of the values of Table \ref{tab:coefficients}.

The $P_{\rm PSD}(\hat{E_1})$ is the D1 energy-dependent PSD correction and is given by
\begin{equation}
P_{\rm PSD}(\hat{E_1}) = 1 - \frac{1}{1727.9 \times \hat{E_1}^{2.53} + 1}
,\end{equation}
where $\hat{E_1}$ is in units of MeV.
This correction applies to a standard PSD selection of 0--110.

$P_{\rm A2}(\hat{E_2})$ gives the transmission probability of the aluminium below the D1 modules
and is computed using
\begin{equation}
P_{\rm A2}(\hat{E_2}) = e^{-\mu_{\rm Al}(\hat{E_2}) \, l_{\rm T23}}
,\end{equation}
where
$\mu_{\rm Al}(\hat{E_2})$ is the energy-dependent interaction coefficient for aluminium in units of
cm$^{-1}$ that is interpolated using a log-log interpolation of the values given in
Table \ref{tab:coefficients} and $l_{\rm between}=0.574$ cm is the thickness of the aluminium plate.

$P_{\rm V23}(\hat{E_2})$ gives the transmission probability of the second and third veto domes
that are situated between the D1 and D2 modules, and is computed using
\begin{equation}
P_{\rm V23}(\hat{E_2}) = e^{-\mu_{\rm Veto}(\hat{E_2}) \, l_{\rm V23}}
,\end{equation}
where
$\mu_{\rm Veto}(\hat{E_2})$ is the energy-dependent interaction coefficient of the Veto domes in
units of cm$^{-1}$ that is interpolated using a log-log interpolation of the values given in
Table \ref{tab:coefficients} and $l_{\rm V23}=3.442$ cm is the combined thickness of the second
and third veto domes.

$P_{\rm D2}(E_{\gamma})$ gives the interaction probability in D2 and is computed using
\begin{equation}
P_{\rm D2}(E_{\gamma}) = 1 - e^{-\mu_{\rm D2}(\hat{E_2}) \, l_{\rm D2}}
,\end{equation}
where
$\mu_{\rm D2}(\hat{E_2})$ is the energy-dependent D2 attenuation coefficient in units of
cm$^{-1}$ that is interpolated using a log-log interpolation of the values given in
Table \ref{tab:coefficients} and $l_{\rm D2}=7.525$ cm is the thickness of the D2 modules.

\begin{figure*}[!th]
\centering
\includegraphics[width=8.8cm]{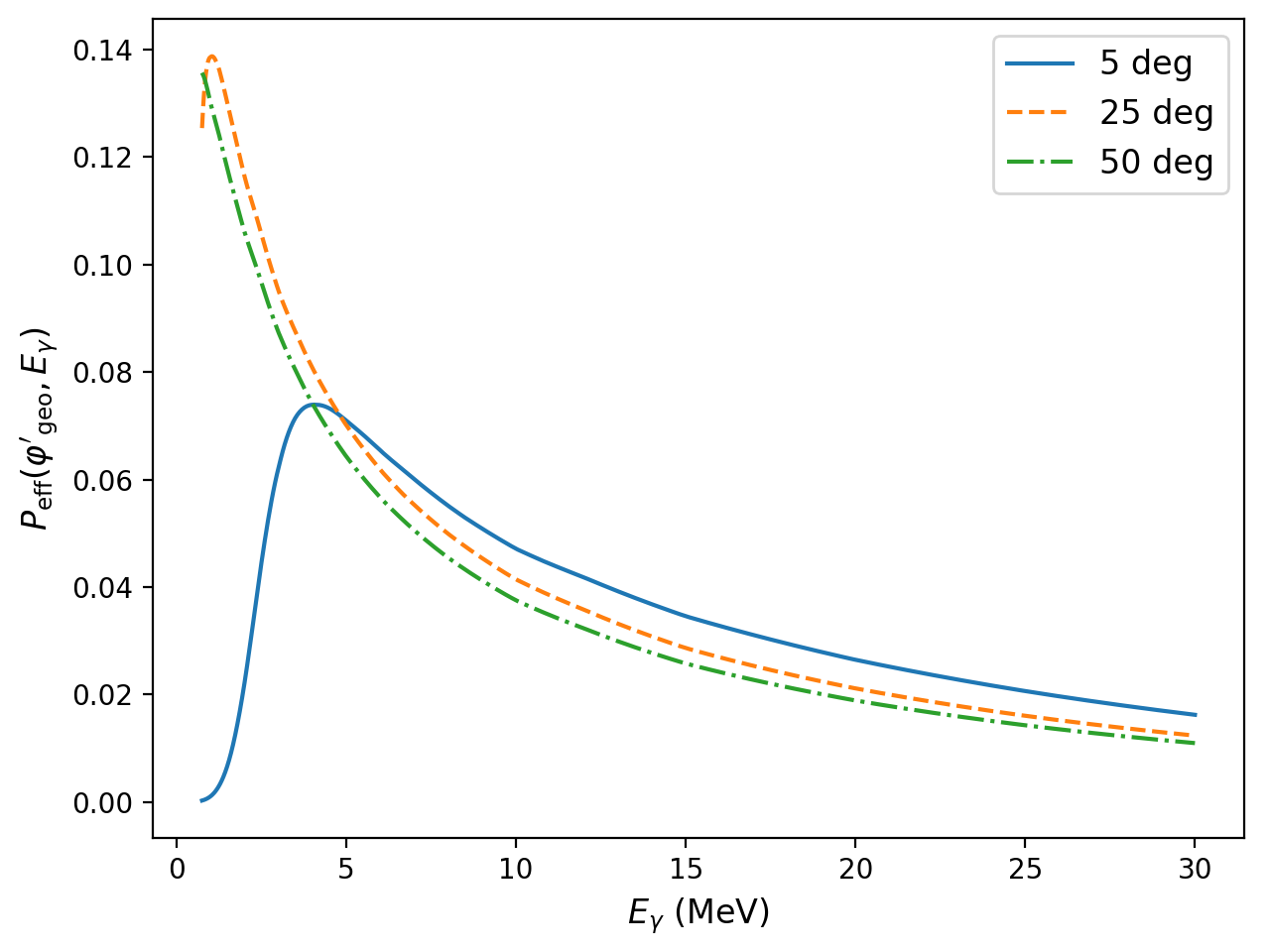}
\includegraphics[width=8.8cm]{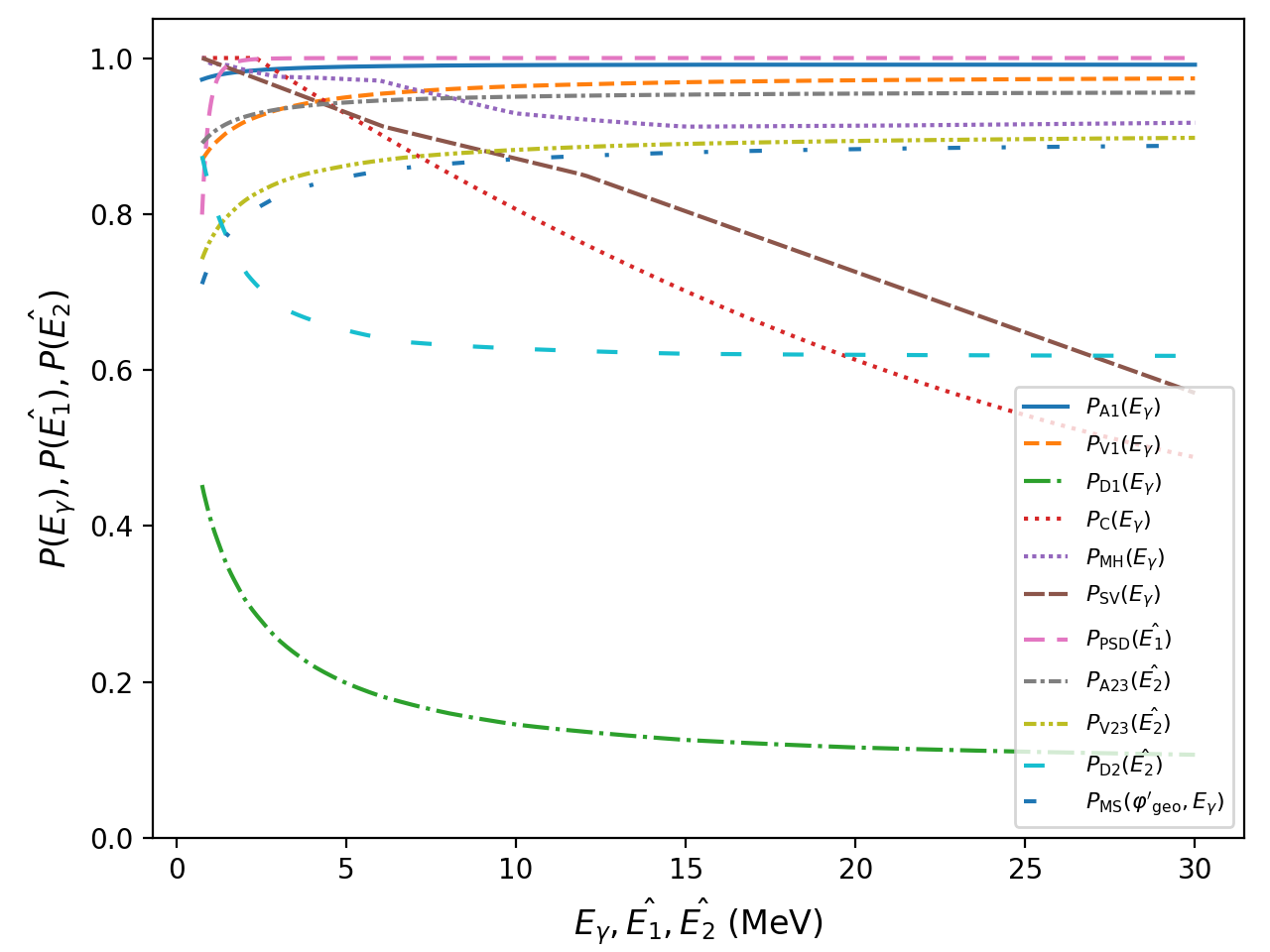}
\caption{
Energy dependence of the efficiency factor $P_{\rm eff}(\varphi'_{\rm geo}, E_{\gamma})$
for the three geometrical scatter angles $\varphi'_{\rm geo}=5\degrees$, $25\degrees$, and $50\degrees$ (left)
and the energy dependence of its components for $\varphi'_{\rm geo}=25\degrees$ (right).
\label{fig:efficiency}
}
\end{figure*}

The $P_{\rm MS}(E_{\gamma},\hat{E_2})$ is the probability that a photon that has interacted in D1
leaves the D1 module without any further interaction.
This probability is computed using
\begin{equation}
P_{\rm MS}(\varphi'_{\rm geo}, E_{\gamma}) =
\frac{
\int_0^{r_{\rm D1}} 
\int_0^{l_{\rm D1}} \mu_{\rm D1}(E_{\gamma}) e^{-\mu_{\rm D1}(E_{\gamma}) z}
\int_0^{\pi} e^{-\mu_{\rm D1}(\hat{E_2}) \, l(r,z,\phi)} \, d\phi \, dz \, dr}
{\pi r^2_{\rm D1} \left( 1 - e^{-\mu_{\rm D1}(E_{\gamma}) \, l_{\rm D1}} \right)}
,\end{equation}
which integrates all possible second interaction locations within a given D1 module assuming vertically
incident photons, where $r_{\rm D1}=13.8$ cm and $l_{\rm D1}=8.5$ cm is the radius and
thickness of a D1 module.
$l(r,z,\phi)$ gives the remaining interaction length for a photon that has interacted at radius $r$,
vertical depth $z$ and azimuth angle $\phi$ of the module and is given by
\begin{equation}
l(r,z,\phi) = \left\{
  \begin{array}{l l}
  \displaystyle
   \frac{l_{\rm D1}-z}{\cos \varphi'_{\rm geo}}, & \mbox{if $l(r,z,\phi) \sin \varphi'_{\rm geo} \le R(r,\phi)$} \\
   \\
   \displaystyle
   \frac{R(r,\phi)}{\sin \varphi'_{\rm geo}}, & \mbox{otherwise} \\
  \end{array}
\right.
\end{equation}
with
\begin{equation}
R(r,\phi) = \sqrt{r^2_{\rm D1} - r^2 \sin^2 \phi} -r \cos \phi .
\end{equation}

Figure \ref{fig:efficiency} illustrates the energy dependence of the efficiency factor
$P_{\rm eff}(\varphi'_{\rm geo}, E_{\gamma})$
and its components.

\subsection{Klein-Nishina contribution}
\label{app:kleinnishina}

\begin{equation}
P_{\rm KN}(\varphi'_{\rm geo}, E_{\gamma}) =
\frac{\int_{\varphi'_{\rm geo,min}}^{\varphi'_{\rm geo,max}}
\sigma_{\rm KN}(\varphi''_{\rm geo}, E_{\gamma}) \, d\varphi''_{\rm geo}}
{\int_{0}^{\pi}
\sigma_{\rm KN}(\varphi''_{\rm geo}, E_{\gamma}) \, d\varphi''_{\rm geo}}
\end{equation}
is the contribution of the Klein-Nishina cross-section
$\sigma_{\rm KN}(\varphi''_{\rm geo}, E_{\gamma})$
to the $\varphi'_{\rm geo}$ bin spanned by $[\varphi'_{\rm geo,min}, \varphi'_{\rm geo,max}]$,
where the integrals are computed using
\begin{equation}
\begin{split}
\int_{\varphi'_{\rm geo,min}}^{\varphi'_{\rm geo,max}}
\sigma_{\rm KN}(\varphi''_{\rm geo}, E_{\gamma}) \, d\varphi''_{\rm geo} = &
\int_0^{E_1^{\rm up}} \sigma_{\rm KN}(E'_1, E_{\gamma}) \, dE'_1 \, - \\
& \int_0^{E_1^{\rm low}} \sigma_{\rm KN}(E'_1, E_{\gamma}) \, dE'_1
\end{split}
\end{equation}
with
\begin{equation}
\begin{split}
\int_0^{E_1} \sigma_{\rm KN}(E'_1, &E_{\gamma}) \, dE'_1 = 
\frac{E_1}{E_{\gamma}} \left( \frac{m_e c^2}{E_1} + 1 \right)^2 + \\
& \left( 2 \left( \frac{m_e c^2}{E_{\gamma}} \right)^2 + 2 \frac{m_e c^2}{E_{\gamma}} - 1 \right) \times
\log \left(1 - \frac{E_1}{E_{\gamma}} \right) - \\
& \frac{1}{2} \left( \frac{E_1}{E_{\gamma}} \right)^2 +
\left( \frac{m_e c^2}{E_{\gamma}} \right)^2
\frac{1}{1 - \frac{E_1}{E_{\gamma}}}
\end{split}
.\end{equation}

\subsection{Module spectral response functions}
\label{app:specrsp}

\begin{figure*}[!th]
\centering
\includegraphics[width=8.8cm]{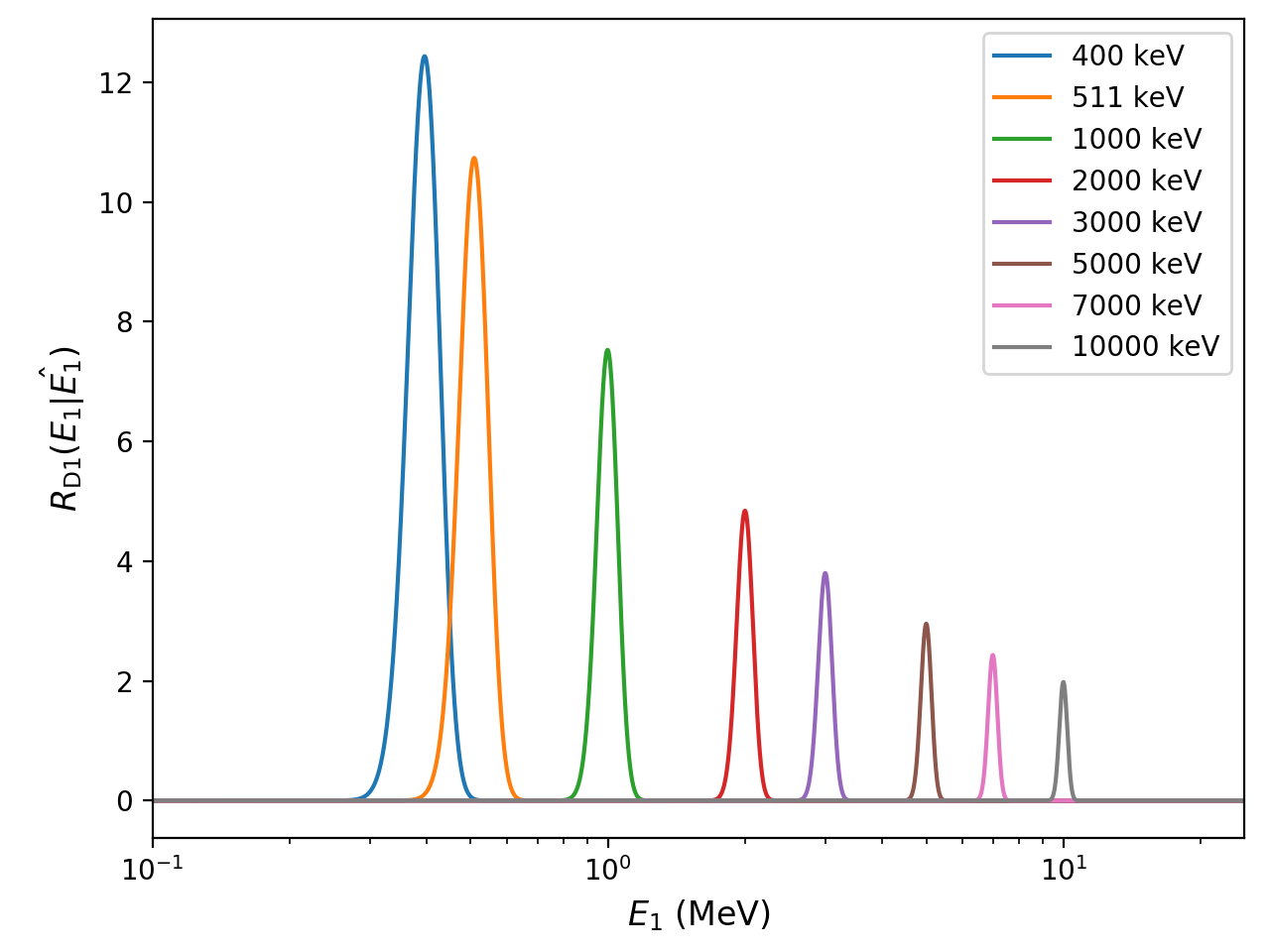}
\includegraphics[width=8.8cm]{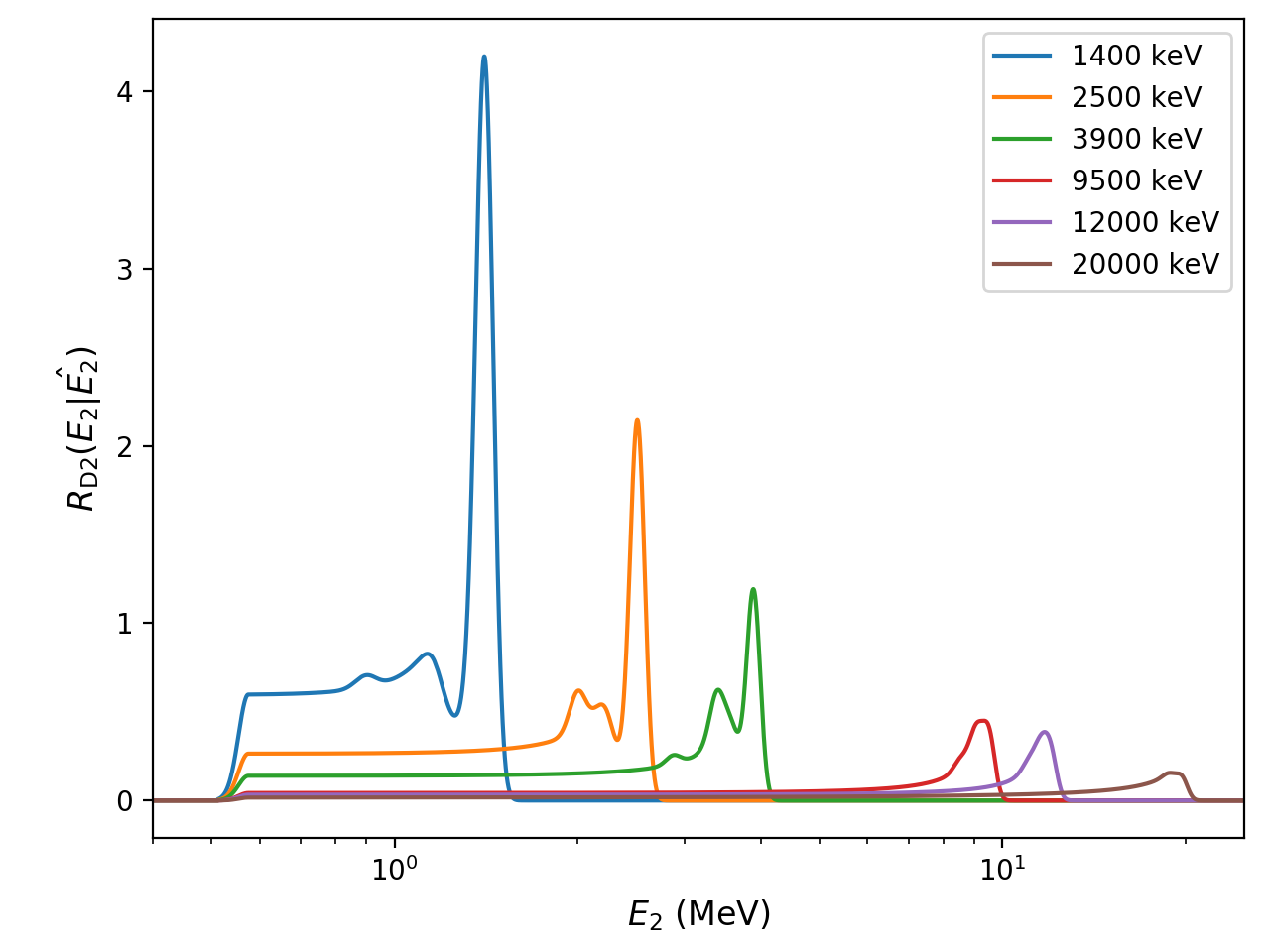}
\caption{
Detector response functions $R_{\rm D1}(E_1|\hat{E_1})$ (left) and $R_{\rm D2}(E_2|\hat{E_2})$
(right) for a selected number of energies that have been chosen to allow for a comparison
with Figs.~3c and 4b in \citet{diehl1992}.
\label{fig:moduleresponse}
}
\end{figure*}

\subsubsection{D1 spectral response}

The spectral response $R_{\rm D1}(E_1|\hat{E_1})$ provides the probability for measuring
an energy $E_1$ when an incident photon with energy $E_{\gamma}$ was Compton scattered
by an angle $\varphi_{\rm geo}$, producing a true energy deposit of $\hat{E_1}$ in the detector
module.
The D1 response is implemented by the {\tt GCOMD1Response} class and the response
evaluation is done using the {\tt GCOMD1Response::operator()} operator.
The spectral D1 response is computed using a Gaussian
\begin{equation}
R_{\rm D1}(E_1|\hat{E_1}) =
\frac{T_1(E_1|\hat{E_1})}{\sigma(\hat{E_1})\sqrt{2\pi}}
\exp \left( -\frac{1}{2} \left( \frac{\mu_{\rm p}(\hat{E_1})-E_1}{\sigma(\hat{E_1})} \right)^2 \right)
,\end{equation}
where $\mu_{\rm p}(\hat{E_1})$ is the position of the Gaussian peak for true energy deposit of
$\hat{E_1}$, with $\mu_{\rm p}(\hat{E_1}) \approx \hat{E_1}$, and $\sigma(\hat{E_1})$ is
the standard deviation of the Gaussian \citep{diehl1992}.
$T_1(E_1|\hat{E_1})$ is a threshold function, defined by
\begin{equation}
T_1(E_1|\hat{E_1}) = \left\{
\begin{array}{l l}
\displaystyle
0, & \mbox{if $E_1 \le E_1^{\rm thres}(\hat{E_1})$} \\
\\
\displaystyle
\frac{E_1 - E_1^{\rm thres}(\hat{E_1})}{\Delta E_1(\hat{E_1})}, & \mbox{if $0 <  E_1-E_1^{\rm thres}(\hat{E_1}) \le \Delta E_1(\hat{E_1})$} \\
\\
\displaystyle
0, & \mbox{if $E_1 > E_1^{\rm max}(\hat{E_1})$} \\
\\
\displaystyle
1, & \mbox{otherwise} \\
\end{array}
\right.
,\end{equation}
where
$E_1^{\rm thres}(\hat{E_1}) = E_1^{\rm min}(\hat{E_1}) - \frac{1}{2} \Delta E_1(\hat{E_1})$.
$\mu_{\rm p}(\hat{E_1})$ and $\sigma(\hat{E_1})$ as well as
$E_1^{\rm min}(\hat{E_1})$, $E_1^{\rm max}(\hat{E_1})$ and $\Delta E_1(\hat{E_1})$
are tabulated as a function of $\hat{E_1}$ in a calibration file of type {\tt SDA} that is included in
GammaLib.
Values for a given energy $\hat{E_1}$ are obtained by linear interpolation of the tabulated values.

The spectral response $R_{\rm D1}(E_1|\hat{E_1})$ is illustrated in the left panel of Fig.~\ref{fig:moduleresponse}
for a selected number of input energies $\hat{E_1}$.
The figure can be compared with Fig.~3c in \citet{diehl1992}, which presents the response
functions as implemented in COMPASS at the time of the CGRO launch for comparable energies.

\subsubsection{D2 spectral response}

The spectral response $R_{\rm D2}(E_2|\hat{E_2})$ provides the probability for measuring
an energy $E_2$ when an photon with energy $\hat{E_2}$ hit a D2 detector module.
The D2 response is implemented by the {\tt GCOMD2Response} class and the response
evaluation is done using the {\tt GCOMD2Response::operator()} operator.
The spectral D2 response is computed using
\begin{equation}
\begin{split}
R_{\rm D2}(E_2|\hat{E_2}) = &
(P_{\rm pp}(E_2|\hat{E_2}) +
P_{\rm se}(E_2|\hat{E_2}) +
P_{\rm de}(E_2|\hat{E_2}) \, + \\
& C_{\rm com}(E_2|\hat{E_2}) +
C_{\rm bkg}(E_2|\hat{E_2})) \times T_2(E_2|\hat{E_2})
\end{split}
,\end{equation}
which is a combination of three Gaussian and two continuum components multiplied by
a threshold function $T_2(E_2|\hat{E_2})$ \citep{diehl1992}.
Here, $\hat{E_2}$ is the energy of the gamma-ray photon incident on a D2 module, and
$E_2$ is the measured energy deposit.

The first Gaussian,
\begin{equation}
P_{\rm pp}(E_2|\hat{E_2}) =
\frac{A_{\rm pp}(\hat{E_2})}{\sigma(\hat{E_2})\sqrt{2\pi}}
\exp \left( -\frac{1}{2} \left( \frac{\mu_{\rm pp}(\hat{E_2})-E_2}{\sigma(\hat{E_2})} \right)^2 \right)
,\end{equation}
describes the so-called photo-peak which represents photons that were fully absorbed in the D2 module.
We note that the position $\mu_{\rm pp}(\hat{E_2})$ of the photo-peak is not necessarily identical to the
incident energy, although below $\sim10$~MeV $\mu_{\rm pp}(\hat{E_2}) \approx \hat{E_2}$ is well satisfied.

The two subsequent Gaussians describe the single and double escape peak, respectively, which are
generated by photons above 1.022~MeV that undergo pair creation resulting in the production of two
511~keV gamma rays.
The escape of one or both of these 511~keV photons from the D2 module leads to characteristic peaks at
energies 0.511~MeV and 1.022~MeV below the photo-peak energy.
The escape peaks are modelled using
\begin{equation}
P_{\rm se}(E_2|\hat{E_2}) =
\frac{A_{\rm se}(\hat{E_2})}{\sigma(\mu_{\rm se}(\hat{E_2}))\sqrt{2\pi}}
\exp \left( -\frac{1}{2} \left( \frac{\mu_{\rm se}(\hat{E_2}) - E_2}{\sigma(\mu_{\rm se}(\hat{E_2}))} \right)^2 \right)
\end{equation}
and
\begin{equation}
P_{\rm de}(E_2|\hat{E_2}) =
\frac{A_{\rm de}(\hat{E_2})}{\sigma(\mu_{\rm de}(\hat{E_2}))\sqrt{2\pi}}
\exp \left( -\frac{1}{2} \left( \frac{\mu_{\rm de}(\hat{E_2}) - E_2}{\sigma(\mu_{\rm de}(\hat{E_2}))} \right)^2 \right)
,\end{equation}
with
$\mu_{\rm se}(\hat{E_2}) = \mu_{\rm pp}(\hat{E_2}) - m_e c^2$
and
$\mu_{\rm de}(\hat{E_2}) = \mu_{\rm pp}(\hat{E_2}) - 2 m_e c^2$,
where $m_e c^2 \approx 511$~keV is the rest energy of the electron.

The first continuum component,
\begin{equation}
\begin{split}
C_{\rm com}(E_2|&\hat{E_2}) =  A_{\rm com}(\hat{E_2}) \, \times \\
& \int_{E_2 - 3 \sigma(E_2)}^{\max (E_2 + 3 \sigma(E_2), E_c)} 
\sigma_{\rm KN}(E|\hat{E_2}) \, f(E|E_2,\hat{E_2}) \, K(E|E_2) \, dE
\end{split}
,\end{equation}
describes the so-called Compton tail which represents photons that underwent a Compton scattering before
escaping the D2 module.
The component is the product of the Klein-Nishina cross-section for Compton scattering,
\begin{equation}
\sigma_{\rm KN}(E|\hat{E_2}) = 
\left( \frac{E/\hat{E_2}}{1-E/\hat{E_2}}
\frac{m_e c^2}{\hat{E_2}} \right)^2 -
\frac{E}{\hat{E_2}} + \frac{1}{1-E/\hat{E_2}}
,\end{equation}
and the probability that the Compton scattered photon does not undergo a second Compton scattering
before escaping, empirically modelled using
\begin{equation}
f(E|\hat{E_2}) = \left \{
  \begin{array}{l l}
  \displaystyle
  \exp \left( -\mu(E|\hat{E_2}) \, l(\hat{E_2}) \right), & \mbox{if $\hat{E_2} > 12.14$ MeV} \\
   \displaystyle
  0, & \mbox{otherwise} \\
  \end{array}
\right .
,\end{equation}
where
\begin{equation}
\mu(E|\hat{E_2}) = 0.72 \, e^{-1.28 (\hat{E_2} - E)^{0.35}} + 0.01 \, (\hat{E_2} - E) + 0.014 \, (\hat{E_2} - E)^{-2.5}
\end{equation}
is the total linear attenuation coefficient in NaI, which is the scintillator material of the D2 modules and
\begin{equation}
l(\hat{E_2}) = 2.9 \ln( \hat{E_2} - 11.14)
\end{equation}
is an empirical path length in the D2 module, where $\hat{E_2}$ is in units of MeV.
The product is convolved with a Gaussian kernel,
\begin{equation}
K(E|E_2) = \frac{1}{\sigma(E_2)\sqrt{2\pi}} \exp \left( -\frac{1}{2} \left( \frac{E - E_2}{\sigma(E_2)} \right)^2 \right)
,\end{equation}
to take the energy resolution of the D2 module into account, where $\sigma(E_2)$ is the standard deviation of the
detector response at the reconstructed energy $E_2$.

The second continuum component,
\begin{equation}
C_{\rm bkg}(E_2|\hat{E_2}) = A_{\rm bkg}(\hat{E_2}) \int_{E_2 - 3 \sigma(E_2)}^{\max (E_2 + 3 \sigma(E_2), \hat{E_2})}
K(E|E_2) \, dE ,
\end{equation}
is a flat background component that is convolved by a Gaussian kernel and that takes 
into account any higher-order scatterings in the D2 modules.

The threshold function is defined by
\begin{equation}
T_2(E_2|\hat{E_2}) = \left \{
\begin{array}{l l}
\displaystyle
\exp{ \left( - \frac{1}{2} \left( \frac{E_2^{\rm thres}(\hat{E_2}) - E_2}
{\sigma_{\rm thres}(\hat{E_2})} \right)^2 \right) }, & \mbox{if $E_2 < E_2^{\rm thres}(\hat{E_2})$} \\
\\
\displaystyle
0, & \mbox{if $E_2 > E_2^{\rm max}(\hat{E_2})$} \\
\\
\displaystyle
1, & \mbox{otherwise} \\
\end{array}
\right .
,\end{equation}
where
$E_2^{\rm thres}(\hat{E_2}) = E_2^{\rm min}(\hat{E_2}) + \frac{1}{2} \Delta E_2(\hat{E_2})$ and
$\sigma_{\rm thres}(\hat{E_2}) = \Delta E_2(\hat{E_2}) / 2 \sqrt{2 \log 2}$

All five response components have amplitude parameters
($A_{\rm pp}$, $A_{\rm se}$, $A_{\rm de}$, $A_{\rm com}$ and $A_{\rm bkg}$)
that are tabulated as a function of incident energy $\hat{E_2}$ in a calibration file of type {\tt SDB}.
In addition, this file contains also tabulated values for the photo-peak energy $\mu_{\rm pp}(\hat{E_2})$,
the standard deviation $\sigma(\hat{E_2})$ and the minimum and maximum threshold energies
$E_2^{\rm min}(\hat{E_2})$ and $E_2^{\rm max}(\hat{E_2})$ and width $\Delta E_2(\hat{E_2})$.
Values for a given energy $\hat{E_2}$ are obtained by linear interpolation of the tabulated values.

The spectral response $R_{\rm D2}(E_2|\hat{E_2})$ is illustrated in the right panel of Fig.~\ref{fig:moduleresponse}
for a selected number of input energies $\hat{E_2}$.
The figure can be compared with Fig.~4b in \citet{diehl1992}, which presents the response
functions as implemented in COMPASS at the time of the CGRO launch for comparable energies.

\subsection{Response convolution}
\label{app:smearing}

The response is convolved with the Gaussian kernel
\begin{equation}
K(\varphi'_{\rm geo}|\varphi_{\rm geo},\theta) =
\frac{1}{\sigma(\varphi'_{\rm geo},\theta)\sqrt{2\pi}}
\exp \left( -\frac{1}{2} \left( \frac{\varphi_{\rm geo} - \varphi'_{\rm geo}}{\sigma(\varphi'_{\rm geo},\theta)} \right)^2 \right)
\end{equation}
to take into account the location uncertainties of the events in the D1 and D2 modules.
The width $\sigma(\varphi'_{\rm geo},\theta)$ of the Gaussian kernel is a function of the scatter angle
$\varphi'_{\rm geo}$ and the zenith angle $\theta$ of the incoming photon, measured with respect
to the COMPTEL pointing axis.
The width
\begin{equation}
\sigma(\varphi'_{\rm geo}\theta) = \sqrt{\sigma_{\rm xy}^2(\varphi'_{\rm geo},\theta) + \sigma_{\rm z}^2(\varphi'_{\rm geo},\theta)}
\end{equation}
is the squared mean of the uncertainty
\begin{equation}
\sigma_{\rm xy}^2(\varphi'_{\rm geo},\theta) =
\frac{\left(\sigma_{\rm x}^2 + \sigma_{\rm y}^2\right)}{h^2}  \, f_{\rm xy}(\varphi'_{\rm geo},\theta)
\end{equation}
in the horizontal direction and the uncertainty
\begin{equation}
\sigma_{\rm z}^2(\varphi'_{\rm geo},\theta) =
\frac{\left(\sigma_{\rm z1}^2 + \sigma_{\rm z2}^2\right)}{h^2}  \, f_{\rm z}(\varphi'_{\rm geo},\theta)
\end{equation}
in the vertical direction, where
$\sigma_{\rm x}=2.3$ cm and $\sigma_{\rm y}=1.96$ cm is the average horizontal location spread in a D1 module with
respect to the D2 module in $x$ and $y$ direction, respectively,
$\sigma_{\rm z1}=2.45$ cm and $\sigma_{\rm z2}=2.17$ cm is the vertical location spread in a D1 and D2 module,
respectively, assuming a homogenous event distribution in the z-direction, and
$h=158$ cm is the distance between the D1 and D2 modules.
The functions
\begin{equation}
\begin{split}
f_{\rm xy}(\varphi'_{\rm geo},\theta) = 
& \cos^2\theta \cos^2 \varphi'_{\rm geo} (4 \sin^2\theta \sin^2 \varphi'_{\rm geo} + \cos^2 \varphi'_{\rm geo} - \sin^2\theta) \, + \\
& \frac{1}{2} \sin^2\theta \, \times \\
& \left( \cos^2\theta \cos^2 \varphi'_{\rm geo} +
\sin^2\theta \sin^2 \varphi'_{\rm geo} \left(1-\frac{3}{4} \cos^2 \varphi'_{\rm geo} \right) \right)
\end{split}
\end{equation}
and
\begin{equation}
\begin{split}
f_{\rm z}(\varphi'_{\rm geo},\theta) =
& \cos^4\theta \sin^2 \varphi'_{\rm geo} \cos^2 \varphi'_{\rm geo} + \\
& \cos^2\theta \sin^2\theta \left( \frac{1}{2} - 3 \sin^2 \varphi'_{\rm geo} \cos^2 \varphi'_{\rm geo}  \right) + \\
& \frac{3}{8} \sin^4\theta \sin^2 \varphi'_{\rm geo} \cos^2 \varphi'_{\rm geo}
\end{split}
\end{equation}
are the geometrical relations that transform from the uncertainties within the modules to uncertainties in
$\varphi'_{\rm geo}$ for a given zenith angle $\theta$.

\begin{figure}[!t]
\centering
\includegraphics[width=8.8cm]{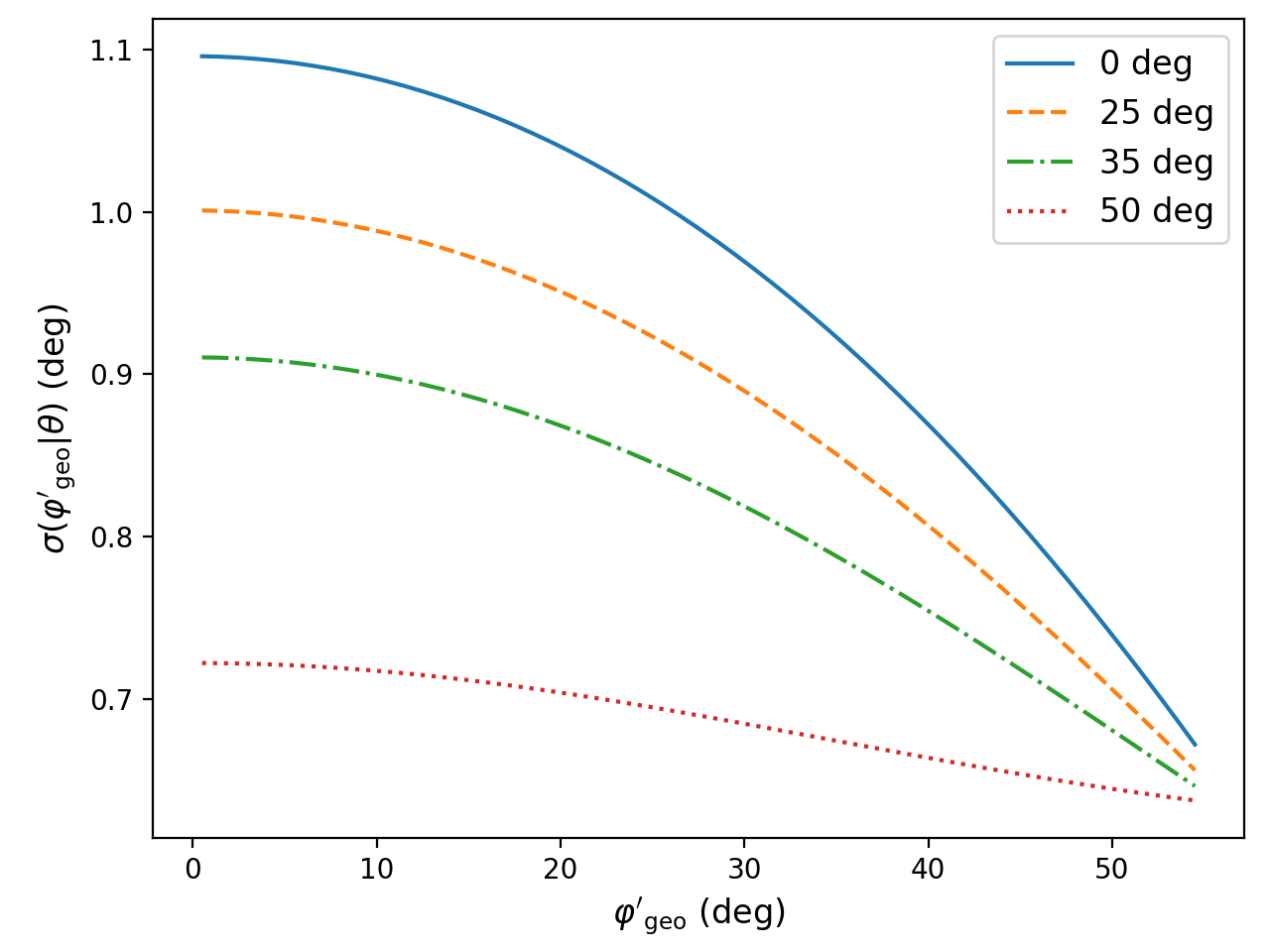}
\caption{
Width $\sigma(\varphi'_{\rm geo},\theta)$ of the Gaussian smoothing kernel as a function of
$\varphi'_{\rm geo}$ for zenith angles $\theta$ of $0\degrees$ (solid blue), $25\degrees$ (dashed orange),
$35\degrees$ (dashed-dotted green), and $50\degrees$ (dotted purple).
\label{fig:siggeo}
}
\end{figure}

Since COMPTEL response functions have no explicit zenith angle dependence, the computations are performed
for an average zenith angle of $\theta=25\degrees$.
The uncertainty introduced by this approximation is illustrated in Fig.~\ref{fig:siggeo}, and amounts to
less than $10\%$ for zenith angle $\theta<35\degrees$.

\section{Carbon footprint of this research}
\label{app:footprint}

\begin{table*}[!t]
\caption{Carbon footprint estimate by emission source of the research work behind this paper.
\label{tab:footprint}}
\centering
\begin{tabular}{c c c}
\hline\hline
Source & Emissions kgCO$_2$e & Comments \\
\hline
Electricity to power office building & $525 \pm 53$ & \\
Heating of office building & $422 \pm 118$ & \\
Water usage in office building & $8 \pm 2$ & \\
Air conditioning in office building & $80 \pm 23$ & \\
Waste management & $209 \pm 106$ & \\
Computing infrastructure & $400 \pm 200$ & Evaluated using {\tt csfootprint} \\
Data flow & $8 \pm 8$ & Includes videoconferencing \\ 
Lunch meals & $171 \pm 103$ & 10\% classical, 10\% flexitarian, 80\% vegetarian  \\
Home-to-office commuting & $89 \pm 36$ & $90\%$ cycling, $5\%$ public transport, $5\%$ car \\
\hline
Total & $1911 \pm 284$ & \\ 
\hline
\end{tabular}
\tablefoot{The footprint is based on a total work time estimate of one FTE or 220 working days.}
\end{table*}

Given the repeated alarms of the scientific community on global warming \citep{ipcc2021, ipcc2022},
biodiversity loss \citep{ipbes2019} and air, water and soil pollution \citep{landrigan2017}, the
environmental impact of any human activity, including obviously research, can no longer be
ignored.
Here we assess the GHG emissions related to our research work, including
contributions from office use, research activities and research environment.
We based our assessment on the carbon footprint estimate of the 
{\em Institut de Recherche en astrophysique et plan\'etologie} \citep[IRAP;][]{martin2022} where most 
of the work for this research was done.
To attribute a share of the total footprint to our work, we estimated that roughly one full time equivalent (FTE) was necessary to accomplish this research, corresponding to 220 working
days.
Consequently, we use the annual emissions per source from \citet{martin2022} divided by
the IRAP staff of 263 persons to assess the contributions from the various sources of
GHG emissions.

The results of our assessment are summarised in Table \ref{tab:footprint}.
We estimate the total carbon footprint of our research work to be $1.9 \pm 0.3$ tCO$_2$e,
composed of $1.2 \pm 0.2$ tCO$_2$e (65\%) for running the office building in which the work
was done, $0.4 \pm 0.2$ tCO$_2$e (21\%) for the computing infrastructure and data flow, and
$0.3 \pm 0.1$ tCO$_2$e (14\%) for lunch meals and home-to-office commuting.

We based our assessment for computing on the carbon footprint estimation implemented in 
ctools, yet when we started this research the relevant functionality was not yet available.
We note, however, that the carbon footprint estimate for the final analyses for this publication 
amounted to 60 kgCO$_2$e, of which 20 kgCO$_2$e were attributed to power consumption 
and 40 kgCO$_2$e to the computing infrastructure.
Making the conservative estimate that ten analysis iterations were necessary for this research
work we obtain a total carbon footprint due to computing of 600 kgCO$_2$e, of which 
200 kgCO$_2$e were due to power consumption and 400 kgCO$_2$e due to the computing 
infrastructure.
To avoid double counting, we disregarded the contribution of power consumption since it is
already accounted for in the electricity to power the office building.

For emission from lunch meals and home-to-office commuting we took the specific
habits of the co-authors weighted by their contribution to the research into account.
We note that our research work did not require any professional travelling.
The footprint for a couple of video conferences that we organised for this research is 
included under the `Data flow' source.

\end{appendix}

\end{document}